\documentclass[12pt]{iopart}

\usepackage{iopams}
\usepackage{graphicx}
\usepackage{subfig}

\begin{document}

\title[Steady-state quantum correlations]{Steady-state quantum correlations of two driven qubits collectively interacting with a vacuum reservoir}

\author{L O Casta\~nos}

\address{Instituto de F\'{\i}sica, Universidad Nacional Aut\'onoma de M\'exico, Apdo. Postal 20-364, M\'exico 01000, M\'exico}
\ead{loccj@yahoo.com}

\begin{abstract}
We consider two two-level atoms fixed at different positions, driven by a monochromatic laser field, and interacting collectively with the vacuum electromagnetic field. A Born-Markov-secular master equation is used to describe the dynamics of the two atoms and their steady-state is obtained analytically for two configurations, one in which the atoms are in equivalent positions and another in which they are not. The steady-state populations of the energy levels of the free atoms, entanglement, quantum discord and degree of mixed-ness are calculated analytically as a function of the laser field intensity and the distance between the two atoms. It is found that driving both atoms with the laser field is inefficient for the generation of steady-state correlations when they are in equivalent positions. On the contrary, inequivalent positions lead to the possibility of considerable steady-state entanglement and left/right quantum discord. It is shown that an $X$-state can be obtained for high laser field intensities for both configurations. This allows the comparison of two measures of quantum discord. The behaviour and relationships between correlations are studied and several limiting cases are investigated.
\end{abstract}


\submitto{\JPA}

\maketitle

\section{Introduction}

For the past two decades there has been an ever growing interest in quantifying and characterizing the correlations in the states of quantum systems. This was initially provoked by the realization that quantum correlations known as entanglement could be harnessed as a resource to bring significant advantage for computing and information processing \cite{benenti}. Now, one of the main interests is to identify which correlations are responsible for these advantages. Moreover, it has been recognized that quantifying correlations in quantum systems is a difficult task and, in general, several measures are needed to capture all of their subtleties.

In order to understand correlations, one generally starts with bipartite systems. Consider two quantum systems $\mathcal{A}$ and $\mathcal{B}$. The state of the composite system $\mathcal{A}+ \mathcal{B}$ can be described by a density operator $\rho_{\mathcal{AB}}$ which contains both classical and quantum correlations. A widely accepted measure of the total correlations in $\rho_{\mathcal{AB}}$ is the \textit{quantum mutual information}. It is defined as
\begin{equation}
\label{QMI}
I(\rho_{\mathcal{AB}}) \ = \ S(\rho_{\mathcal{A}}) + S(\rho_{\mathcal{B}}) - S(\rho_{\mathcal{AB}}) \ ,
\end{equation}
where $\rho_{\mathcal{A}}$ ($\rho_{\mathcal{B}}$) is the density operator of $\mathcal{A}$ ($\mathcal{B}$) and $S(\rho) = -\mbox{Tr}[\rho \mbox{log}_{2}(\rho) ]$ is the von Neumann entropy of $\rho$  (the trace is taken in the state space where $\rho$ is a density operator). It has been found that $I(\rho_{\mathcal{AB}})$ is the distance (as measured by the relative entropy) of $\rho_{\mathcal{AB}}$ to its closest product state $\rho_{\mathcal{A}}\otimes \rho_{\mathcal{B}}$ \cite{Modi}. It has also been shown that $I(\rho_{\mathcal{AB}})$ measures the asymptotically minimal amount of local noise one has to add to turn $\rho_{\mathcal{AB}}$ into a product state \cite{Winter}. Finally, $I(\rho_{\mathcal{AB}})$ is the maximum amount of information that Alice can send secretly to Bob using $\rho_{\mathcal{AB}}$ as a one-time pad \cite{Schumacher}.

Once the total correlations in $\rho_{\mathcal{AB}}$ have been quantified, it is natural to ask whether these can be clearly divided into a classical and quantum part. Several axioms have been proposed as requirements of a measure of the classical correlations $C^{cl}(\rho_{\mathcal{AB}})$ in $\rho_{\mathcal{AB}}$ \cite{VedralC}. These consist in being zero for product states, being invariant under local unitary transformations and non-increasing under local operations, and being equal to $S(\rho_{\mathcal{A}}) = S(\rho_{\mathcal{B}})$ for pure states $\rho_{\mathcal{AB}}$. Also, it has been  pointed out that a measure of the classical correlations should quantify the correlation between $\mathcal{A}$ and $\mathcal{B}$ instead of a property of only one of them, that is, it should be symmetric under interchange of $\mathcal{A}$ and $\mathcal{B}$. Taking these axioms as a basis, the following measures for the classical correlations in $\rho_{\mathcal{AB}}$ have been proposed \cite{VedralC}:
\begin{eqnarray}
\label{CorrelacionClasica}
C_{\mathcal{B}}^{cl}(\rho_{\mathcal{AB}}) \ = \ S(\rho_{\mathcal{A}}) - \mbox{min}_{ \{ B_{i} \} } \ \sum_{i}p_{i}S(\rho_{\mathcal{A}}^{i}) \ , \cr
C_{\mathcal{A}}^{cl}(\rho_{\mathcal{AB}}) \ = \ S(\rho_{\mathcal{B}}) - \mbox{min}_{ \{ A_{j} \} } \ \sum_{j}p_{j}'S(\rho_{\mathcal{B}}^{j}) \ ,
\end{eqnarray}
where the minimum is taken over all sets $\{ B_{i} \}$  ($\{ A_{j} \}$)  of one-dimensional orthogonal projectors that sum up to the identity and that constitute measurements performed only on $\mathcal{B}$ ($\mathcal{A}$). Also,
\begin{equation}
\rho_{\mathcal{A}}^{i} \ = \ \frac{1}{p_{i}}\mbox{Tr}_{\mathcal{B}}\left[ (\mathbb{I}\otimes B_{i})\rho_{\mathcal{AB}}(\mathbb{I}\otimes B_{i}^{\dagger}) \right] \ ,
\end{equation} 
with 
\begin{equation}
p_{i} \ = \ \mbox{Tr}_{\mathcal{AB}}\left[ \ (\mathbb{I}\otimes B_{i})\rho_{\mathcal{AB}}(\mathbb{I}\otimes B_{i}^{\dagger}) \ \right] \ ,
\end{equation}
is the density operator of $\mathcal{A}$ after obtaining the result associated with $B_{i}$ in a measurement of $\mathcal{B}$ (similar equations hold for $\rho_{\mathcal{B}}^{j}$ and $p_{j}'$). These measures satisfy the axioms mentioned above and are equal to zero if and only if $\rho_{\mathcal{AB}} = \rho_{\mathcal{A}}\otimes\rho_{\mathcal{B}}$. Hence, it is in accordance with the accepted idea that only product states are devoid of correlations. Nevertheless, these measures are dependent on which system is measured, that is, $C_{\mathcal{A}}^{cl}(\rho_{\mathcal{AB}}) \not= C_{\mathcal{B}}^{cl}(\rho_{\mathcal{AB}})$ in general. Thus it depends on the properties of each subsystem.

More recently, other measures and notions of classical correlations have been proposed \cite{Modi}, \cite{Winter}. In particular, it has been proposed that the classical correlations should be quantified by the minimum distance (as measured by the relative entropy) between a classically correlated state associated with $\rho_{\mathcal{AB}}$ and a product state \cite{Modi}. This last proposal has the advantage of placing all correlations (total, quantum, and classical) on an equal footing that allows a direct comparison between all of them \cite{Modi}.

Given that the classical part of the correlations is not yet clearly quantified, let us now turn to quantum correlations. Quantum states $\rho_{\mathcal{AB}}$ are normally divided into separable or entangled. Let us remember that $\rho_{\mathcal{AB}}$ is separable if it can be expressed in the form
\begin{equation}
\rho_{\mathcal{AB}} \ = \ \sum_{j} p_{j} \rho_{\mathcal{A},j} \otimes \rho_{\mathcal{B},j} \ ,
\end{equation}
with $\rho_{\mathcal{A},j}$ ($\rho_{\mathcal{B},j}$) density operators of $\mathcal{A}$ ($\mathcal{B}$), and $p_{j} \in[0,1]$ such that $\sum_{j}p_{j} = 1$. If $\rho_{\mathcal{AB}}$ is not a separable state, then it is an entangled one. It was thought that entanglement embodied all the quantum correlations in $\rho_{\mathcal{AB}}$, and that separable states were purely classical. Nevertheless, it has been realized that entanglement is not the only aspect of quantum correlations, since some separable states may still present non-classical correlations \cite{Modi}-\cite{Luo}. One way to measure the non-classicality of the correlations in $\rho_{\mathcal{AB}}$ is to use the quantum discord \cite{Zurek}. This quantity is defined to be the difference between the quantum mutual information (\ref{QMI}) and the  correlations (\ref{CorrelacionClasica}):
\begin{eqnarray}
\label{QD}
D^{Q}_{\mathcal{A}}(\rho_{\mathcal{AB}}) \ = \ I(\rho_{\mathcal{AB}}) - C_{\mathcal{A}}^{cl}(\rho_{\mathcal{AB}}) \ , \cr
D^{Q}_{\mathcal{B}}(\rho_{\mathcal{AB}}) \ = \ I(\rho_{\mathcal{AB}}) - C_{\mathcal{B}}^{cl}(\rho_{\mathcal{AB}}) \ ,
\end{eqnarray}
$D^{Q}_{\mathcal{A}}(\rho_{\mathcal{AB}})$ ($D^{Q}_{\mathcal{B}}(\rho_{\mathcal{AB}})$) is usually referred to as the left (right) quantum discord. It has been shown that $D_{\mathcal{A}}^{Q}(\rho_{\mathcal{AB}}), \ D_{\mathcal{B}}^{Q}(\rho_{\mathcal{AB}})$ are always non-negative \cite{Zurek}. In fact, $D_{\mathcal{A}}^{Q}(\rho_{\mathcal{AB}}) = 0$ if and only if
\begin{equation}
\label{Discordia1}
\rho = \sum_{k} p_{k}|\psi_{k}\rangle\langle \psi_{k}|\otimes \rho_{k} \ ,
\end{equation}
where $\{ |\psi_{k}\rangle \}$ is an orthonormal basis of $\mathcal{A}$, $\rho_{k}$ are density operators of $\mathcal{B}$, and $p_{k}$ are non-negative numbers such that $\sum_{k}p_{k} = 1$ \cite{Zurek}. Similarly, $D^{Q}_{\mathcal{B}}(\rho_{\mathcal{AB}}) = 0$ if and only if
\begin{equation}
\label{Discordia2}
\rho = \sum_{k} p_{k}\rho_{k} \otimes |\psi_{k}\rangle\langle \psi_{k}| \ ,
\end{equation}
where $\{ |\psi_{k}\rangle \}$ is an orthonormal basis of $\mathcal{B}$, $\rho_{k}$ are density operators of $\mathcal{A}$, and $p_{k}$ are non-negative numbers such that $\sum_{k}p_{k} = 1$.

In general $D^{Q}_{\mathcal{A}}(\rho_{\mathcal{AB}}) \not= D^{Q}_{\mathcal{B}}(\rho_{\mathcal{AB}})$. The quantum discord $D^{Q}_{\mathcal{B}}(\rho_{\mathcal{AB}})$ can be interpreted to be a measure of the information of $\mathcal{A}$ in $\rho_{\mathcal{AB}}$ that cannot be extracted by performing measurements only on $\mathcal{B}$ \cite{Zurek}. Hence, if $D^{Q}_{\mathcal{B}}(\rho_{\mathcal{AB}})$ is large, a lot of information is destroyed by any measurement on $\mathcal{B}$ alone; if $D^{Q}_{\mathcal{B}}(\rho_{\mathcal{AB}})$ is small, almost all the information about $\mathcal{A}$ in $\rho_{\mathcal{AB}}$ can be obtained by measurements only on $\mathcal{B}$ \cite{Zurek}.

Once quantum discord is recognized as a measure of non-classical correlations, it is natural to ask if it
measures the same correlations as entanglement. Given that some separable mixed states have non-zero quantum discord and separable states by definition do not have entanglement, it is concluded that entanglement and quantum discord are in general different quantities. Nevertheless, they do coincide when $\rho_{\mathcal{AB}}$ is a pure state or a mixture of Bell states \cite{Rau}. Furthermore, it has been found that in general the quantum discord of $\rho_{\mathcal{AB}}$ is not simply the sum of some measure of the entanglement in $\rho_{\mathcal{AB}}$ and some other non-classical correlation \cite{Rau}, \cite{Luo}.

The study of the dynamics of quantum correlations is intimately linked to the study of open quantum systems, that is, quantum systems which are coupled to other quantum systems called environment \cite{Breuer}. When the environment has many degrees of freedom (a reservoir), it normally induces a loss of coherence in the system of interest that ultimately has a detrimental effect on the correlations present in it. For example, entanglement may decay exponentially with time or may vanish completely in a finite time (called \textit{entanglement sudden death}) \cite{SuddenDeath}-\cite{chinos2}. Moreover, the environment can also affect the degree of mixed-ness of the state of the system of interest \cite{Breuer}, and this in turn also affects the entanglement since it is known that a mixed state cannot have an arbitrary degree of entanglement \cite{MaxEntanglement}.

Recently, the open system dynamics of both the quantum discord and classical correlations have started to be studied \cite{Maziero}-\cite{nonMarkovian}. In particular it has been found that under Markovian environments quantum discord may be more robust than entanglement \cite{Boas}. For non-Markovian environments
and independent reservoirs for each part of a bipartite system, the quantum discord may vanish only at discrete instants whereas the entanglement can disappear during a finite time interval. For a common reservoir, quantum discord and entanglement can behave very differently with sudden birth of the former but not of the latter \cite{nonMarkovian}.

It is the purpose of this article to study the entanglement, quantum discord, classical correlations, and degree of mixed-ness in the steady-state of the following open quantum system: two two-level atoms (qubits in the jargon of quantum information) at fixed positions driven by a monochromatic laser field and interacting collectively with all the modes of the quantum electromagnetic field. The latter will be assumed to be in the vacuum state. In the notation used above, $\mathcal{A}$ will be one of the atoms (say, the atom at position $\mathbf{r}_{1}$) and $\mathcal{B}$ will be the other atom (say, the atom at position $\mathbf{r}_{2}$). Here we also have a third party $\mathcal{C}$ playing a decisive role in the dynamics: the vacuum electromagnetic field which we will consider as a reservoir. The system is open because we are interested in studying only the subsystem $\mathcal{A}+\mathcal{B}$ of the complete system $\mathcal{A}+\mathcal{B}+\mathcal{C}$. Since the two atoms are interacting collectively with the reservoir, $\mathcal{C}$ acts as a medium that can allow quantum correlations to be formed between the two atoms for some time. In order that these correlations have the possibility of  being long-lived and not being ultimately destroyed by the reservoir, the two atoms will be driven by a laser field.

We will investigate in detail two configurations for the two atoms. In the first, one of the atoms will be placed in a position where the laser electric field is zero, while the other atom will located at a position where it is not zero. In the second, each atom will be placed in a position where the driving electric field takes the same value. These two configurations will allow us to study how the correlations change when the atoms are in equivalent/non-equivalent positions and to determine what is the effect when both atoms are being driven by the laser field. It is important to mention that we will neglect the coherent dipole-dipole interaction between the two atoms that results of the collective interaction with the reservoir. We will only keep the dissipative collective interaction with the reservoir. This will allow us to calculate a steady-state density operator of the two atoms analytically, as well as, all the aforementioned correlations in it. Furthermore, this will allow us to identify which effects are due solely to the dissipative interaction. In future work we will add the dipole-dipole interaction.

Using the same system as ours, but without the driving field, the dynamical generation of entanglement between two atoms due to the collective interaction with the reservoir has been recently studied \cite{hindu},\cite{Davidovich}. It was found that, as a result of the interaction through the reservoir, the system develops non-negligible entanglement (as measured by the concurrence) for a period of time if the atoms are close enough and if initially one is in the excited state and the other is in the ground state. Also, it was reported that entanglement initially present in the system of two atoms is more robust when the two atoms are close when compared to the case where the atoms are far apart.
The entanglement present in a system of two driven non-identical two-level atoms in a special configuration (namely one atom is located at a node while the other atom is placed at an antinode of a driving laser field with a standing wave cosine structure) has also been studied \cite{dissipation}. The steady-state density operator of the system was obtained numerically and the concurrence was evaluated. It was found that the two atom system decays to a stationary entangled state only when the Rabi frequency equals the difference between the two atoms' transition frequencies.

The present article is organized as follows. In Section II we summarize some results on measures of correlations for two qubit systems. In Section III the system of interest is described and the master equation governing the dynamics of the two atoms is established. In Section IV, the steady-state density operator is calculated analytically for the case of only one atom being driven by the laser field. The populations of the eigenstates of the free Hamiltonian of the two atoms, the degree of entanglement, the quantum discord, and the degree of mixed-ness of the two atoms are evaluated as functions of the laser field intensity and of the distance between the atoms. In Section V, the steady-state density operator of the two atoms is calculated analytically for the case of equal coupling strengths. The same analysis of Section IV is done for this new configuration. Furthermore, the results of both configurations are compared. The conclusions are given in Section VI.

\section{Quantifying correlations in two qubit systems}

In this article we will be considering two two-level atoms (qubits) which we will number by $1$ and $2$. In the following sections these labels will correspond to the atom at position $\mathbf{r}_{1}$ and to the atom at position $\mathbf{r}_{2}$, respectively. In terms of the notation of the Introduction, $1$ will replace $\mathcal{A}$, while $2$ will replace $\mathcal{B}$. We will now show how to calculate all correlations by simple formulas. We will only give the algorithms and refer the interested reader to the original articles for the proofs.

The kets $|j:+\rangle$ \ and \ $|j:-\rangle$ \ will denote the excited and ground states of the $j$th atom ($j=1,2$), respectively. In the following we will be making constant use of the triplet-singlet basis
\begin{eqnarray}
\mathsf{B}  \ = \  \left\{ \ |1,1\rangle, \ |1,0\rangle, \ |1,-1\rangle, \ |0,0\rangle \ \right\} \ ,
\end{eqnarray}
for the state space of the two atoms:
\begin{eqnarray}
\label{acoplada}
 |1,1 \rangle = |1:+\rangle \otimes |2:+\rangle \ , \cr
 |1,0 \rangle = \frac{1}{\sqrt{2}} \left( \ |1:+\rangle \otimes |2:-\rangle + |1:-\rangle \otimes |2:+\rangle \ \right) \ , \cr
 |1,-1 \rangle = |1:-\rangle \otimes |2:-\rangle \ , \cr
 |0,0 \rangle = \frac{1}{\sqrt{2}} \left( \ |1:+\rangle \otimes |2:-\rangle - |1:-\rangle \otimes |2:+\rangle \ \right) \ .
\end{eqnarray}
We shall also use the usual tensor product basis
\begin{eqnarray}
\mathsf{B}'  \ = \ \left\{ \ |+,+\rangle, \ |+,-\rangle, \ |-,+\rangle, \ |-,-\rangle \ \right\} \ ,
\end{eqnarray}
where
\begin{eqnarray}
\label{desacoplada}
 |+,+ \rangle = |1:+\rangle \otimes |2:+\rangle \ , \cr
 |+,- \rangle = |1:+\rangle \otimes |2:-\rangle  \ , \cr
 |-,+ \rangle = |1:-\rangle \otimes |2:+\rangle \ , \cr
 |-,- \rangle = |1:-\rangle \otimes |2:-\rangle \ .
\end{eqnarray}
Furthermore, we will denote the density operator of the two atoms by $\rho_{12}$.

We are interested in quantifying the degree of entanglement of the system of two atoms. We will use the concurrence $C$, which can be calculated as \cite{Plenio}:
\begin{equation}
 \label{concurrencia}
C = \mbox{max}\{ 0, \sqrt{\lambda_{1}} - \sqrt{\lambda_{2}} -\sqrt{\lambda_{3}} -\sqrt{\lambda_{4}} \} \ ,
\end{equation}
where $\lambda_{1} \geq \lambda_{2} \geq \lambda_{3} \geq \lambda_{4}$ are the eigenvalues of the matrix
\begin{equation}
\label{matrizConcurrencia}
\mathbb{M} \ = \ [ \rho_{12} ]_{\mathsf{B}'}( \sigma_{y} \otimes \sigma_{y} ) [ \rho_{12} ]_{\mathsf{B}'}^{*}  ( \sigma_{y} \otimes \sigma_{y} ) \ .
\end{equation}
Here $\sigma_{y}$ is the well-known Pauli matrix, $[ \rho_{12} ]_{\mathsf{B}'}$ is the matrix representation of $\rho_{12}$ in the basis $\mathsf{B}'$, and $[ \rho_{12} ]_{\mathsf{B}'}^{*}$ is the element-wise complex conjugate of the density matrix $[ \rho_{12} ]_{\mathsf{B}'}$. Note that any basis of the state space of the two atoms related to the tensor product basis by an orthogonal change of coordinates matrix can be used for the matrix representations (for example, $\mathsf{B}$ could be used instead of $\mathsf{B}'$). The concurrence takes values between $0$ and $1$. It is $1$ when the atoms are in a maximally entangled state, while it is zero when the atoms are in a separable state.

Evaluation of the quantum discord given by (\ref{QD}) in general requires considerable numerical minimization. Although a method to calculate easily the classical correlation and quantum discord for a general two-qubit $X$-state has been proposed \cite{Rau}, it has been found to be unreliable because it does not take into account all of the restrictions in the minimization problem \cite{FallaRau}. Hence, we will show here how to calculate numerically the right quantum discord.

The first step is to express $\rho_{12}$ as a linear combination of Pauli operators and tensor products of Pauli operators:
\begin{equation}
\label{ExpansionRho12}
\rho_{12} \ = \ \frac{1}{4}\left( \ \mathbb{I} + \mathbf{x}\cdot\bar{\sigma}_{1} + \mathbf{y}\cdot\bar{\sigma}_{2} + \sum_{i,j=1}^{3} T_{ij} \sigma_{1i}\otimes\sigma_{2j} \ \right) \ .
\end{equation}
Here $\mathbb{I}$ is the identity operator, $\bar{\sigma}_{j}$ is the vector of Pauli operators of atom $j$ and $\sigma_{1i}\otimes\sigma_{2j}$ is the tensor product of the Pauli operator $\sigma_{1i}$ of atom $1$ and the Pauli operator $\sigma_{2j}$ of atom $2$. Notice that $\mathbf{x}$ is the Bloch vector of the density operator $\rho_{1} = \mbox{Tr}_{2}[\rho_{12}]$ of atom $1$, while $\mathbf{y}$ is the Bloch vector of the density operator $\rho_{2} = \mbox{Tr}_{1}[\rho_{12}]$ of atom $2$.

The next step is to see what is the structure of a one-dimensional orthogonal projector $P_{0}$ in the state space of atom $2$:
\begin{equation}
\label{P0}
\left[ \ P_{0} \ \right]_{\mathsf{B}_{2}}
\ = \
\left(
\begin{array}{cc}
a              &    be^{-i\phi} \\
be^{i\phi}     &    1-a
\end{array}
\right) \ ,
\end{equation}
where $\mathsf{B}_{2} = \{ |2:+\rangle , \ |2:-\rangle  \}$, $[ P_{0} ]_{\mathsf{B}_{2}}$ is the matrix representation of $P_{0}$ with respect to the basis $\mathsf{B}_{2}$ of the state space of atom $2$, \ $a\in [0,1]$, \ $\phi \in [0, 2\pi )$ and $b = \sqrt{a(1-a)}$.

With (\ref{P0}) at hand, we notice that 
\begin{eqnarray}
\label{P0SigmaP0}
P_{0} \sigma_{x} P_{0} &=& 2 b \cdot \mbox{cos}(\phi) P_{0}  \ , \cr
P_{0} \sigma_{y} P_{0}  &=& 2 b \cdot \mbox{sin}(\phi) P_{0}  \ , \cr
P_{0} \sigma_{z} P_{0}  &=& (2a-1) \cdot P_{0}  \ .
\end{eqnarray}
We now define
\begin{equation}
\label{gammaC}
\bar{\gamma} \ \equiv \ \left( \gamma_{1}, \ \gamma_{2} , \ \gamma_{3}  \right) \ \equiv \ \left( 2b\cdot \mbox{cos}(\phi), \ 2b\cdot \mbox{sin}(\phi) , \ 2a -1  \right) \ .
\end{equation}

Using (\ref{ExpansionRho12}) and (\ref{P0SigmaP0}) we now see how $\rho_{12}$ transforms when a measurement on atom $2$ is performed and the result associated with $P_{0}$ is obtained: 
\begin{eqnarray}
\label{PoRhoPo}
\frac{1}{p_{0}} (\mathbb{I}\otimes P_{0})\rho_{12} (\mathbb{I}\otimes P_{0})
&=& \frac{1}{4p_{0}}\left( \ \mu \mathbb{I} \ + \ \sum_{i=1}^{3}\nu_{i}\sigma_{i} \ \right)\otimes P_{0}  \ , \\
&=& \rho_{P_{0}} \otimes P_{0} \ , 
\end{eqnarray}
where
\begin{eqnarray}
p_{0} &\equiv& \mbox{Tr}\left[ \ (\mathbb{I}\otimes P_{0})\rho_{12} (\mathbb{I}\otimes P_{0}) \ \right] \ = \ \frac{1}{2}\mu \ .
\end{eqnarray}
Here we have introduced the quantities
\begin{eqnarray}
\label{definicionesPo}
\mu \ \equiv \ 1 + \mathbf{y}\cdot\bar{\gamma} , \qquad \nu_{i} \ \equiv \ x_{i} + \sum_{j=1}^{3}\mathbb{T}_{ij}\gamma_{j} , \qquad \bar{\nu} \ \equiv \ \left( \nu_{1}, \nu_{2}, \nu_{3} \right) \ .
\end{eqnarray}
Note that $\rho_{P_{0}}$ is a density operator of atom $1$. Since $P_{0}$ is an orthogonal projector in the state space of atom $2$ it follows that 
\begin{eqnarray}
\label{azarosoNotar2}
\rho_{P_{0}} \ = \ \mbox{Tr}_{2}\left\{ \ \frac{1}{p_{0}}(\mathbb{I}\otimes P_{0})\rho_{12} (\mathbb{I}\otimes P_{0}) \ \right\}  \ .
\end{eqnarray}

We now calculate the von Neumann entropy of $\rho_{P_{0}}$
\begin{eqnarray}
\label{SPoRhoPo}
S\left( \ \rho_{P_{0}} \right)
&=& -\frac{1}{2}\left( 1 + \frac{|\bar{\nu}|}{\mu} \right) \mbox{log}_{2}\left[ \ \frac{1}{2}\left( \ 1 + \frac{|\bar{\nu}|}{\mu} \ \right) \ \right] \ \cr
&& \cr
&& -\frac{1}{2}\left( 1 - \frac{|\bar{\nu}|}{\mu} \right) \mbox{log}_{2}\left[ \ \frac{1}{2}\left( \ 1 - \frac{|\bar{\nu}|}{\mu} \ \right) \ \right] \ \ .
\end{eqnarray}

With results (\ref{ExpansionRho12})-(\ref{SPoRhoPo}) at hand we are all set to determine a formula for the right classical correlations $C_{2}^{cl}(\rho_{12})$. Let $\{ |\psi\rangle , |\phi\rangle \}$ be a complete set of one-dimensional orthogonal projectors (that is, an orthogonal basis) for the state space of atom $2$. Using the basis $\mathsf{B}_{2} = \{ |2:+\rangle, \ |2:-\rangle  \}$ it is straightfoward to show that the matrix representations of $|\psi\rangle$ and $|\phi\rangle$ have the form
\begin{eqnarray}
\label{MatPsiPhiCC}
\left[ \ |\psi\rangle\langle\psi | \ \right]_{\mathsf{B}_{2}}
&=&
\left(
\begin{array}{cc}
|\alpha|^{2}                                 &    |\alpha|\sqrt{1 - |\alpha|^{2}}e^{-i\Phi} \cr
|\alpha|\sqrt{1 - |\alpha|^{2}}e^{i\Phi}     &    1-|\alpha|^{2}
\end{array}
\right) \ , \cr
&& \cr
&& \cr
\left[ \ |\phi\rangle\langle\phi | \ \right]_{\mathsf{B}_{2}}
&=&
\left(
\begin{array}{cc}
1-|\alpha|^{2}                                &    -|\alpha|\sqrt{1 - |\alpha|^{2}}e^{-i\Phi} \cr
-|\alpha|\sqrt{1 - |\alpha|^{2}}e^{i\Phi}     &    |\alpha|^{2}
\end{array}
\right) \ .
\end{eqnarray}
where $|\alpha| \in [0,1]$, \ $\Phi \in [0, 2\pi )$. Notice that if we take $a=|\alpha|^{2}$ and $\phi = \Phi$ in all the results for $P_{0}$ we will get the corresponding results for $|\psi\rangle\langle\psi |$. Similarly, taking $a=1-|\alpha|^{2}$ and $\phi = \Phi +\pi$ we get the corresponding results for $|\psi\rangle\langle\psi |$. If we define $\rho_{\psi} = \rho_{P_{0}}$ when we identify $P_{0}$ with  $|\psi\rangle\langle\psi |$, and we take $\rho_{\phi} = \rho_{P_{0}}$ when we identify $P_{0}$ with  $|\phi\rangle\langle\phi |$, we find using (\ref{ExpansionRho12})-(\ref{SPoRhoPo}) that
\begin{eqnarray}
\label{entropiaCondicional2Qubits}
&& p_{\psi} S(\rho_{\psi}) \ + \ p_{\phi} S(\rho_{\phi}) \ , \cr
&& \cr
&=& -\frac{\mu_{+}}{4}\left( 1 + \frac{|\bar{\nu}_{+}|}{\mu_{+}} \right) \mbox{log}_{2}\left[ \ \frac{1}{2}\left( \ 1 + \frac{|\bar{\nu}_{+}|}{\mu_{+}} \ \right) \ \right] \cr
&& -\frac{\mu_{+}}{4}\left( 1 - \frac{|\bar{\nu}_{+}|}{\mu_{+}} \right) \mbox{log}_{2}\left[ \ \frac{1}{2}\left( \ 1 - \frac{|\bar{\nu}_{+}|}{\mu_{+}} \ \right) \ \right] \ \cr
&& -\frac{\mu_{-}}{4}\left( 1 + \frac{|\bar{\nu}_{-}|}{\mu_{-}} \right) \mbox{log}_{2}\left[ \ \frac{1}{2}\left( \ 1 + \frac{|\bar{\nu}_{-}|}{\mu_{-}} \ \right) \ \right] \cr
&& -\frac{\mu_{-}}{4}\left( 1 - \frac{|\bar{\nu}_{-}|}{\mu_{-}} \right) \mbox{log}_{2}\left[ \ \frac{1}{2}\left( \ 1 - \frac{|\bar{\nu}_{-}|}{\mu_{-}} \ \right) \ \right] \ , 
\end{eqnarray}
where we have defined
\begin{eqnarray}
\label{DefinicionesEntropiaCondicional}
&& \bar{\gamma}_{\psi} \ \equiv \ (\gamma_{\psi 1}, \ \gamma_{\psi 2}, \ \gamma_{\psi 3}) \ , \cr
&& \cr
&& \gamma_{\psi 1} \ \equiv \ 2\cdot\mbox{cos}(\Phi)|\alpha|\sqrt{1-|\alpha|^{2}} \ , \cr
&& \cr
&& \gamma_{\psi 2}  \ \equiv \ 2\cdot\mbox{sen}(\Phi)|\alpha|\sqrt{1-|\alpha|^{2}} \ , \cr
&& \cr
&& \gamma_{\psi 3} \ \equiv \ 2|\alpha|^{2}-1 \ , \cr
&& \cr
&& \bar{\nu}_{\pm} \ \equiv \ (\nu_{\pm 1}, \ \nu_{\pm 2}, \ \nu_{\pm 3}) \ , \cr
&& \cr
&& \nu_{\pm i} \ \equiv \ x_{i} \ \pm \  \mathbb{T}_{(i)} \cdot \bar{\gamma}_{\psi} \ , \cr
&& \cr
&& \mu_{\pm} \ \ \equiv \ 1 \ \pm \ \mathbf{y}\cdot \bar{\gamma}_{\psi} \ .
\end{eqnarray}
Also, $\mathbb{T}_{(i)}$ is the $i$-th row of the matrix $\mathbb{T}$ whose component $i,j$ is $T_{ij}$.

To calculate the classical correlation $C_{2}^{cl}(\rho_{12})$ we have to minimize (\ref{entropiaCondicional2Qubits}) over all complete sets of one-dimensional orthogonal projectors $\{ \ |\psi\rangle, |\phi\rangle \ \}$.  From (\ref{MatPsiPhiCC}) we note that the set $\{ \ |\psi\rangle , \ |\phi\rangle \ \}$ is determined by two real parameters: $|\alpha| \in [0,1]$, \ $\Phi \in [0, 2\pi )$. Then $C_{2}^{cl}(\rho_{12})$ can be calculated as follows:
\begin{eqnarray}
\label{CorrelacionClasica2Qubits}
C^{cl}_{2}(\rho_{12}) \ = \ S(\rho_{1}) \ - \ \mbox{min}_{|\alpha|\in[0,1], \ \Phi\in [0,2\pi]} \ \left[ \ p_{\psi} S(\rho_{\psi}) \ + \ p_{\phi} S(\rho_{\phi}) \ \right] \ ,
\end{eqnarray}
where $p_{\psi} S(\rho_{\psi}) \ + \ p_{\phi} S(\rho_{\phi})$ is defined in (\ref{entropiaCondicional2Qubits}). We note that the minimum always exists because the minimization process is done over a compact set. Also notice that $\Phi \in [0,2\pi )$ was extended to $\Phi \in [0,2\pi ]$. This was done to have a compact set and is of no consequence since we are interested in the absolute minimum of a periodic function of period $2\pi$. Also notice that once $C_{2}^{cl}(\rho_{12})$ has been calculated, the right quantum discord $D_{2}^{Q}(\rho_{12})$ follows easily by substracting $C_{2}^{cl}(\rho_{12})$ from the quantum mutual information $I(\rho_{12})$.

Now suppose that we make measurements only on the atom at position $\mathbf{r}_{1}$, that is, we want to calculate the left classical correlations $C_{1}^{cl}(\rho_{12})$. Instead of going through the whole process of calculating conditions analogous to those in (\ref{entropiaCondicional2Qubits})-(\ref{CorrelacionClasica2Qubits}), we only need to find the matrix representation of the density operator $\rho_{12}$ with respect to the basis $\mathsf{B}''$ in which the order of atoms $1$ and $2$ has been interchanged
\begin{eqnarray}
\label{BaseReordenada}
\mathsf{B}'' &=& \Big\{ \ |2:+\rangle\otimes |1:+\rangle , \ |2:+\rangle\otimes |1:-\rangle , \cr
&& \ |2:-\rangle\otimes |1:+\rangle , \ |2:-\rangle\otimes |1:-\rangle   \ \Big\} \ ,
\end{eqnarray}
and apply the procedure described above in (\ref{entropiaCondicional2Qubits})-(\ref{CorrelacionClasica2Qubits}) to obtain the quantum discord $D_{1}^{Q}(\rho_{12})$ and the classical correlations $C_{1}^{cl}(\rho_{12})$. It is easy to find such matrix representation. Suppose that the density operator $\rho_{12}$ has the following matrix representation with respect to $\mathsf{B}'$
\begin{equation}
\label{RhoO}
[\rho_{12}]_{\mathsf{B}'} \ = \
\left(
\begin{array}{cccc}
R_{11} & R_{12} & R_{13} & R_{14} \cr
R_{21} & R_{22} & R_{23} & R_{24} \cr
R_{31} & R_{32} & R_{33} & R_{34} \cr
R_{41} & R_{42} & R_{43} & R_{44}
\end{array}
\right) \ .
\end{equation}
Then the matrix representation of $\rho_{12}$ with respect to $\mathsf{B}''$ has the following form:
\begin{equation}
\label{RhoN}
[\rho_{12}]_{\mathsf{B}''} \ = \
\left(
\begin{array}{cccc}
R_{11} & R_{13} & R_{12} & R_{14} \cr
R_{31} & R_{33} & R_{32} & R_{34} \cr
R_{21} & R_{23} & R_{22} & R_{24} \cr
R_{41} & R_{43} & R_{42} & R_{44}
\end{array}
\right) \ .
\end{equation}
Note that $[\rho_{12}]_{\mathsf{B}''}$ is obtained by interchanging columns $2$ and $3$ and rows $2$ and $3$ of $[\rho_{12}]_{\mathsf{B}'}$.

For a general two-qubit density matrix the quantum discord has to be calculated numerically by a procedure such as the one described above. Therefore, several alternative measures have been proposed. One in particular calculates the distance of $\rho_{12}$ to the set of zero discord states $\Omega_{0}$ given by (\ref{Discordia1}) if measurements are made on atom $1$ or given by (\ref{Discordia2}) if measurements are carried out on atom $2$ \cite{VedralD}. It was found that \cite{VedralD}:
\begin{equation}
\label{DiscordiaQVedral}
D_{1}^{(2)}(\rho_{12}) \ = \ \mbox{min}_{\chi \in \Omega_{0}} || \rho_{12} - \chi ||_{F}^{2} \  = \ \frac{1}{4}\left( \ \mathbf{x}\mathbf{x}^{T} + || \mathbb{T} ||_{F}^{2} - k_{\mbox{max}} \ \right) \ .
\end{equation}
Here $\mathbf{x}$ is a real column vector whose three components are given by $x_{i} = \mbox{Tr}(\rho_{12} \sigma_{1i}\otimes\mathbb{I})$, $\mathbb{T}$ is a $3 \times 3$ real matrix whose components are given by $\mathbb{T}_{ij} = \mbox{Tr}(\rho_{12} \sigma_{1i}\otimes\sigma_{2j})$, and $k_{\mbox{max}}$ is the largest eigenvalue of the matrix $K = \mathbf{x}\mathbf{x}^{T} + \mathbb{T}\mathbb{T}^{T}$. Note that $\sigma_{ij}$ is the $j$-th Pauli matrix of atom $i$. Also, the distance is measured using the usual Hilbert-Schmidt-Frobenius norm:
\begin{equation}
\label{normaFrobenius}
|| A ||_{F}^{2} \ = \ \mbox{Tr}(A^{\dagger}A) \ ,
\end{equation}
with $A$ a linear operator in the state space of the system (it could also be a square matrix). $D_{1}^{(2)}(\rho_{12})$ is called the geometric measure of left discord. Using (\ref{RhoO}) and (\ref{RhoN}) one can determine the geometric measure of right quantum discord $D_{2}^{(2)}(\rho_{12})$.

To quantify the degree of mixed-ness we will use the linear entropy $S_{L}$ \cite{Breuer} defined as
\begin{equation}
 \label{EntropiaLineal}
S_{L}(\rho_{12}) \ = \ 1 -\mbox{Tr}\left[ \rho_{12}^{2} \right] \ .
\end{equation}
Recall that $S_{L}(\rho_{12}) = 0$ if $\rho_{12}$ is a pure state, while $S_{L}(\rho_{12}) = 3/4$ if $\rho_{12}$ is a maximum mixed state.

\section{Two driven qubits collectively interacting with a vacuum reservoir}

We consider two identical two-level atoms with transition frequency $\omega_{A}$ at fixed positions $\mathbf{r}_{1}$ and $\mathbf{r}_{2}$, driven by a classical monochromatic laser field, and interacting with all the modes of the quantum electromagnetic field. In the following we will refer to the latter as the reservoir and, in some occasions, to the atom at position $\mathbf{r}_{j}$ as atom $j$.

The Hamiltonian of the system in the long-wavelength approximation and in the electric dipole representation is
\begin{equation}
 \label{HamiltonianoCompleto}
 H \ = \ H_{0} + V + V_{AL}(t)
\end{equation}
where $H_{0}$ is the free hamiltonian of the two atoms and of the reservoir
\begin{equation}
\label{HamiltonianoH0}
H_{0} \ = \ \frac{\hbar \omega_{A}}{2} \sigma_{3} \ + \ \sum_{j} \hbar \omega_{j}\left( a_{j}^{\dagger}a_{j} + \frac{1}{2} \right) \ ,
\end{equation}
$V$ is the electric dipole interaction between the two atoms and the electric field $\mathbf{E}(\mathbf{r})$ of the reservoir
\begin{equation}
\label{HamiltonianoV}
V \ = \ -\mathbf{d}_{1} \cdot \mathbf{E}(\mathbf{r}_{1}) \ - \mathbf{d}_{2} \cdot \mathbf{E}(\mathbf{r}_{2}) \ ,
\end{equation}
and $V_{AL}(t)$ is the electric dipole interaction in the rotating-wave-approximation between the two atoms and the classical monochromatic electric field $\mathbf{E}_{L}(\mathbf{r},t)$ of frequency $\omega_{L}$
\begin{equation}
 \label{VAL}
 V_{AL}(t)
 \ = \  - \sum_{j=1}^{2} \hbar G(\mathbf{r}_{j})\left( \sigma_{+j}e^{-i \omega_{L}t} + \sigma_{-j}e^{i \omega_{L}t} \right) \ .
\end{equation}
Recall that $|j:+\rangle$ \ and \ $|j:-\rangle$ \ are the excited and ground states of the $j$th atom ($j=1,2$), respectively, and  $\sigma_{3} = (\sigma_{31} + \sigma_{32})$ \ with \ $\sigma_{3j} = |j:+\rangle\langle j:+| - |j:-\rangle\langle j:-|$ \ is the inversion operator. Furthermore, $\mathbf{d}_{j}$ is the dipole operator of the $j$th atom and, since the atoms are identical and have two levels, can be expressed as
\begin{equation}
\label{dipoloD}
\mathbf{d}_{j} \ = \ \mathbf{d}_{01} \sigma_{+j} \ + \ \mathbf{d}_{01}^{*} \sigma_{-j} \ ,
\end{equation}
where \ $\sigma_{\pm j} = |j:\pm \rangle\langle j: \mp|$ \ are the transition operators for the $j$-th atom and \ $\mathbf{d}_{01} = \langle 1:+ | \mathbf{d}_{1}|1:- \rangle$. Notice that the dipoles have been taken to be parallel to each other. The quantum electric field $\mathbf{E}(\mathbf{r})$ at position $\mathbf{r}$ is given by its expansion in terms of plane waves
\begin{equation}
\label{campoE}
\mathbf{E}(\mathbf{r}) \ = \ i \sum_{j} \sqrt{ \frac{\hbar \omega_{j}}{2 \epsilon_{0} V}} a_{j}e^{i\mathbf{k}_{j}\cdot \mathbf{r}} \hat{\epsilon}_{j} + h.c. \ ,
\end{equation}
where $V$ is the quantization volume, $a_{j}(a_{j}^{\dagger})$ is the annihilation (creation) operator of a photon in mode $j$, and $\sum_{j}$ is a sum over all the modes of the quantum electromagnetic field. The wave and polarization vectors of mode $j$ are $\mathbf{k}_{j}$ and $\hat{\epsilon}_{j}$, respectively, while $\omega_{j} = c k_{j}$ is the corresponding angular frequency.

We have taken the driving electric field $\mathbf{E}_{L}(\mathbf{r},t)$ of the form
\begin{equation}
 \label{EL}
 \mathbf{E}_{L}(\mathbf{r},t) \ = \ g(\mathbf{r})\left( \mathbf{\mathcal{E}}_{L}e^{-i \omega_{L}t} + \mathbf{\mathcal{E}}_{L}^{*}e^{i \omega_{L}t} \right)
\end{equation}
with $g(\mathbf{r})$ a real-valued function describing the spatial structure of the field and $\mathbf{\mathcal{E}}_{L}$ a constant complex vector which contains the polarization of the electric field. Also, $G(\mathbf{r}) = g(\mathbf{r})(\mathbf{d}_{01}\cdot\mathbf{\mathcal{E}}_{L})/\hbar$ is assumed to be a real quantity, and in the following we shall denote $G(\mathbf{r}_{j})$ by $G_{j}$. For example, (\ref{EL}) could describe a stationary wave with polarization independent of position by taking $g(\mathbf{r}) = \mbox{cos}(\mathbf{k}_{L}\cdot \mathbf{r})$.

We will denote the density operator of the complete system (two atoms plus quantum electromagnetic field) by $\rho(t)$, while $\rho_{12}(t)$ will denote the density operator of the two atoms. Recall that $\rho_{12}(t)$ is the reduced density operator obtained by tracing $\rho(t)$ over the reservoir degrees of freedom. Also, we assume that the initial state of the system is of the form
\begin{equation}
 \label{EstadoInicialCF}
 \rho(t=0) \ = \ \rho_{12}(t=0)\otimes| \mathbf{0} \rangle\langle \mathbf{0}| \ ,
\end{equation}
that is, the system will initially be in a product state in which the reservoir is in the vacuum state.

When the interaction with the vacuum reservoir is neglected ($V$ is zero in (\ref{HamiltonianoCompleto})), $\rho_{12}(t)$ is determined by von Neumann's equation:
\begin{equation}
 \label{classicalField}
 i\hbar \frac{d}{dt}\rho_{12}(t) \ = \ \left[ \ \frac{\hbar \omega_{A}}{2}\sigma_{3} + V_{AL}(t), \ \rho_{12}(t)  \right] \ .
\end{equation}
On the other hand, when there is no driving field ($V_{AL}(t)$ is zero in (\ref{HamiltonianoCompleto})), the dynamics of the density operator $\rho_{12}(t)$ of the two atoms can be described by a Born-Markov-Secular master equation \cite{tanas}:
\begin{equation}
\label{EcMaestra}
\frac{d}{dt} \rho_{12}(t) \ = \ - \frac{i}{\hbar}\left[ \frac{\hbar \omega_{A}}{2}\sigma_{3} + H_{Ls}, \ \rho_{12}(t) \right] \ + \ \mathcal{D}[\rho_{12}(t)] \ .
\end{equation}
Here $H_{Ls}$ is the Lamb-shift Hamiltonian which can be expressed as a sum of two parts: one that describes an
effective coherent dipole-dipole interaction $h_{dd}$ between the two atoms and another that represents a shift $\epsilon_{Ls}$ of the  energy levels of the free Hamiltonian of the two atoms \cite{tanas}. The dissipator $\mathcal{D}$ is given by
\begin{equation}
\label{disipador}
\mathcal{D}( \rho ) \ = \ \mathcal{D}_{1}( \rho ) \ + \ \mathcal{D}_{2}( \rho ) \ + \ \mathcal{D}_{3}( \rho ) \ ,
\end{equation}
where $\mathcal{D}_{j}$ is the dissipator for the $j$th atom interacting with the reservoir ($j=1,2$)
\begin{equation}
\label{disipador1}
\mathcal{D}_{j}(\rho) \ = \
\gamma_{1} \left( \sigma_{-j} \rho \sigma_{+j} - \frac{1}{2}\left\{ \ \sigma_{+j}\sigma_{-j} , \rho \ \right\} \right) \qquad (j=1,2),
\end{equation}
and $\mathcal{D}_{3}$ is the part that describes the dissipative collective interaction of the two atoms with the reservoir
\begin{eqnarray}
\label{disipador3}
\mathcal{D}_{3}( \rho ) &=&  \frac{3}{2} \gamma_{1} F(|\mathbf{r}_{1}-\mathbf{r}_{2}|) \Big( \sigma_{-1} \rho \sigma_{+2} - \frac{1}{2}\left\{ \sigma_{+2}\sigma_{-1} , \rho \right\} \cr
&& + \sigma_{-2} \rho \sigma_{+1} - \frac{1}{2}\left\{ \sigma_{+1}\sigma_{-2} , \rho \right\} \ \Big) \ .
\end{eqnarray}
Here $\{ \cdot , \cdot \}$ is the anti-commutator and $\gamma_{1}$ is equal to the spontaneous emission rate of a two-level atom interacting with all the modes of the electromagnetic field
\[
 \gamma_{1} = \frac{1}{4 \pi \epsilon_{0}} \cdot \frac{4 |\mathbf{d}_{01}|^{2} \omega_{A}^{3}}{3 \hbar c^{3}} \ .
\]
The function $F(|\mathbf{r}_{1}-\mathbf{r}_{2}|)$ is defined by
\begin{equation}
\label{F}
F(|\mathbf{r}_{1}-\mathbf{r}_{2}|)
\equiv \frac{d_{\bot}^2}{|\mathbf{d}_{01}|^{2}} \cdot \frac{\mbox{sin}(x)}{x} + \left(3\frac{d_{\bot}^2}{|\mathbf{d}_{01}|^{2}} -2 \right)\frac{\mbox{cos}(x) - \frac{\mbox{sin}(x)}{x}}{x^{2}} \ ,
\end{equation}
with
\[
 x = \frac{\omega_{A}}{c}|\mathbf{r}_{1} - \mathbf{r}_{2}| \ ,
\]
and $d_{\bot}^2$ the square of the norm of the projection of $\mathbf{d}_{01}$ onto the plane perpendicular to $\mathbf{r}_{1} - \mathbf{r}_{2}$. In the following we will denote $F(|\mathbf{r}_{1}-\mathbf{r}_{2}|)$ by $F_{12}$ \cite{notacion}. We note that $F_{12}$ is an oscillatory function whose absolute maximum and minimum values are $2/3$ and $-0.2237$, and that they occur only at the points \ $\omega_{A}|\mathbf{r}_{1} -\mathbf{r}_{2}|/c = 0$ \ (independent of the value of $d_{\bot}^2$) and \ $\omega_{A}|\mathbf{r}_{1} -\mathbf{r}_{2}|/c = 4.233$ \ (with $d_{\bot}^2 = |\mathbf{d}_{01}|^{2}$), respectively.

Applying the approximation of independent rates of variation \cite{Cohen}, the master equation for $\rho_{12}(t)$ which takes into account the interaction of the two atoms with the vacuum reservoir and with the classical field is given by $(t \geq 0)$
\begin{equation}
 \label{EcMaestraCF}
 \frac{d}{dt}\rho_{12}(t) \ = \ -{i \over \hbar}\left[ \frac{\hbar \omega_{A}}{2}\sigma_{3} + H_{Ls} + V_{AL}(t), \ \rho_{12}(t) \right] + \mathcal{D}\left[ \rho_{12}(t) \right] \  .
\end{equation}
In the following we will assume that the energy shift $\epsilon_{Ls}$ has been incorporated in $\hbar \omega_A$ and we will neglect the coherent dipole-dipole interaction $h_{dd}$ between the two atoms which arises from the collective interaction with the reservoir. In general, $h_{dd}$ is negligible for distant atoms and for particular distances between the atoms \cite{dissipation}, but it can become important when the atoms are very close \cite{tanas}. Later on, we
will make explicit when the effects of $h_{dd}$ can be important. Furthermore, these will be taken into account in future work.

Passing to the interaction picture (IP) defined by the unitary transformation $U_{0}(t,0) = \mbox{exp}[-i\omega_{A}\sigma_{3}t/2]$ we obtain the master equation that will be taken as the model under study for the rest of the article:
\begin{equation}
 \label{EcMaestraCFIP}
 \frac{d}{dt}\rho_{12}(t) \ = \ -{i \over \hbar}\left[  V_{AL}^{I}(t), \rho_{12}(t) \right] + \mathcal{D}\left[ \rho_{12}(t) \right] \ \ \ \ (t \geq 0) \ ,
\end{equation}
where
\begin{equation}
 \label{VALI}
V_{AL}^{I}(t) \ = \ - \sum_{j=1}^{2} \hbar G(\mathbf{r}_{j})\left( \sigma_{+j}e^{-i\delta_{\mbox{\tiny{L}}}t} + \sigma_{-j}e^{i \delta_{\mbox{\tiny{L}}}t}  \right) \ ,
\end{equation}
$\delta_{\mbox{\tiny{L}}} = \omega_{L}-\omega_{A}$, and $\rho_{12}(t)$  is now the IP density operator of the two atoms.

To study the dynamics of the system we will calculate the steady-state solution $\rho_{12}^{ST}$ of (\ref{EcMaestraCFIP}) for two special configurations which will be analyzed in the following. Recall that $\rho_{12}^{ST}$ is a steady-state solution of (\ref{EcMaestraCFIP}) if $\rho_{12}^{ST}$ is not explicitly time dependent and
\begin{equation}
 \label{SEstacionaria}
   -{i \over \hbar}\left[  V_{AL}^{I}(t), \rho_{12}^{ST} \right] + \mathcal{D}\left[ \rho_{12}^{ST} \right] = 0 \ .
\end{equation}
We note that in general (\ref{EcMaestraCFIP}) has no steady-state solutions if $\delta_{\mbox{\tiny{L}}} \not= 0$, since density operators $\rho_{12}^{ST}$ that satisfy (\ref{SEstacionaria}) have non-diagonal time dependent elements if $\delta_{\mbox{\tiny{L}}} \not= 0$. This will be shown explicitly in the following. Also notice that other steady-state solutions may appear in other interaction pictures.

\section{$G(\mathbf{r}_{2}) = 0$}

In this part we will assume that one of the atoms is fixed in a position where the classical electric field (\ref{EL}) is zero, while the other atom is fixed in a position where it is not zero. Therefore, we take $G_{2} = G(\mathbf{r}_{2}) = 0$. Notice that increasing (decreasing) the intensity of the laser field increases (decreases) $|G_{1}| = |G(\mathbf{r}_{1})|$ (see the definition of $G_{1}$ following (\ref{EL})). Therefore, $G_{1}$ can be made to vary by increasing or decreasing the intensity of the electric field (\ref{EL}). Furthermore, the distance $|\mathbf{r}_{1}-\mathbf{r}_{2}|$ between the atoms can also be varied independently of $G_{1}$. For example, one of the atoms could be placed at a node of a stationary wave to have $G_{2} = 0$, the other atom could be placed anywhere else (except at a position where $G_{1} = 0$), and the intensity of the field could be varied to have $|G_{1}| = |G(\mathbf{r}_{1})|$ take on any positive value.

It is found that the populations of the solution $\rho_{12}^{ST}$ of (\ref{SEstacionaria}) in the triplet-singlet basis (\ref{acoplada}) are given by
\begin{eqnarray}
 \label{RhoAEstG20}
 \langle 1,1 | \rho_{12}^{ST} |1,1 \rangle
 &=& \frac{18}{\kappa}\left[ \ 16F_{12}^{2}\bar{G}_{1}^{6} + 9(4-F_{12}^{2})F_{12}^{2}\bar{G}_{1}^{4} \ \right] \ , \cr
 \langle 1,0 | \rho_{12}^{ST} |1,0 \rangle
 &=& \frac{1}{\kappa}\left[ \ \mu_{-}\bar{G}_{1}^{6} + \nu_{-}\bar{G}_{1}^{4} + \eta_{-}\bar{G}_{1}^{2}\ \right] \ , \cr
 \langle 1,-1 | \rho_{12}^{ST} |1,-1 \rangle
 &=& 1 - \langle 1,1 | \rho_{12}^{ST} |1,1 \rangle - \langle 1,0 | \rho_{12}^{ST} |1,0 \rangle - \langle 0,0 | \rho_{12}^{ST} |0,0\rangle \ ,  \cr
 \langle 0,0 | \rho_{12}^{ST} |0,0 \rangle
 &=& \frac{1}{\kappa}\left[ \ \mu_{+}\bar{G}_{1}^{6} + \nu_{+}\bar{G}_{1}^{4} +  \eta_{+}\bar{G}_{1}^{2}\ \right] \ ,
\end{eqnarray}
where
\begin{eqnarray}
\bar{G}_{1} &=& G_{1}/\gamma_{1} \ , \cr
\kappa
&=& \bar{G}_{1}^{6}(2048 + 1152F_{12}^{2}) + \bar{G}_{1}^{4}(2560 + 5184F_{12}^{2} - 1296F_{12}^{4}) \cr
&& + \bar{G}_{1}^{2}(864 + 144F_{12}^{2} + 486F_{12}^{4}) \cr
&& + \frac{9}{8}(2+3F_{12})^{2}(2-3F_{12})^{2}(4-F_{12}^{2})\ , \cr
\mu_{\pm}
&=& 32*(16 \pm 12F_{12} + 9F_{12}^{2}) \ , \cr
\nu_{\pm}
&=& 18*(32 \pm 48F_{12} + 68F_{12}^{2} - 9F_{12}^{4}) \ , \cr
\eta_{\pm}
&=& 9*(2 \pm 3F_{12})^{2}(4 - F_{12}^{2}) \ .
\end{eqnarray}
On the other hand, the coherences of $\rho_{12}^{ST}$ are given by
\begin{eqnarray}
\label{RhoAEst2}
\langle 0,0 | \rho_{12}^{ST} |1,1 \rangle
&=& ie^{i \delta_{\mbox{\tiny{L}}}t}\frac{9\sqrt{2}}{\kappa}\bar{G}_{1}^{3}F_{12}\Big[ \ 3(2+3F_{12})(4-F_{12}^{2}) \cr
&& + 8(4 +2F_{12}-3F_{12}^{2})\bar{G}_{1}^{2} \ \Big] \ , \cr
\langle 1,1 | \rho_{12}^{ST} |0,0 \rangle
&=&
\langle 0,0 | \rho_{12}^{ST} |1,1 \rangle^{*} \ , \cr
\langle 1,0 | \rho_{12}^{ST} |1,1 \rangle
&=& ie^{i \delta_{\mbox{\tiny{L}}}t}\frac{9\sqrt{2}}{\kappa}\bar{G}_{1}^{3}F_{12}\Big[ \ 3(2-3F_{12})(4-F_{12}^{2}) \cr
&& + 8(4-2F_{12}-3F_{12}^{2})\bar{G}_{1}^{2} \ \Big] \ , \cr
\langle 1,1 | \rho_{12}^{ST} |1,0 \rangle
&=&
\langle 1,0 | \rho_{12}^{ST} |1,1 \rangle^{*} \ , \cr
\langle 1,-1 | \rho_{12}^{ST} |1,0 \rangle
&=& -ie^{i \delta_{\mbox{\tiny{L}}}t}\frac{\sqrt{2}}{4\kappa}\bar{G}_{1}\Big[ \ 32(32 -12F_{12} -18F_{12}^{2} + 27F_{12}^{3})\bar{G}_{1}^{4} \cr && + 36(32 -24F_{12} -4F_{12}^{2} + 54F_{12}^{3} -9F_{12}^{4})\bar{G}_{1}^{2} \  \cr
&& + 9(2 - 3F_{12})^{2}(2+3F_{12})(4-F_{12}^{2}) \ \Big] \ , \cr
\langle 1,0 | \rho_{12}^{ST} |1,-1 \rangle
&=& \langle 1,-1 | \rho_{12}^{ST} |1,0 \rangle^{*} \ , \cr
\langle 1,-1 | \rho_{12}^{ST} |0,0 \rangle
&=& -ie^{i \delta_{\mbox{\tiny{L}}}t}\frac{\sqrt{2}}{4 \kappa}\bar{G}_{1}\Big[ \ 32(32 +12F_{12} -18F_{12}^{2} - 27F_{12}^{3})\bar{G}_{1}^{4} \cr
&& + 36(32 +24F_{12} -4F_{12}^{2} - 54F_{12}^{3} +9F_{12}^{4})\bar{G}_{1}^{2} \cr
&& + 9(2 + 3F_{12})^{2}(2 -3F_{12})(4-F_{12}^{2}) \ \Big] \ , \cr
\langle 0,0 | \rho_{12}^{ST} |1,-1 \rangle
&=&
\langle 1,-1 | \rho_{12}^{ST} |0,0 \rangle^{*} \ , \cr
\langle 1,-1 | \rho_{12}^{ST} |1,1 \rangle
&=& \frac{3}{2\kappa}e^{i2\delta_{\mbox{\tiny{L}}}t}F_{12}\bar{G}_{1}^{2}\Big[ \ -256\bar{G}_{1}^{4} - 9(64)F_{12}^{2}\bar{G}_{1}^{2} \cr 
&& + 9(16 -40F_{12}^{2} +9F_{12}^{4})  \ \Big] \ , \cr
\langle 1,1 | \rho_{12}^{ST} |1,-1 \rangle
&=&
\langle 1,-1 | \rho_{12}^{ST} |1,1 \rangle^{*} \ , \cr
\langle 0,0 | \rho_{12}^{ST} |1,0 \rangle
&=& \frac{\bar{G}_{1}^{2}}{\kappa}\Big[ \ 512\bar{G}_{1}^{4}  +36(16 + 16F_{12}^{2} - 9F_{12}^{4})\bar{G}_{1}^{2} \cr
&& + 9(16 -40F_{12}^{2} + 9F_{12}^{4}) \ \Big] \ , \cr
\langle 1,0 | \rho_{12}^{ST} |0,0 \rangle
&=& \langle 0,0 | \rho_{12}^{ST} |1,0 \rangle^{*} \ .
\end{eqnarray}
Note that some coherences of $\rho_{12}^{ST}$ depend explicitly on time. Hence, in the IP we are working, a steady-state solution (\ref{SEstacionaria}) exists only for the resonant case ($\delta_{\mbox{\tiny{L}}} = 0$). The steady-state solution in (\ref{RhoAEstG20}) and (\ref{RhoAEst2}) would also be approximately valid for small enough detuning $\delta_{\mbox{\tiny{L}}}$, for instance for $| \delta_{\mbox{\tiny{L}}} t | \ll 1$. In the following we will restrict to the case where the (exact) steady-state solution exists, that is, we will take $\delta_{\mbox{\tiny{L}}} = 0$.

Using the following argument it can be shown that the density operator $\rho_{12}(t)$ for the two atoms tends to the state $\rho_{12}^{ST}$ above for any initial state $\rho_{12}(0)$ if $G_{2} =0$ and there is resonance ($\delta_{\mbox{\tiny{L}}} = 0$).
To that end, we take $G_{2} = 0$ and $\delta_{\mbox{\tiny{L}}} = 0$, and we choose the triplet-singlet basis (\ref{acoplada}) for the state space of the two atoms. (\ref{EcMaestraCFIP}) with the initial condition $\rho_{12}(0)$ can then be reexpressed as an initial value problem (IVP) of the form
\begin{eqnarray}
\label{sistemaEquivalenteG20}
\dot{\mathbf{x}}(t) \ = \ \mathbb{A}\mathbf{x}(t) \ , \qquad \mathbf{x}(0) = \mathbf{x}_{0} \ ,
\end{eqnarray}
where $\mathbf{x}(t)$ is a $16$ component column complex vector associated with $\rho_{12}(t)$, $\mathbb{A}$ is a $16 \times 16$ constant complex matrix, and $\mathbf{x}_{0}$ is a $16$ component column complex vector defined by $\rho_{12}(0)$. The IVP in (\ref{sistemaEquivalenteG20}) has a unique solution of the form \cite{Coddington}
\begin{eqnarray}
\label{xAx}
\mathbf{x}(t) \ = \ c_{1}e^{-\lambda_{1}t}\mathbf{v}_{1}(t) + ... + c_{16}e^{-\lambda_{16}t}\mathbf{v}_{16}(t) \ ,
\end{eqnarray}
where $-\lambda_{j}$ are the complex eigenvalues of $\mathbb{A}$, $\mathbf{v}_{j}(t)$ are vectors whose components are polynomials in $t$ of degree less than the algebraic multiplicity of the corresponding eigenvalue $-\lambda_{j}$, and $c_{j}$ are complex constants determined by $\mathbf{x}_{0}$. When an eigenvalue $-\lambda_{j}$ has an algebraic multiplicity equal to its geometric multiplicity, the corresponding vectors $\mathbf{v}_{j}$ reduce to linearly independent eigenvectors associated with $-\lambda_{j}$. It can be shown that zero is an eigenvalue with algebraic multiplicity equal to one. If necessary rearranging the order, we take $\lambda_{1}=0$. Hence, $\mathbf{v}_{1}$ is an eigenvector of $\mathbb{A}$ associated with $\lambda_{1}=0$. Furthermore, the rest of the eigenvalues have negative real part (that is, Re($\lambda_{j}$) $> 0$ for $j \not= 1$). It then follows that
\begin{eqnarray}
\mathbf{x}(t) \ \rightarrow \ c_{1}e^{-\lambda_{1}t}\mathbf{v}_{1}(t) = c_{1}\mathbf{v}_{1} \ \ \ \mbox{if} \ t \rightarrow +\infty .
\end{eqnarray}
Now $c_{1}\mathbf{v}_{1}$ is associated with $\rho_{12}^{ST}$ above. Hence, for any initial condition $\rho_{12}(t)$ tends to $\rho_{12}^{ST}$ as $t \rightarrow + \infty$.

There are several limiting cases of interest. First, if $\bar{G}_{1} \rightarrow 0$, then it is seen from (\ref{RhoAEstG20}) and (\ref{RhoAEst2}) that $\rho_{12}^{ST} \rightarrow |1,-1\rangle\langle 1,-1|$. This result is expected since the atoms are located at different positions and, without the driving field, the reservoir ultimately leaves the two atoms in their respective ground states.

Another limiting case of more interest occurs when the classical electric field is very intense ($|\bar{G}_{1}| \rightarrow +\infty$). From (\ref{RhoAEstG20})-(\ref{RhoAEst2}) we see that the matrix representation of $\rho_{12}^{ST}$ in the \textit{tensor product basis} $\mathsf{B}'$ (\ref{desacoplada}) takes the form of an $X$-state
\begin{equation}
\label{XstateG20}
[\rho_{12}^{ST}]_{\mathsf{B}'} \ = \
\left(
\begin{array}{cccc}
\frac{\frac{9}{4}F_{12}^{2}}{16+9F_{12}^{2}}  &  0                                     &
0                                             &  -\frac{3F_{12}}{16+9F_{12}^{2}} \cr
0                                             & \frac{1}{2} - \frac{\frac{9}{4}F_{12}^{2}}{16+9F_{12}^{2}} & -\frac{3F_{12}}{16+9F_{12}^{2}}               & 0 \cr
0                                             & - \frac{3F_{12}}{16+9F_{12}^{2}}       & \frac{\frac{9}{4}F_{12}^{2}}{16+9F_{12}^{2}}  & 0 \cr
-\frac{3F_{12}}{16+9F_{12}^{2}}               & 0                                      &
0                                             &  \frac{1}{2} - \frac{\frac{9}{4}F_{12}^{2}}{16+9F_{12}^{2}}
\end{array}
\right) \ .
\end{equation}
In this case the corresponding populations of the \textit{triplet-singlet basis} $\mathsf{B}$ (\ref{acoplada}) take the form
\begin{eqnarray}
\label{poblacionesG20inf}
\langle 1,1|\rho_{12}^{ST}|1,1 \rangle &=& \frac{\frac{9}{4}F_{12}^{2}}{16 +9F_{12}^{2}} \ , \cr
\langle 1,0|\rho_{12}^{ST}|1,0 \rangle &=& \frac{1}{4} - \frac{3F_{12}}{16 + 9F_{12}^{2}} \ , \cr
\langle 1,-1|\rho_{12}^{ST}|1,-1 \rangle &=& \frac{1}{2} - \frac{\frac{9}{4}F_{12}^{2}}{16+9F_{12}^{2}} \ , \cr
\langle 0,0|\rho_{12}^{ST}|0,0 \rangle &=& \frac{1}{4} + \frac{3F_{12}}{16 + 9F_{12}^{2}} \ .
\end{eqnarray}
Noting that $F_{12}$ takes on its absolute maximum value of $2/3$ when $|\mathbf{r}_{1}-\mathbf{r}_{2}|=0$ (independent of the value of $d_{\bot}^2$) and its absolute minimum value of $-0.2237$ when $|\mathbf{r}_{1}-\mathbf{r}_{2}|=4.233$ and $d_{\bot}^2 = |\mathbf{d}_{01}|^{2}$, we can establish bounds on the populations in (\ref{poblacionesG20inf}). The populations of the states $|1,1\rangle$ and $|0,0\rangle$ cannot grow more than $1/20$ and $1/4 + 1/10$, respectively, and these maximum values are achieved when the atoms are very close together and the electric field is very intense. At the same time the populations of the states $|1,-1\rangle$ and $|1,0\rangle$ take on their minimum values of $1/2-1/20$ and $1/4 - 1/10$. We note that the dissipative collective interaction of the two atoms with the vacuum reservoir is responsible for the non-zero population of the state $|1,1\rangle$ and that this interaction is not so effective, since this population can only grow to a relatively small value of $1/20$ for high laser field intensities.

Now lets take the limit of the two atoms very far apart ($|\mathbf{r}_{1}-\mathbf{r}_{2}| \rightarrow +\infty$). From (\ref{RhoAEstG20}) and (\ref{RhoAEst2}) we get the following density matrix in the \textit{tensor product basis} $\mathsf{B}'$ (\ref{desacoplada}):
\begin{eqnarray}
\label{poblacionesr12inf}
[\rho_{12}^{ST}]_{\mathsf{B}'} \ = \
\left(
\begin{array}{cccc}
0 & 0                                                           & 0 & 0 \cr
0 & \frac{4\bar{G}_{1}^{2}}{1+8\bar{G}_{1}^{2}}                 & 0 & i\frac{2\bar{G}_{1}}{1 + 8\bar{G}_{1}^{2}} \cr
0 & 0                                                           & 0 & 0 \cr
0 & -i\frac{2\bar{G}_{1}}{1 + 8\bar{G}_{1}^{2}} & 0 & \frac{1 + 4\bar{G}_{1}^{2}}{1+8\bar{G}_{1}^{2}} \cr
\end{array}
\right) \ .
\end{eqnarray}
Note that in this case the system behaves as if the two atoms were interacting with independent reservoirs and, since only one atom is being driven by the laser field, the population of the state $|1,1\rangle = |+,+\rangle$ is zero.

A final limiting case of interest occurs when the laser field intensity is very weak when compared to the spontaneous emission rate of a single atom ($|\bar{G}_{1}| \ll 1$). From (\ref{RhoAEstG20})-(\ref{RhoAEst2}) we find to second order in $\bar{G}_{1}$ that the steady-state density matrix in the \textit{triplet-singlet basis} $\mathsf{B}$ (\ref{acoplada}) takes the form
\begin{eqnarray}
\label{RhoG20G1c}
&& [\rho_{12}^{ST}]_{\mathsf{B}} \cr
&=&
\left(
\begin{array}{cccc}
 0  &  0                                                               &  \frac{3F_{12}\bar{G}_{1}^{2}}{1 - \left(\frac{3}{2}F_{12}\right)^{2}}                 & 0 \cr
 0  & \frac{2\bar{G}_{1}^{2}}{\left(1 + \frac{3}{2}F_{12}\right)^{2}}  &  i \frac{\sqrt{2}\bar{G}_{1}}{1 + \frac{3}{2}F_{12}}                                   &  \frac{2\bar{G}_{1}^{2}}{1-\left(\frac{3}{2}F_{12} \right)^{2} } \cr
  \frac{3F_{12}\bar{G}_{1}^{2}}{1 - \left(\frac{3}{2}F_{12}\right)^{2}} &  -i \frac{\sqrt{2}\bar{G}_{1}}{1 + \frac{3}{2}F_{12}} & 1- \frac{2\bar{G}_{1}^{2}}{\left(1 + \frac{3}{2}F_{12}\right)^{2}} - \frac{2\bar{G}_{1}^{2}}{\left(1 - \frac{3}{2}F_{12}\right)^{2}} &-i \frac{\sqrt{2}\bar{G}_{1}}{1 - \frac{3}{2}F_{12}} \cr
  0& \frac{2\bar{G}_{1}^{2}}{1-\left(\frac{3}{2}F_{12} \right)^{2} }  & i \frac{\sqrt{2}\bar{G}_{1}}{1- \frac{3}{2}F_{12}} & \frac{2\bar{G}_{1}^{2}}{\left(1 - \frac{3}{2}F_{12}\right)^{2}}
 \end{array}
 \right) \ .
\end{eqnarray}
 Notice that $\bar{G}_{1}^{2}$ must be sufficiently small in order for (\ref{RhoG20G1c}) to make sense, since it can be seen that the population of the state $|0,0\rangle$ (component $4,4$ above) diverges as the distance between the atoms tends to zero. In fact, it must occur that
\begin{eqnarray}
\label{cotaG1}
\bar{G}_{1}^{2} \ \leq \ \frac{1}{4} \cdot \frac{\left( 1 - \frac{3}{2}F_{12} \right)^{2} \left( 1 + \frac{3}{2}F_{12} \right)^{2}}{1 + \left( \frac{3}{2}F_{12} \right)^{2}} \ ,
\end{eqnarray}
in order that $\langle 1,0 | \rho_{12}^{ST} |1,0 \rangle + \langle 0,0 | \rho_{12}^{ST} |0,0 \rangle \leq 1$. If one uses the bound in (\ref{cotaG1}) it follows from (\ref{RhoG20G1c}) that
\begin{eqnarray}
\label{cotas00}
\langle 1,0 | \rho_{12}^{ST} |1,0 \rangle
&=& \frac{1}{2} - \frac{\frac{3}{2}F_{12}}{1 + \left(\frac{3}{2}F_{12}\right)^{2}} \rightarrow 0 \qquad \mbox{as} \ |\mathbf{r}_{1} - \mathbf{r}_{2}| \rightarrow 0^{+}\cr
\langle 0,0 | \rho_{12}^{ST} |0,0 \rangle
&=& \frac{1}{2} + \frac{\frac{3}{2}F_{12}}{1 + \left(\frac{3}{2}F_{12}\right)^{2}} \rightarrow 1 \qquad \mbox{as} \ |\mathbf{r}_{1} - \mathbf{r}_{2}| \rightarrow 0^{+}.
\end{eqnarray}
Hence, for a weak driving field and small distance between the atoms, the level $|0,0\rangle$ is much more populated than the level $|1,0\rangle$. It must be kept in mind that the effects of the coherent dipole-dipole interaction $h_{dd}$ could modify deeply the results of this limiting case.

\begin{figure}[htbp]
  \centering
  \subfloat[]{\label{ConcG20a11}\includegraphics[scale=0.4]{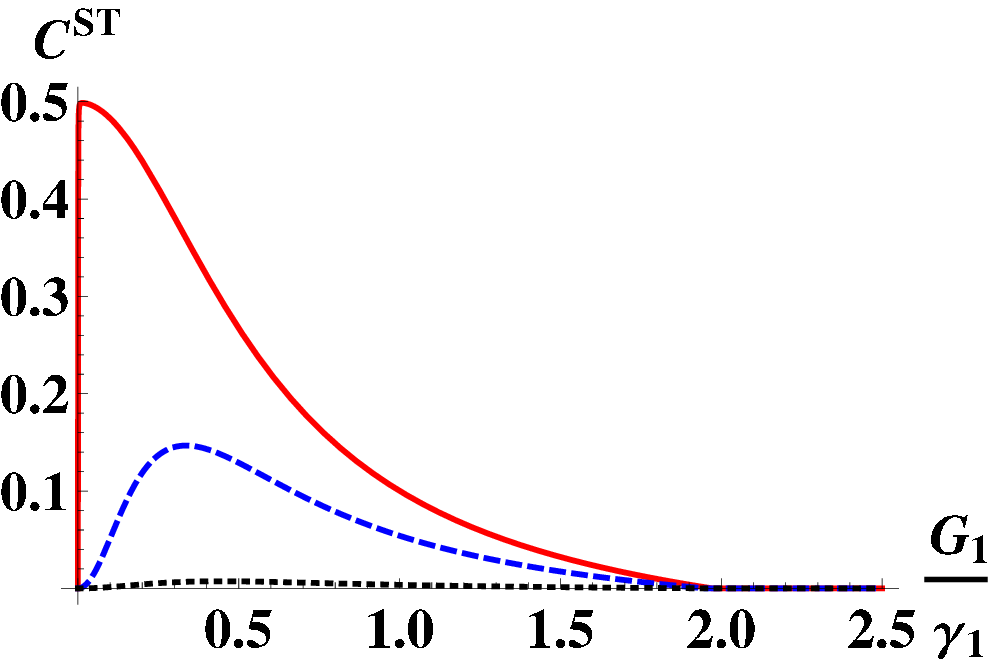}} \hspace{1cm}
  \subfloat[]{\label{ConcG20a10}\includegraphics[scale=0.4]{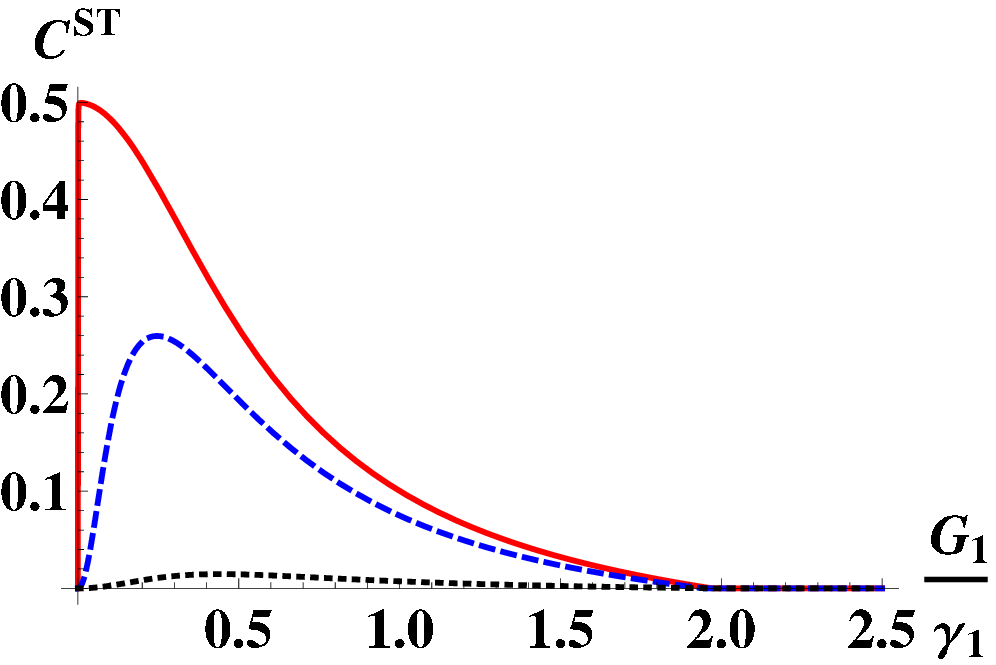}} \hspace{1cm}
  \subfloat[]{\label{CloseUpG20}\includegraphics[scale=0.4]{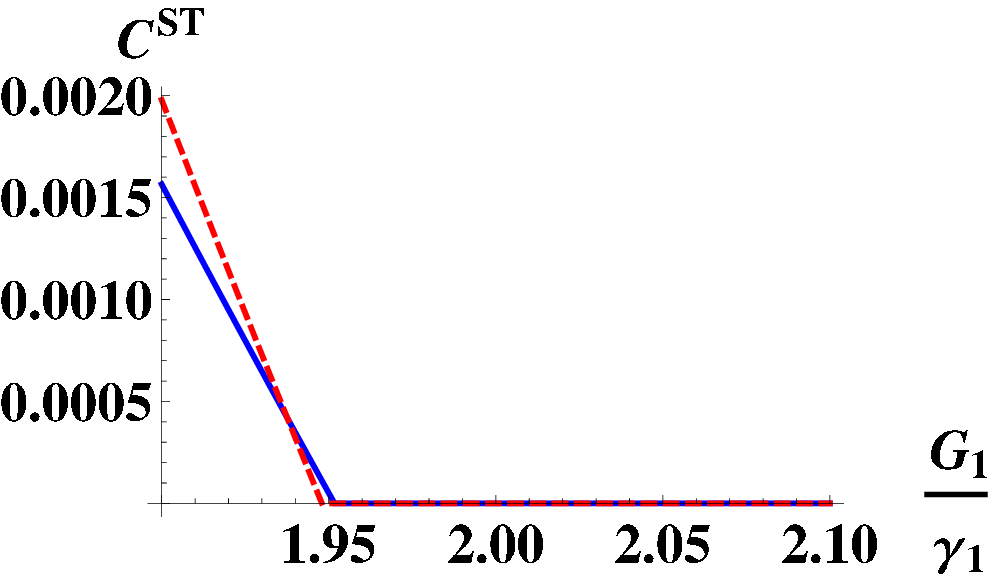}}\\
   \caption{(Color online) Case $G_{2} = 0$: (\ref{ConcG20a11}) and (\ref{ConcG20a10}) show the steady-state concurrence $C^{ST}$ of the two atoms as a function of $\bar{G}_{1} = G_{1}/\gamma_{1}$ when $d_{\bot}^2 = |\mathbf{d}_{01}|^{2}$ and $d_{\bot}^2 = 0$, respectively. Results are shown when the distance between the two atoms is $\lambda_{A}/100$ (red-solid line), $\lambda_{A}/4$ (blue-dashed line), and $\lambda_{A}$ (black-dotted line) with $\lambda_{A}$ the wavelength associated with the atomic transition. (\ref{CloseUpG20}) shows a close-up of the entanglement sudden death when the distance between the two atoms is $\lambda_{A}/4$. This is shown for the cases $d_{\bot}^2 = |\mathbf{d}_{01}|^{2}$ (blue-solid line) and $d_{\bot}^2 = 0$ (red-dashed line). }
  \label{concurrenciaCasosG20}
\end{figure}

We now discuss the degree of entanglement of $\rho_{12}^{ST}$ as measured by the concurrence $C^{ST}$, figure (\ref{concurrenciaCasosG20}). We observe that as the intensity of the laser field increases ($\bar{G}_{1}$ increases), $C^{ST}$ increases, takes on a maximum value, then decreases and finally dies abruptly, figure (\ref{concurrenciaCasosG20}). This \textit{sudden death} of entanglement as a function of the laser field intensity is found for most values of the distance between the atoms. Notice that $C^{ST}$ takes on large values only when the distance between the atoms is less than the wavelength associated with the atomic transition and when the intensity of the laser field is such that $\bar{G}_{1} = G_{1}/\gamma_{1} \leq 1$. This reflects the fact that the atoms can  become entangled by exchanging spontaneously emitted photons only when they are not so far apart. Notice that in the region where the atoms are very close the dipole-dipole interaction $h_{dd}$ could be important.

\begin{figure}[htbp]
  \centering
  \subfloat[]{\label{ConcG20a1100}\includegraphics[scale=0.4]{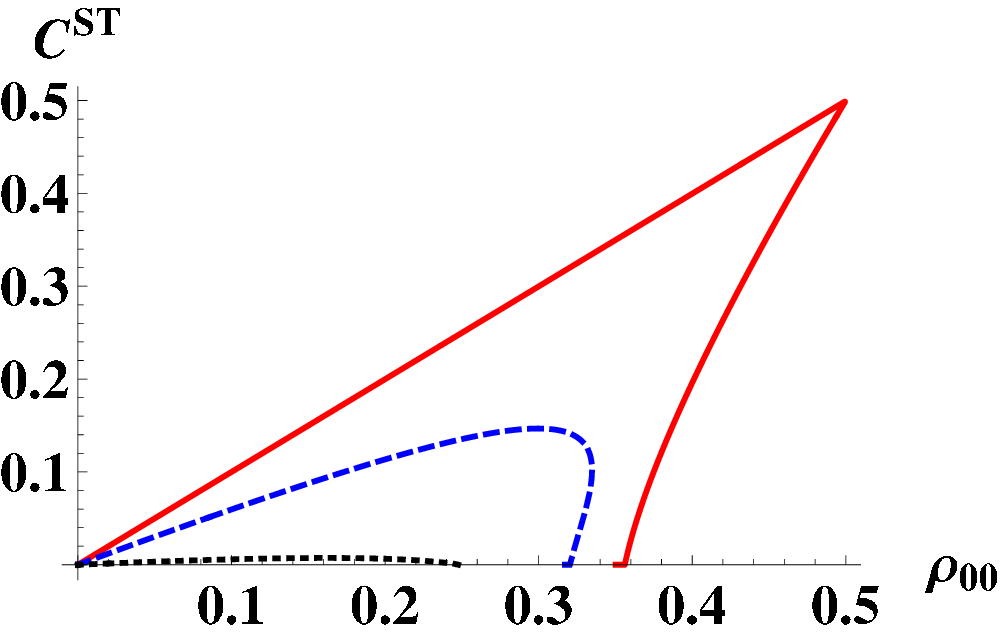}} \hspace{1cm}
  \subfloat[]{\label{ConcG20a1000}\includegraphics[scale=0.4]{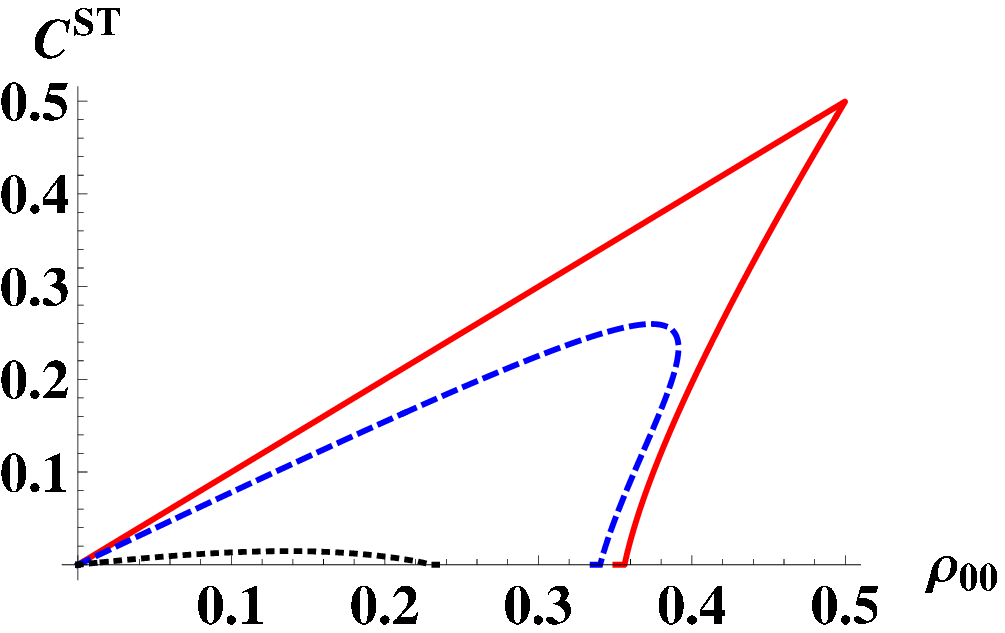}} \hspace{1cm}
  \subfloat[]{\label{ConcG20CloseUp00}\includegraphics[scale=0.4]{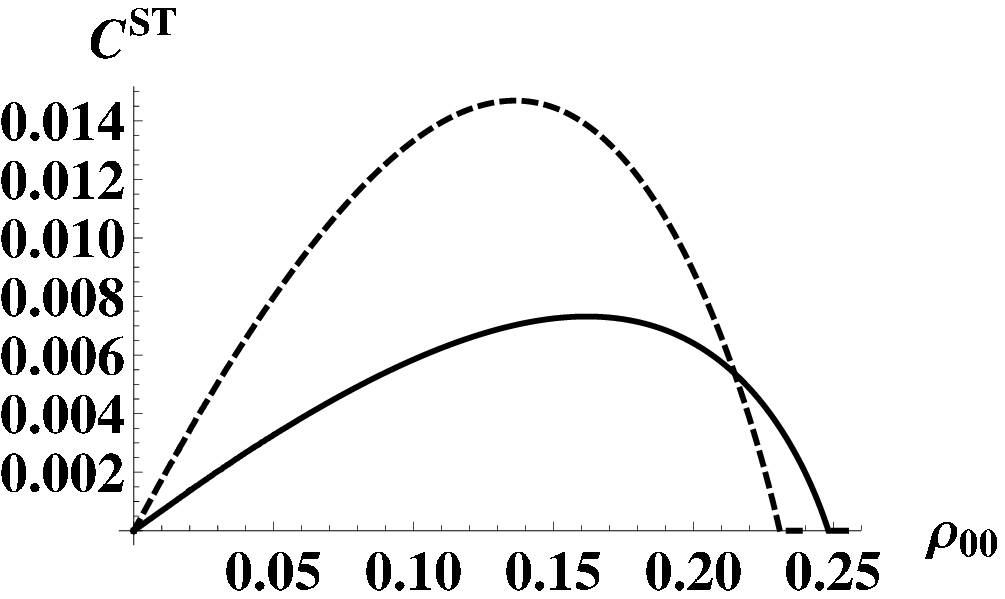}}\\
   \caption{(Color online) Case $G_{2} = 0$: (\ref{ConcG20a1100}) and (\ref{ConcG20a1000}) show the steady-state concurrence $C^{ST}$ of the two atoms as a function of the population $\rho_{00}$ of the antisymmetric state $|0,0\rangle$ when $d_{\bot}^2=|\mathbf{d}_{01}|^{2}$ and $d_{\bot}^2 = 0$, respectively. Results are shown when the distance between the two atoms is $\lambda_{A}/100$ (red-solid line), $\lambda_{A}/4$ (blue-dashed line), and $\lambda_{A}$ (black-dotted line) with $\lambda_{A}$ the wavelength associated with the atomic transition. (\ref{ConcG20CloseUp00}) shows a close-up when the distance between the two atoms is $\lambda_{A}$. This is shown for the cases $d_{\bot}^2 = |\mathbf{d}_{01}|^{2}$ (black-solid line) and $d_{\bot}^2 = 0$ (black-dashed line).}
  \label{ConcY00}
\end{figure}

For the model under consideration, to understand the behaviour of the steady-state concurrence we consider the case of a weak driving field. The concurrence of the density operator in (\ref{RhoG20G1c}) is easily calculated to be
\begin{eqnarray}
\label{concG20G1c}
C^{ST} \ = \ \frac{3}{2}|F_{12}| \cdot\left( \ \langle 0,0 | \rho_{12}^{ST} |0,0 \rangle + \langle 1,0 | \rho_{12}^{ST} |1,0 \rangle \ \right) \ ,
\end{eqnarray}
to second order in $\bar{G}_{1}$. Hence, we observe that for weak driving fields the steady-state concurrence is determined by the populations of the states $|0,0\rangle$ and $|1,0\rangle$. If, in addition, the atoms are close, from (\ref{cotas00}) we observe that the population of the level $|0,0\rangle$ dominates over that of the state $|1,0\rangle$. Hence, the population of $| 0,0\rangle$ and the function $F_{12}$ determine the behaviour of the steady-state concurrence. This can be observed in figure (\ref{ConcY00}) where it is seen that the population of $|0,0\rangle$ is responsible to a large extent of the behaviour of the concurrence not only for weak laser field intensities. From that figure we observe that, when the distance between the atoms is less than a wavelength associated with the atomic transition, the concurrence grows with the population of the level $|0,0\rangle$ and is maximized when the latter is maximized. As the population of the state $|0,0\rangle$ decreases to its asymptotic value (that is, when $\bar{G}_{1} \rightarrow + \infty$), the concurrence decreases to zero. Notice that this behaviour is not exhibited when the two atoms are separated by a wavelength of the atomic transition, figure (\ref{ConcG20CloseUp00}). The reason for this can be explained using (\ref{cotas00}) and (\ref{concG20G1c}) for a weak driving field. When the atoms are more separated, $F_{12}$ tends to $0$ and we find from (\ref{cotas00}) that the populations of the levels $|0,0\rangle$ and $|1,0\rangle$ tend to be the same. Therefore, from (\ref{concG20G1c}) we observe that the concurrence is determined by both of these populations and not only one of them as in the case when the atoms are close. Why the concurrence disappears when the laser field becomes very intense can be explained by the asymptotic $X$-form of the density matrix in the product state basis, (\ref{XstateG20}). There we observe that the laser field has reached such a high magnitude that many coherences have decreased to zero and that the atomic transitions are saturated. The latter is taken to mean that the populations have reached their asymptotic value. Furthermore, the concurrence of this particular $X$-state can be easily calculated to be zero.

\begin{figure}[htbp]
  \centering
  \vspace{1.5cm}
  \subfloat[]{\label{Contoura11G20}\includegraphics[scale=0.5]{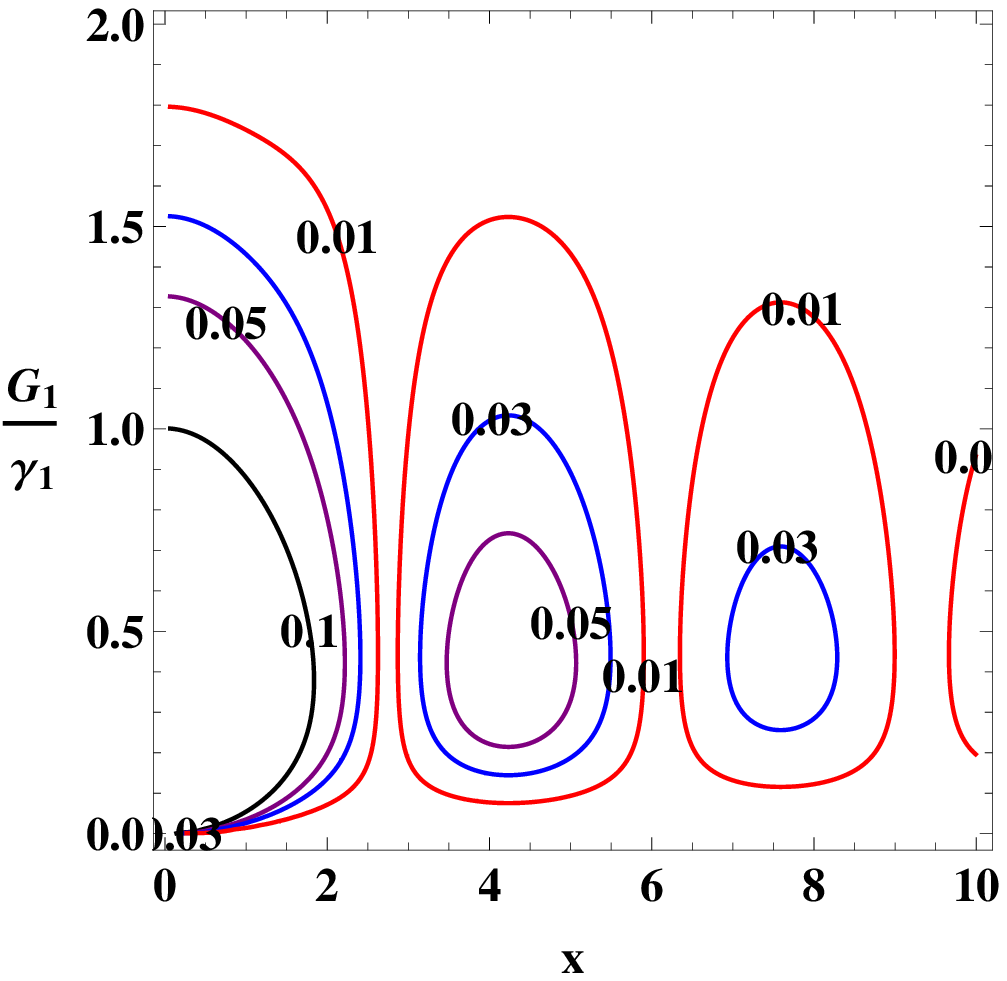}} \hspace{1cm}
  \subfloat[]{\label{Contoura10G20}\includegraphics[scale=0.5]{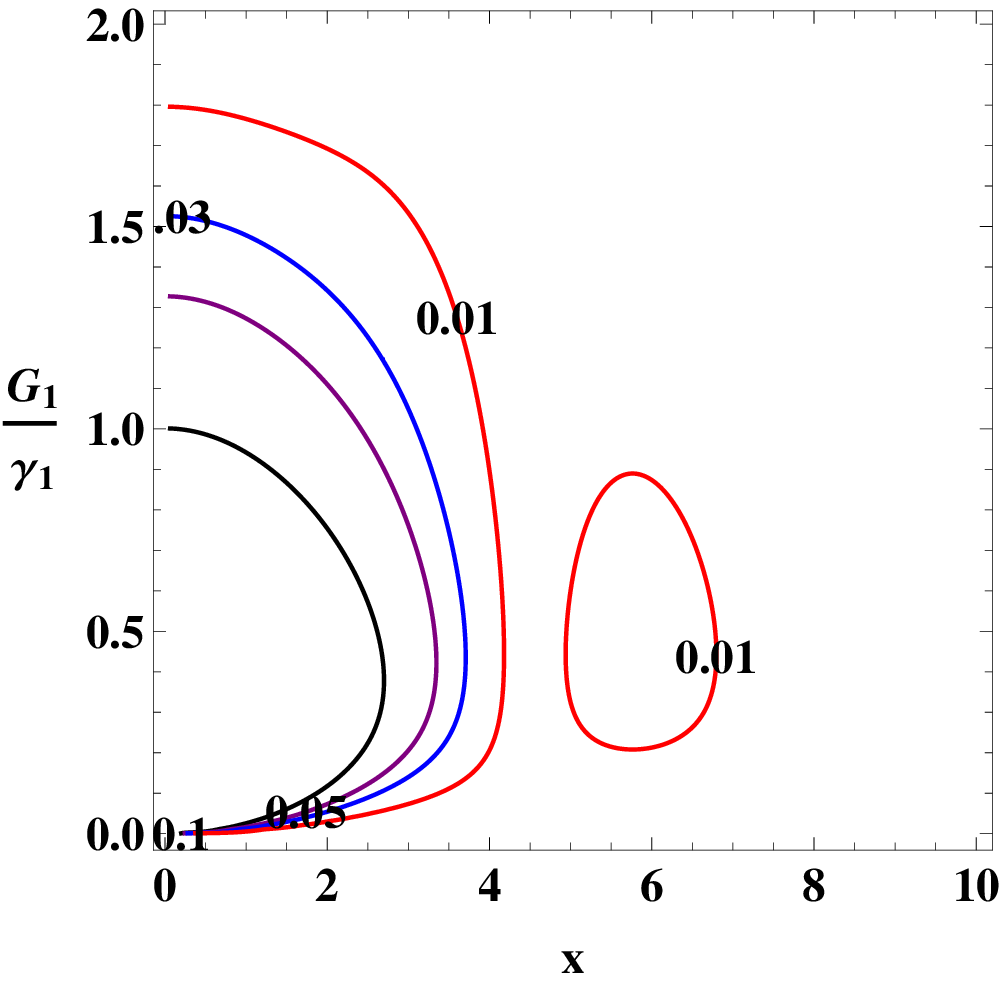}}\\
  \caption{(Color online) Case $G_{2} = 0$: (\ref{Contoura11G20}) and (\ref{Contoura10G20}) show contour plots of the steady-state concurrence $C^{ST}$ as a function of $x = \omega_{A}|\mathbf{r}_{1}-\mathbf{r}_{2}|/c$ (horizontal axis) and $\bar{G}_{1}$ (vertical axis) for the cases $d_{\bot}^2=|\mathbf{d}_{01}|^{2}$ and $d_{\bot}^2 = 0$, respectively. Contours for $C^{ST} = 0.1, 0.05, 0.03, 0.01$ are drawn.}
  \label{contourG20}
\end{figure}

Also note that the decay of $C^{ST}$ with the distance between the two atoms is slower when $d_{\bot}^2 = 0$ (that is, the dipole moments of the two atoms are parallel to the line that joins them), and that $C^{ST}$ exhibits an oscillatory behaviour as a function of this distance that is more noticeable when $d_{\bot}^2 = |\mathbf{d}_{01}|^{2}$, figures (\ref{concurrenciaCasosG20}) and (\ref{contourG20}). This behaviour is easily explained in the case of a weak laser field intensity ($\bar{G}_{1}$ is small). From (\ref{concG20G1c}) we see that $C^{ST}$ depends on $|F_{12}|$ which is an oscillatory function of the distance between the two atoms, (\ref{F}). It can be seen that the $F_{12}$ has oscillations of greater amplitude when $d_{\bot}^2 = |\mathbf{d}_{01}|^{2}$ and that these oscillations are smoothed out when $d_{\bot}^2 = 0$. This is the reason why $C^{ST}$ has a greater oscillatory behaviour when $d_{\bot}^2 =|\mathbf{d}_{01}|^{2}$ than in the case where $d_{\bot}^2 = 0$.

In the preceding paragraphs we have established that for any distance between the two atoms the entanglement present in the system disappears as soon as the laser field intensity is high enough (that is, $\bar{G}_{1}$ has achieved a sufficiently high value). So it is natural to ask whether all quantum correlations disappear also. Since $\rho_{12}^{ST}$ has the form of an $X$-state in the limit $\bar{G}_{1} \rightarrow + \infty$  (\ref{XstateG20}), we can calculate numerically very easily with (\ref{entropiaCondicional2Qubits})-(\ref{CorrelacionClasica2Qubits}) the quantum discord of $\rho_{12}^{ST}$ when measurements are carried out on one of the two atoms.

We will first consider the case in which the measurements are made on the atom at position $\mathbf{r}_{2}$. It is found that the steady-state quantum mutual information $I(\rho_{12}^{ST})$ is given by:
\begin{eqnarray}
\label{QMIG201}
I(\rho_{12}^{ST})
&=& \frac{-4 + \sqrt{y}}{2 \sqrt{y}} \mbox{log}_{2}(-4 + \sqrt{y}) \ - \ \frac{\frac{9}{2}F_{12}^{2}}{y}\mbox{log}_{2}\left( \frac{9}{2}F_{12}^{2} \right) \cr
&& + \ \frac{1}{2}\mbox{log}_{2}(y) \ - \ \frac{16 + \frac{9}{2}F_{12}^{2}}{y}\mbox{log}_{2}\left( 16 + \frac{9}{2}F_{12}^{2} \right) \cr
&&  + \frac{ 4 + \sqrt{y}}{2 \sqrt{y}} \mbox{log}_{2}(4 + \sqrt{y}) \ - \ 1 \ ,
\end{eqnarray}
where \ $y = 16+9F_{12}^{2}$. We calculate numerically the right quantum discord.

\begin{figure}[htbp]
  \centering
  \subfloat[]{\label{QDcompletoa11G20X}\includegraphics[scale=0.5]{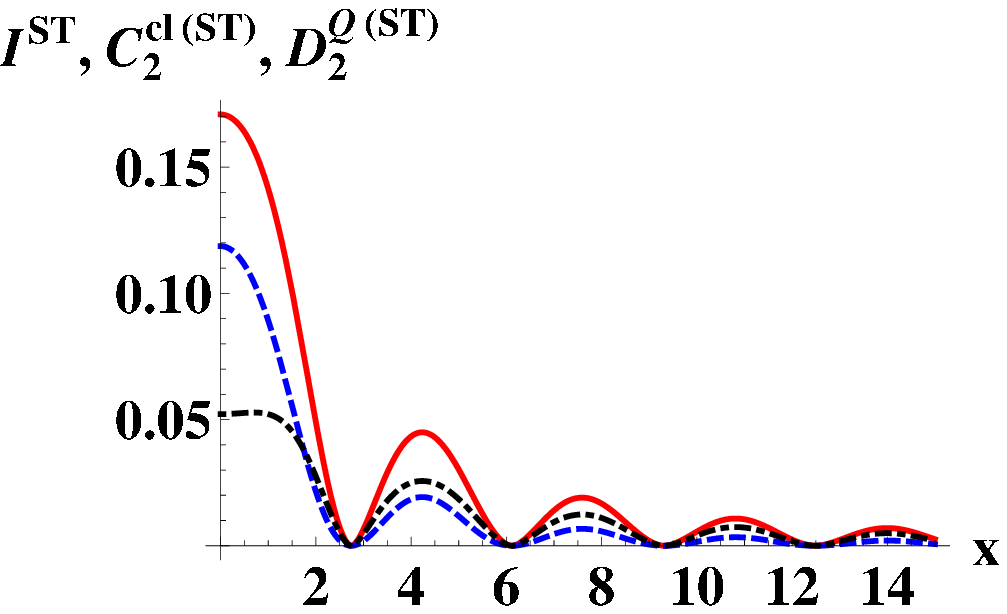}} \hspace{1.5cm}
  \subfloat[]{\label{QDcompletoa10G20X}\includegraphics[scale=0.5]{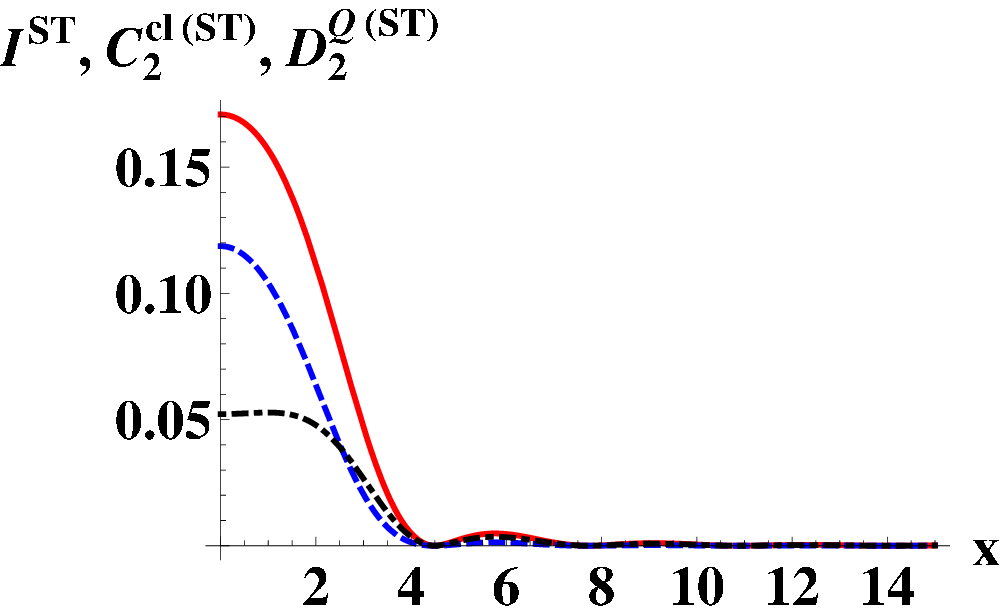}}\\
   \caption{(Color online) Case $G_{2} = 0$:  (\ref{QDcompletoa11G20X}) and (\ref{QDcompletoa10G20X}) show the steady-state quantum mutual information $I^{ST}$ (red-solid line), classical correlations $C_{2}^{cl(ST)}$ (blue-dashed lined), and quantum discord $D_{2}^{Q(ST)}$ (black-dot-dashed line) as a function of $x= \omega_{A}|\mathbf{r}_{1}-\mathbf{r}_{2}|/c$ in the limit $|\bar{G}_{1}| \rightarrow + \infty$. Here measurements are performed on the atom at position $\mathbf{r}_{2}$. (\ref{QDcompletoa11G20X}) illustrates the case $d_{\bot}^2 = |\mathbf{d}_{01}|^{2}$, while (\ref{QDcompletoa10G20X}) shows the case $d_{\bot}^2 = 0$. }
  \label{QDG20X}
\end{figure}

Figure (\ref{QDG20X}) shows the quantum mutual information, right classical correlations, and right quantum discord as a function of $x= \omega_{A}|\mathbf{r}_{1}-\mathbf{r}_{2}|/c$. The first observation is that $\rho_{12}^{ST}$ has both quantum and classical correlations even though the entanglement is long gone (from figure (\ref{concurrenciaCasosG20}) we see that it died abruptly around $\bar{G}_{1} = 2$). In particular, the quantum correlations are larger than the classical ones except for $x \leq 2$. Moreover, notice that the three correlations oscillate as a function of the distance between the atoms and that this oscillatory behaviour is much more pronounced when $d_{\bot}^2 = |\mathbf{d}_{01}|^{2}$. Also, the quantum discord is exactly equal to zero for certain distances. The oscillatory behaviour and the exact location of the discrete zeros of the quantum discord $D_{2}^{Q}(\rho_{12}^{ST})$ are determined by $F_{12}$ (see below for a transparent justification of this). Finally, notice that $D_{2}^{Q}(\rho_{12}^{ST}) = I(\rho_{12}^{ST}) = C_{2}^{cl}(\rho_{12}^{ST}) = 0$ when $|\mathbf{r}_{1} - \mathbf{r}_{2}| \rightarrow + \infty$. Hence, $\rho_{12}^{ST}$ has no correlations at all when the atoms are far apart. This is expected, since the atoms do not interact with each other through the reservoir in this case.

It is of great interest to compare the right quantum discord with the geometric measure of quantum discord given in (\ref{DiscordiaQVedral}) to see how well the latter estimates the exact result. Using that equation, the right geometric discord $D_{2}^{(2)}(\rho_{12}^{ST})$ is easily calculated to be
\begin{equation}
\label{QDVG20X}
D_{2}^{(2)}(\rho_{12}^{ST}) \ = \ \frac{36F_{12}^{2}}{(16 +9F_{12}^{2})^{2}} \ .
\end{equation}

\begin{figure}[htbp]
  \centering
  \subfloat[]{\label{comparaciona11}\includegraphics[scale=0.5]{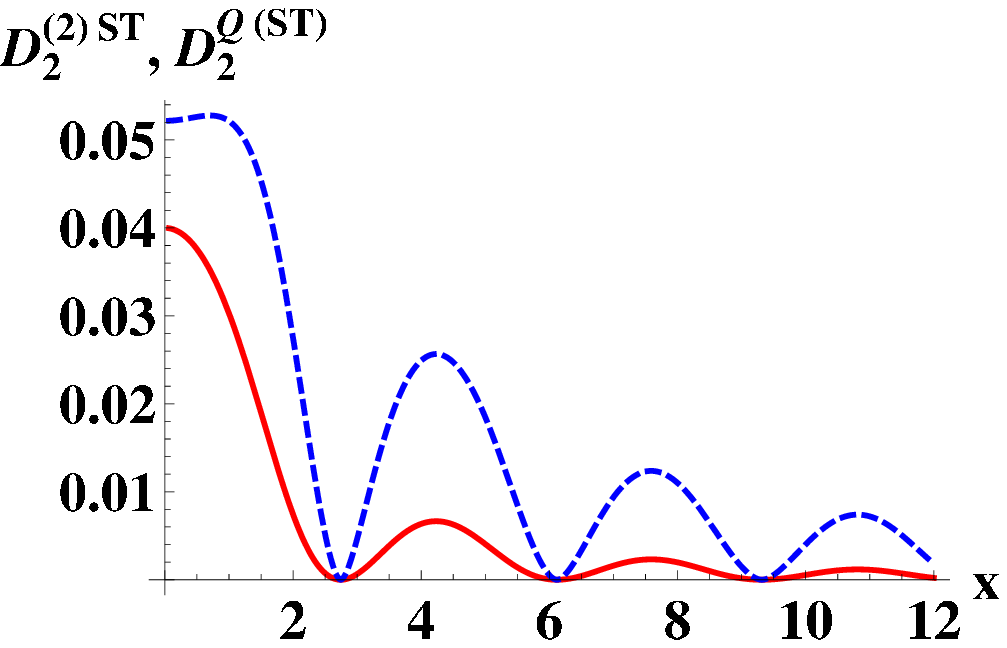}} \hspace{1cm}
  \subfloat[]{\label{comparaciona10}\includegraphics[scale=0.5]{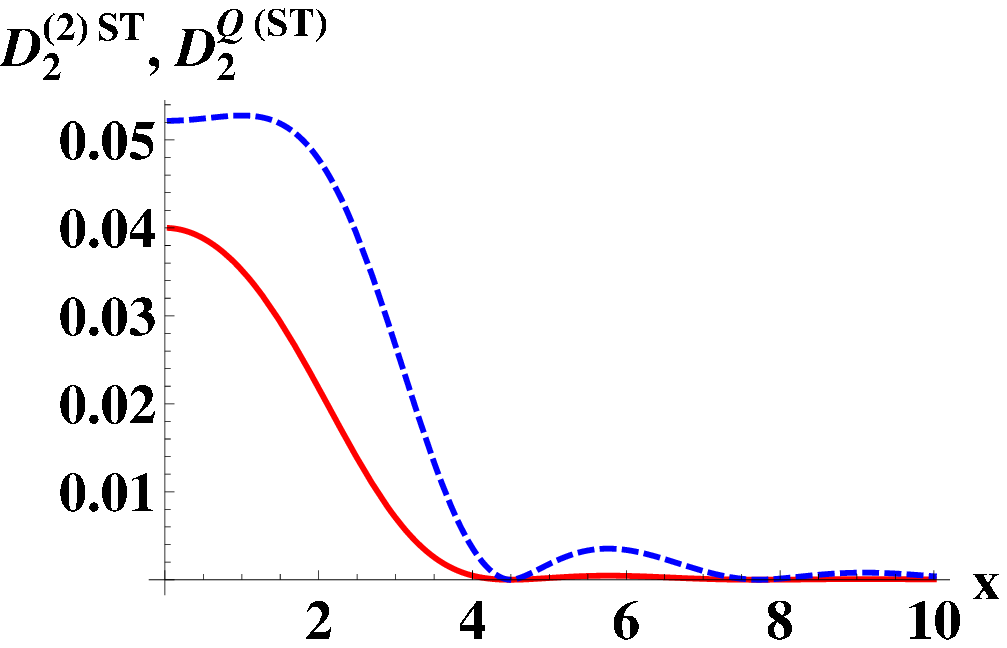}}\\
   \caption{(Color online) Case $G_{2} = 0$: (\ref{comparaciona11}) and (\ref{comparaciona10}) show the steady-state exact quantum discord $D_{2}^{Q}(\rho_{12}^{ST})$ (blue-dashed line) and the geometric measure of quantum discord $D_{2}^{(2)}(\rho_{12}^{ST})$ (red-solid line) as a function of $x= \omega_{A}|\mathbf{r}_{1}-\mathbf{r}_{2}|/c$ in the limit $\bar{G}_{1} \rightarrow + \infty$. Measurements are performed on the atom at position $\mathbf{r}_{2}$. (\ref{comparaciona11}) illustrates the case $d_{\bot}^2 = |\mathbf{d}_{01}|^{2}$, while (\ref{comparaciona10}) shows the case $d_{\bot}^2 = 0$.}
  \label{comparacionG20}
\end{figure}

From figure (\ref{comparacionG20}) we observe that there is good qualitative agreement between the estimate $D_{2}^{(2)}(\rho_{12}^{ST})$ and the exact $D_{2}^{Q}(\rho_{12}^{ST})$. In particular, they have the same zeros. Moreover, we note from (\ref{QDVG20X}) that both the oscillatory behaviour and the zeros of $D_{2}^{(2)}(\rho_{12}^{ST})$ are due to $F_{12}$. Hence, it follows that the discrete zeros of the exact quantum discord $D_{2}^{Q}(\rho_{12}^{ST})$ are exactly those of $F_{12}$, (\ref{F}). Also note that the decrease in the amplitude of the oscillations of $D_{2}^{Q}(\rho_{12}^{ST})$ when $d_{\bot}^2 = 0$ is due to the fact that the oscillations of $F_{12}$ are greatly decreased when $d_{\bot}^2 = 0$. Nevertheless, notice that the exact quantum discord is considerably larger than that given in (\ref{QDVG20X}), so, in general, the quantitative agreement between $D_{2}^{Q}(\rho_{12}^{ST})$ and $D_{2}^{(2)}(\rho_{12}^{ST})$ is not so good (except in determining the zeros of $D_{2}^{Q}(\rho_{12}^{ST})$).

\begin{figure}[htbp]
  \centering
  \subfloat[]{\label{QDVa11G20A2}\includegraphics[scale=0.5]{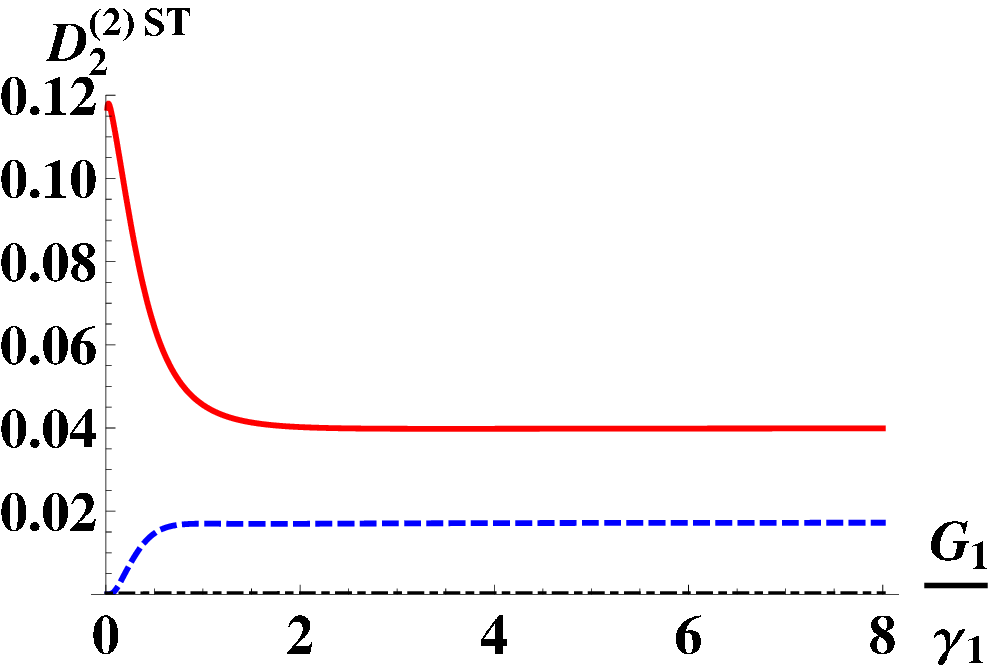}}\hspace{1cm}
  \subfloat[]{\label{QDVa10G20A2}\includegraphics[scale=0.5]{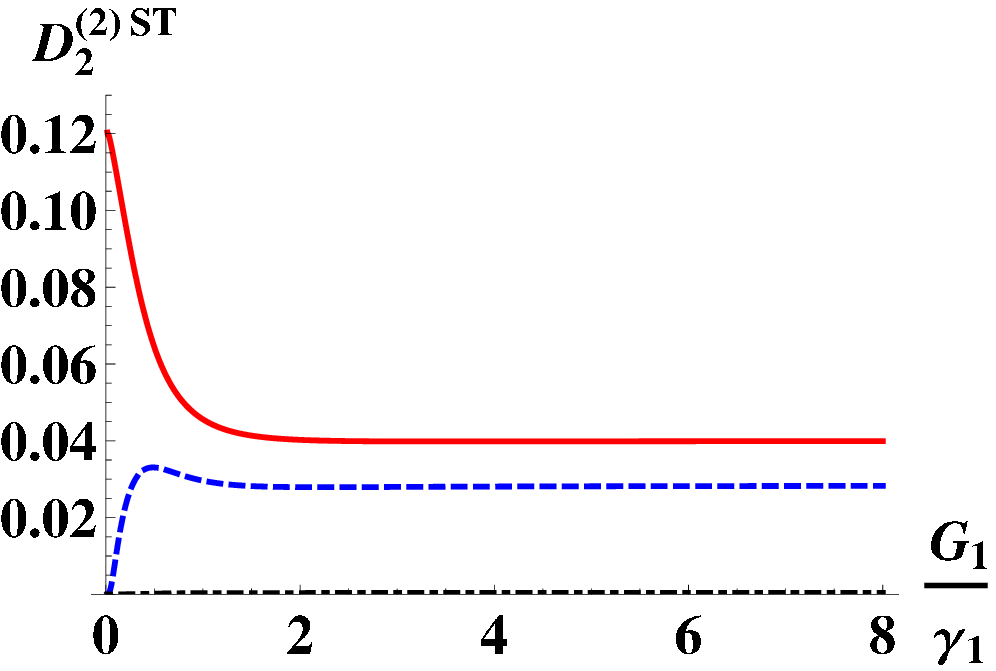}}\\
   \caption{(Color online) Case $G_{2} = 0$: (\ref{QDVa11G20A2}) and (\ref{QDVa10G20A2}) show the steady-state geometric discord $D_{2}^{(2)ST}$ as a function of $\bar{G}_{1} = G_{1}/\gamma_{1}$ for a distance between the two atoms equal to $\lambda_{A}/100$ (red-solid line), $\lambda_{A}/4$ (blue-dashed line), $\lambda_{A}$ (black-dot-dashed line) where $\lambda_{A}$ is the wavelength associated with the atomic transition. (\ref{QDVa11G20A2}) illustrates the case $d_{\bot}^2 = |\mathbf{d}_{01}|^{2}$, while (\ref{QDVa10G20A2}) shows the case $d_{\bot}^2=0$.}
  \label{QDVedralG20X}
\end{figure}

\begin{figure}[htbp]
  \centering
  \vspace{2cm}
  \subfloat[]{\label{QDVa11G20contour}\includegraphics[scale=0.45]{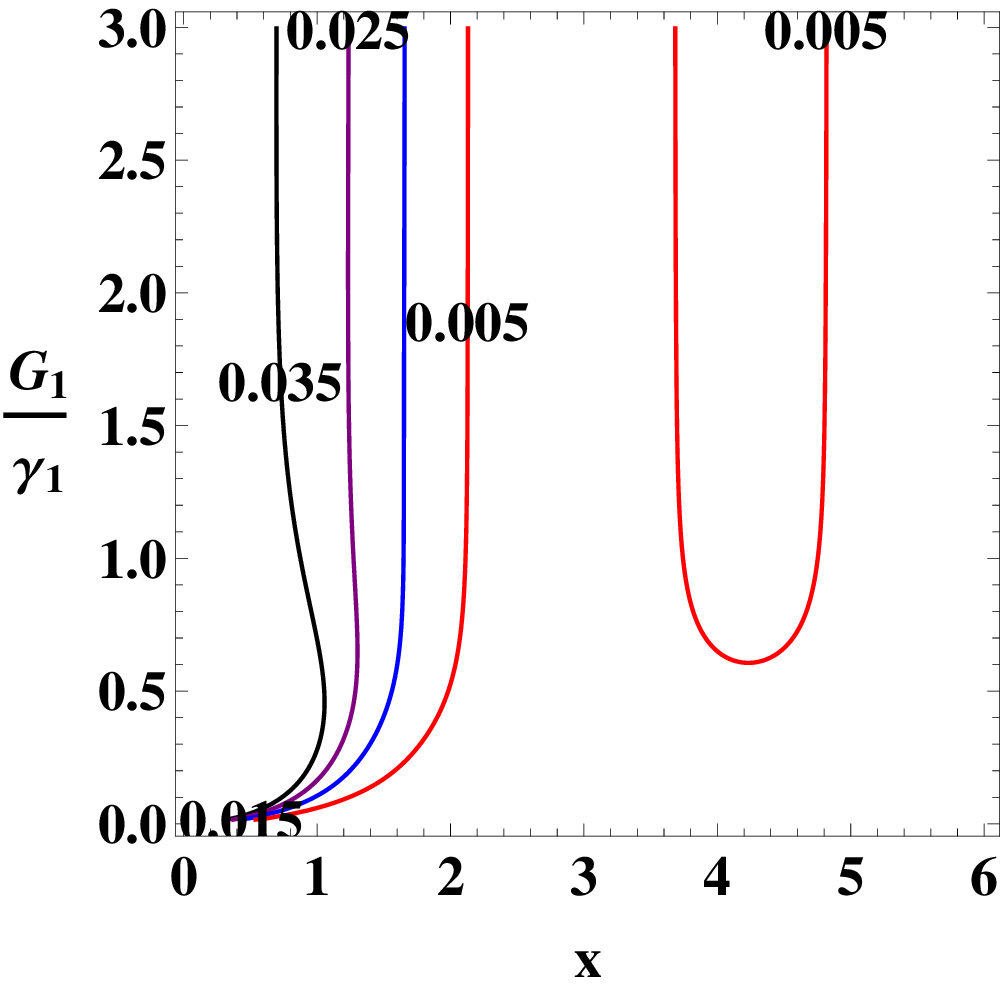}}\hspace{1.5cm}
  \subfloat[]{\label{QDVa10G20contour}\includegraphics[scale=0.5]{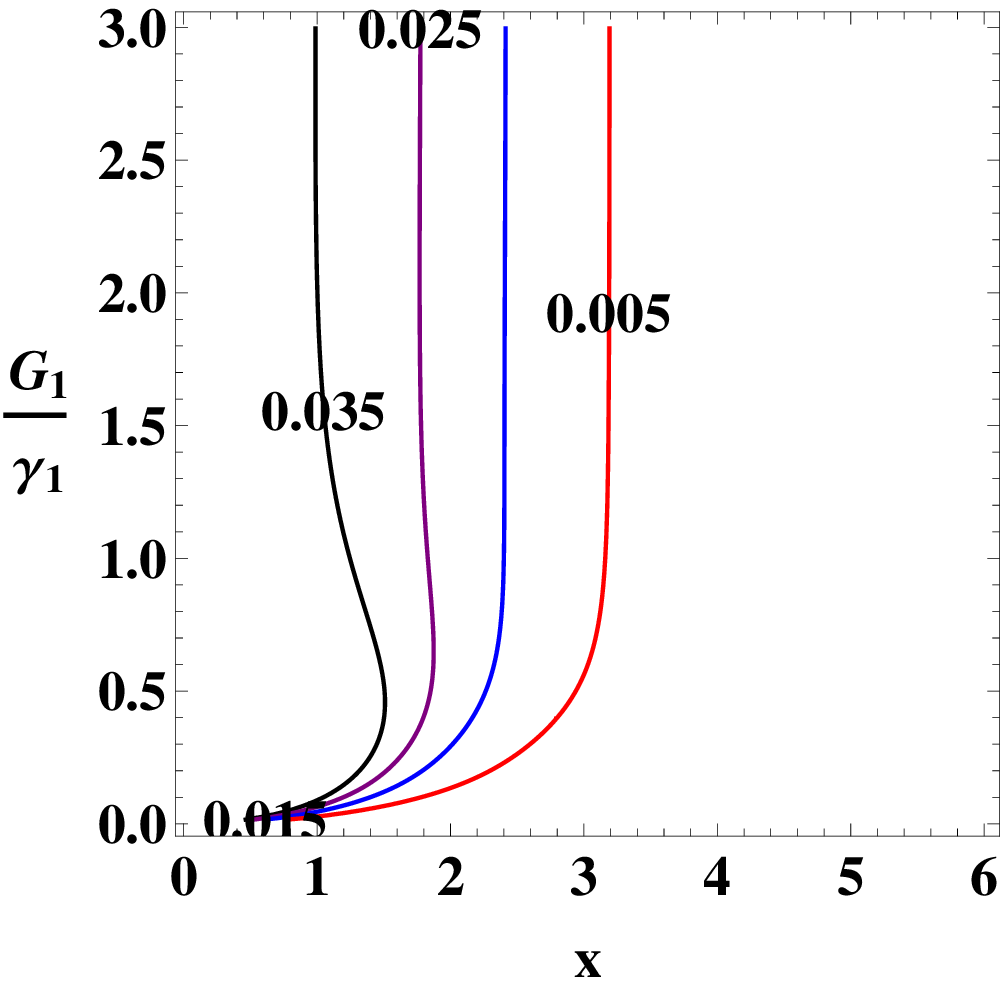}}\\
   \caption{(Color online) Case $G_{2} = 0$: (\ref{QDVa11G20contour}) and (\ref{QDVa10G20contour}) show contour plots of the steady-state geometric discord $D_{2}^{(2)}(\rho_{12}^{ST})$ as a function of $x = \omega_{A}|\mathbf{r}_{1}-\mathbf{r}_{2}|/c$ (horizontal axis) and $\bar{G}_{1}$ (vertical axis) for the cases $d_{\bot}^2 = |\mathbf{d}_{01}|^{2}$ and $d_{\bot}^2 = 0$, respectively. Contours for $D_{2}^{(2)}(\rho_{12}^{ST}) = 0.035, 0.025, 0.015, 0.005$ are drawn.}
  \label{QDVcontour1}
\end{figure}

The geometric discord in (\ref{DiscordiaQVedral}) allows us to calculate easily an estimate of the quantum discord for the general case (\ref{RhoAEstG20})-(\ref{RhoAEst2}). Figure (\ref{QDVedralG20X}) shows that for a fixed distance between the atoms the geometric discord quickly achieves the stationary value given in (\ref{QDVG20X}). Notice that the decay of the geometric discord with increasing distance between the atoms is slower when $d_{\bot}^2 = 0$. This is seen more easily in the contour plots of the geometric discord shown in figure (\ref{QDVcontour1}). In these figures it is also observed that, as a function of the distance between the atoms, the geometric discord has oscillations of a greater magnitude when $d_{\bot}^2 = |\mathbf{d}_{01}|^{2}$. We saw above the these are due to $F_{12}$.

We now consider the case in which the measurements are performed on the atom at position $\mathbf{r}_{1}$. In the limiting case where $|\bar{G}_{1}| \rightarrow + \infty$, $\rho_{12}^{ST}$ takes the form of an $X$-state (\ref{XstateG20}). The quantum mutual information $I(\rho_{12}^{ST})$ is exactly the same as before (this quantity is independent of who is being measured), but we find (numerically) that the quantum discord $D_{1}^{Q}(\rho_{12}^{ST})$ is now zero and the classical correlations $C_{1}^{cl}(\rho_{12}^{ST})$ are equal to $I(\rho_{12}^{ST})$. Hence, all information of the atom at position $\mathbf{r}_{2}$ contained in the correlations of $\rho_{12}^{ST}$ can be extracted by measuring the state of the atom at $\mathbf{r}_{1}$ \cite{Zurek}. This contrasts greatly with the case considered above, that is, with the case were the measurements are performed on the atom at $\mathbf{r}_{2}$. There we found that $\rho_{12}^{ST}$ presents non-negligible quantum discord. Thus, when measurements are performed only on the atom at $\mathbf{r}_{2}$, $\rho_{12}^{ST}$ is perturbed and there is information that cannot be extracted \cite{Zurek}. Furthermore, we have here another example were both the quantum discord and the classical correlations are not symmetric: $D_{1}^{Q}(\rho_{12}^{ST}) \not= D_{2}^{Q}(\rho_{12}^{ST})$ and $C_{1}^{cl}(\rho_{12}^{ST}) \not= C_{2}^{cl}(\rho_{12}^{ST})$.

\begin{figure}[htbp]
  \centering
  \subfloat[]{\label{QDVAtomo1a11G20}\includegraphics[scale=0.5]{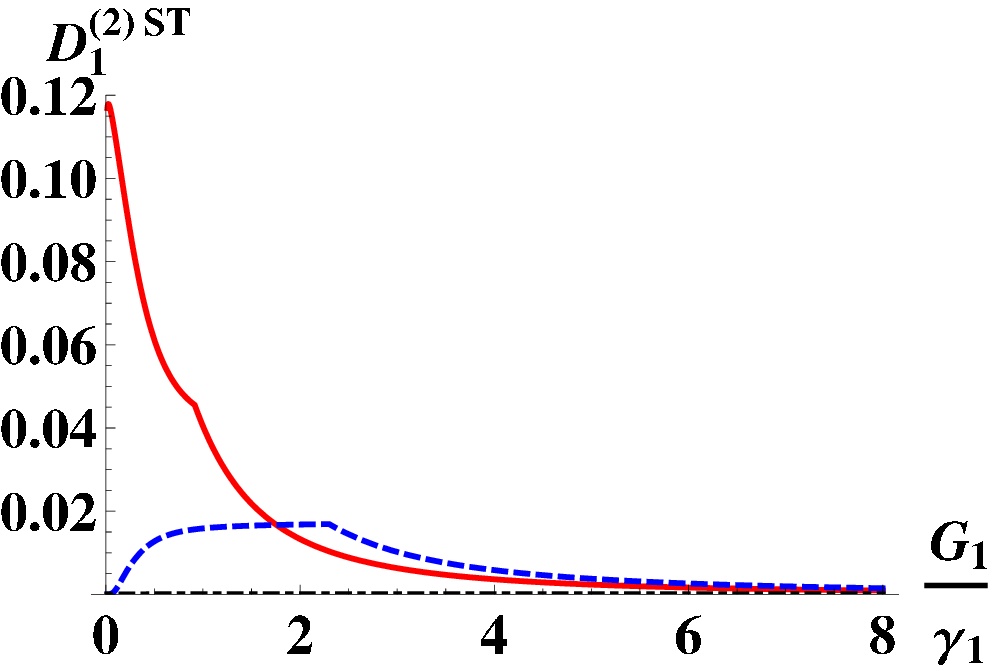}}\hspace{1cm}
  \subfloat[]{\label{QDVAtomo1a10G20}\includegraphics[scale=0.5]{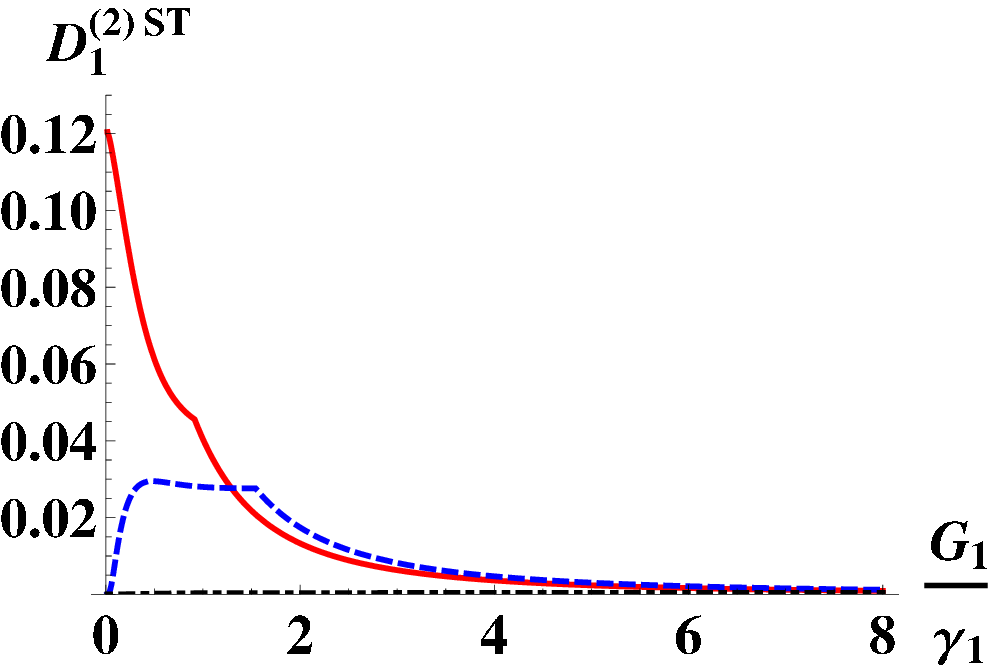}}\\
   \caption{(Color online) Case $G_{2} = 0$: (\ref{QDVAtomo1a11G20}) and (\ref{QDVAtomo1a10G20}) illustrate the steady-state geometric discord $D_{1}^{(2)ST}$ as a function of $\bar{G}_{1} = G_{1}/\gamma_{1}$ when the distance between the atoms is $\lambda_{A}/100$ (red-solid line), $\lambda_{A}/4$ (blue-dashed line), and $\lambda_{A}$ (black-dot-dashed line) where $\lambda_{A}$ is the wavelength associated with the atomic transition. (\ref{QDVAtomo1a11G20}) shows the case $d_{\bot}^2 = |\mathbf{d}_{01}|^{2}$, while (\ref{QDVAtomo1a10G20}) illustrates the case $d_{\bot}^2=0$.}
  \label{QDVAtomo1G20bis}
\end{figure}

\begin{figure}[htbp]
  \centering
  \vspace{1.5cm}
  \subfloat[]{\label{QDVa11G20contour1}\includegraphics[scale=0.5]{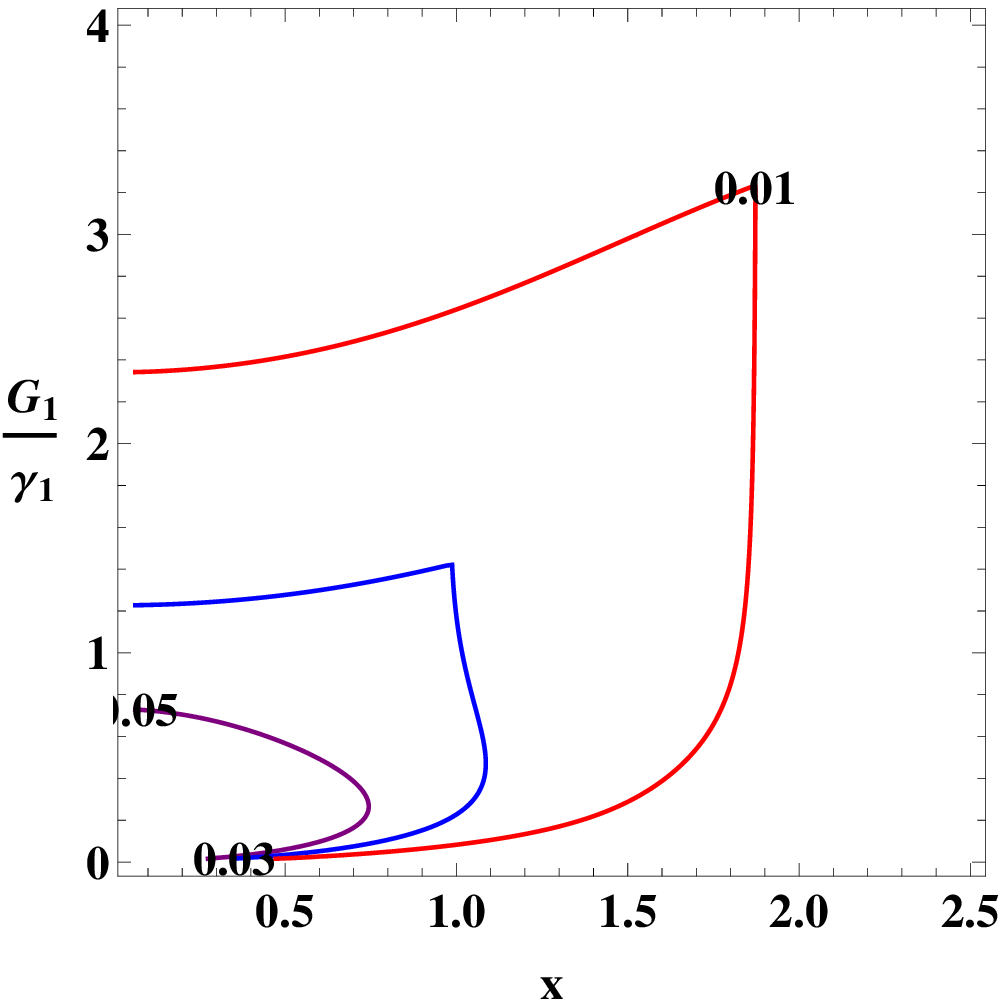}}\hspace{1cm}
  \subfloat[]{\label{QDVa10G20contour1}\includegraphics[scale=0.5]{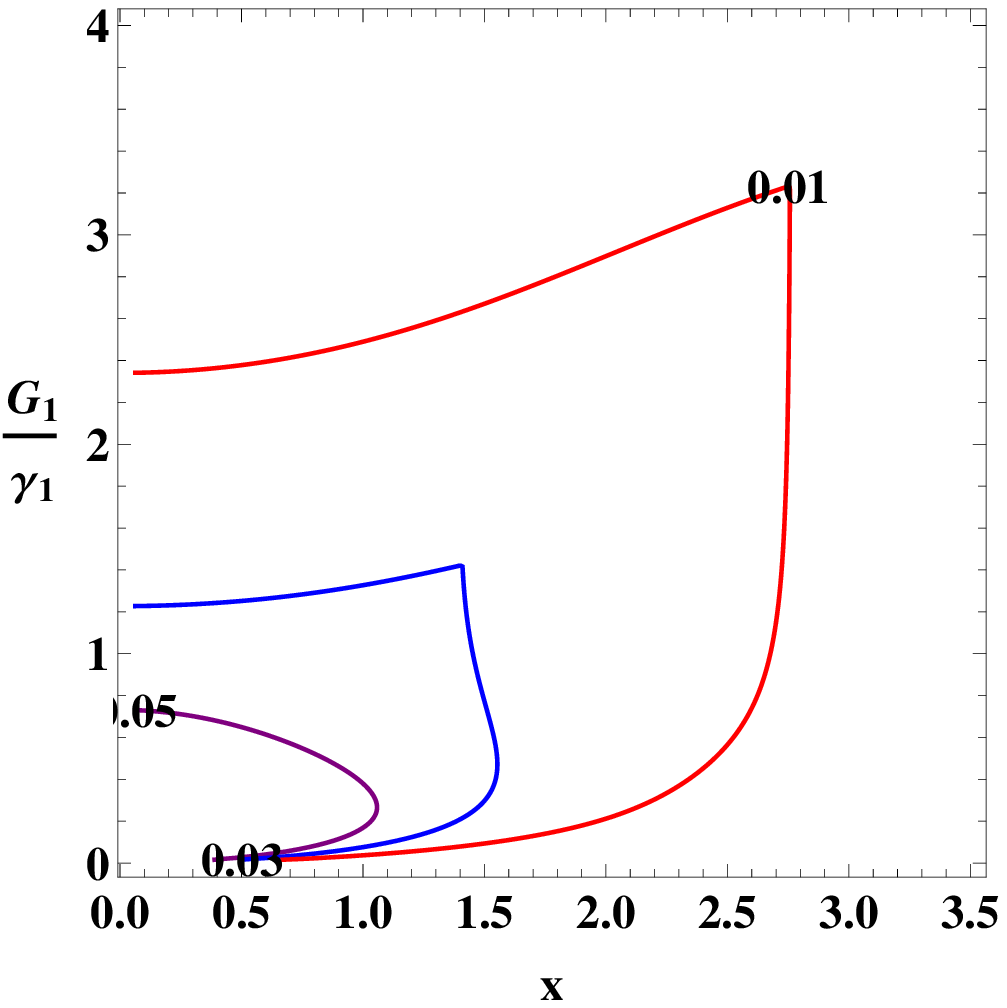}}\\
   \caption{(Color online) Case $G_{2} = 0$: (\ref{QDVa11G20contour1}) and (\ref{QDVa10G20contour1}) show contour plots of the geometric discord $D_{1}^{(2)}(\rho_{12}^{ST})$ as a function of $x = \omega_{A}|\mathbf{r}_{1}-\mathbf{r}_{2}|/c$ (horizontal axis) and of $\bar{G}_{1} = G_{1}/\gamma_{1}$ (vertical axis) for the cases $d_{\bot}^2 = |\mathbf{d}_{01}|^{2}$ and $d_{\bot}^2=0$, respectively. Contours for $D_{1}^{(2)}(\rho_{12}^{ST}) = 0.05, 0.03, 0.01$ are drawn.}
  \label{QDVcontour2}
\end{figure}

Again the use of the geometric discord (\ref{DiscordiaQVedral}) allows us to give an estimate of the quantum discord in the general case (\ref{RhoAEstG20})-(\ref{RhoAEst2}). From figure (\ref{QDVAtomo1G20bis}) we note that the geometric discord as a function of $\bar{G}_{1}$ is smooth by parts, since it exhibits edges. Also notice that it tends to zero as $\bar{G}_{1}$ increases, a result in accordance with that obtained above $D_{1}^{Q}(\rho_{12}^{ST}) = 0$ in the limit $\bar{G}_{1} \rightarrow 0$. Contour plots of $D_{1}^{(2)}(\rho_{12}^{ST})$ are also shown in figure (\ref{QDVcontour2}). Notice that the $D_{1}^{(2)}(\rho_{12}^{ST})$ decays more slowly as a function of the distance between atoms for $d_{\bot}^2 = 0$ when compared to $d_{\bot}^2 = | \mathbf{d}_{01}|^{2}$, and that it is drastically different to $D_{2}^{(2)}(\rho_{12}^{ST})$ (compare figures (\ref{QDVcontour1}) and (\ref{QDVcontour2})).

\begin{figure}[htbp]
  \centering
  \subfloat[]{\label{SLa11G20}\includegraphics[scale=0.5]{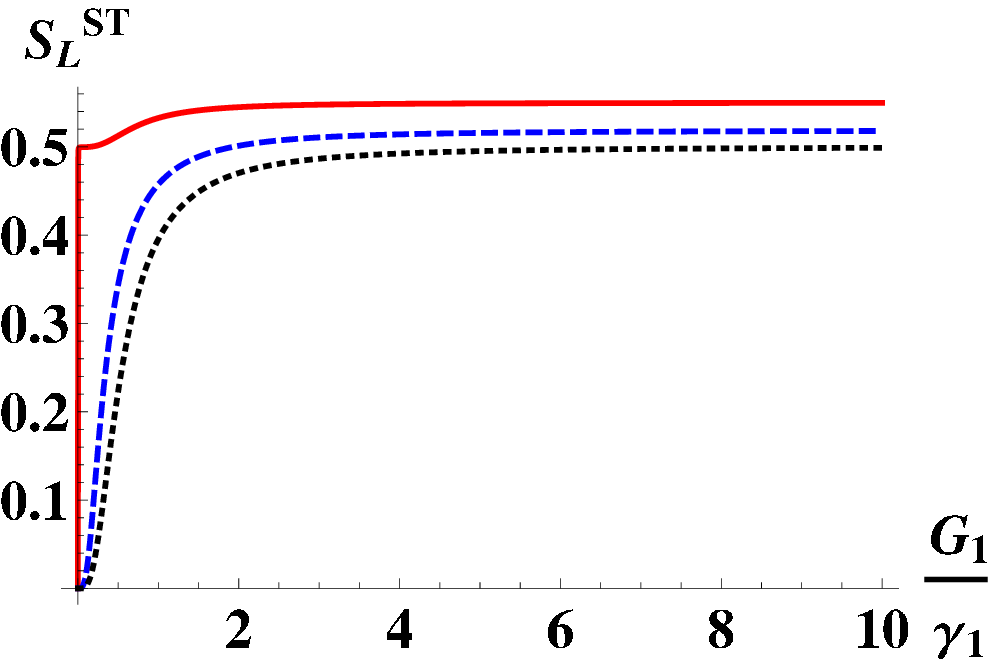}}\hspace{1cm}
  \subfloat[]{\label{SLa10G20}\includegraphics[scale=0.5]{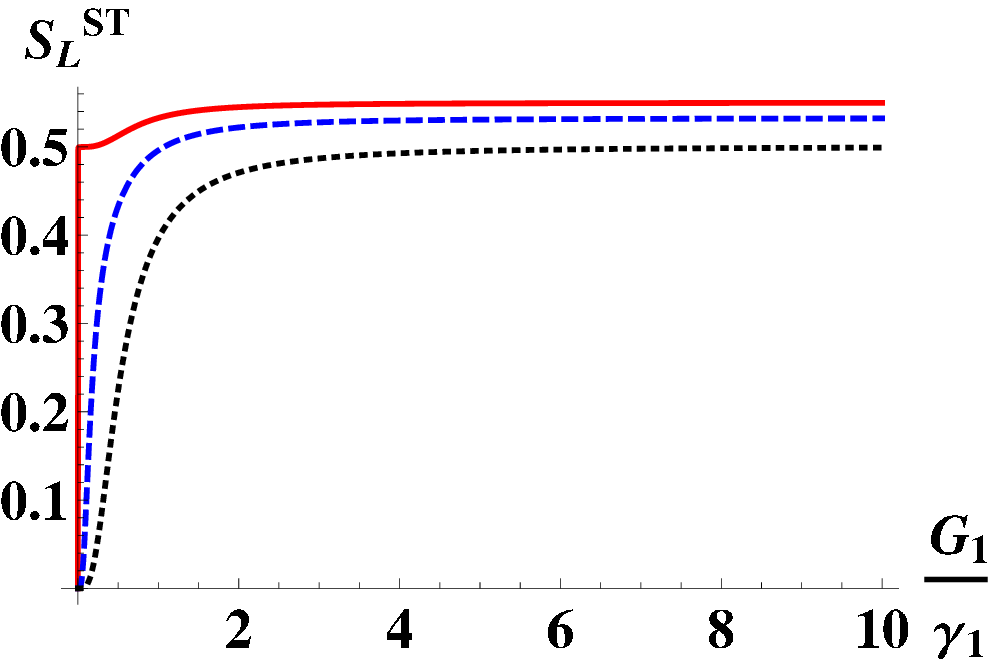}}\\
   \caption{(Color online) Case $G_{2} = 0$: (\ref{SLa11G20}) and (\ref{SLa10G20}) illustrate the steady-state linear entropy $S_{L}^{ST}$ as a function of $\bar{G}_{1} = G_{1}/\gamma_{1}$ for $d_{\bot}^{2} = |d_{01}|^{2}$ and $d_{\bot}^{2} = 0$, respectively. Results are shown when the distance between the two atoms is $\lambda_{A}/100$ (red-solid line), $\lambda_{A}/4$ (blue-dashed line), and $\lambda_{A}$ (black-dotted line) where $\lambda_{A}$ is the wavelength associated with the atomic transition.}
  \label{SLG20}
\end{figure}

We now turn to discuss the degree of mixed-ness of the $\rho_{12}^{ST}$, figure (\ref{SLG20}). We observe that as the intensity of the laser field is increased ($\bar{G}_{1}$ increases), $S_{L}^{ST}$ increases until it acquires an asymptotic value which is easily calculated from (\ref{XstateG20}):
\begin{equation}
\label{SLG20A}
S_{L}^{ST} \ = \ \frac{3}{4} - \frac{4}{16 + 9F_{12}^{2}} \ .
\end{equation}
Notice that the system is never in a maximum mixed state (as measured by the linear entropy) and that the maximum value $11/20$ of $S_{L}^{ST}$ occurs when the atoms are very close. Also, the linear entropy takes an asymptotic value of $1/2$ when the atoms are far apart. This behaviour is understandable since the dissipative interaction between the two atoms by means of the reservoir is strongest when the atoms are close together, and only one atom is affected by the laser field when the distance between the two atoms is large. Finally, we note that the variations of $S_{L}^{ST}$ are smoothed out when $d_{\bot}^2 = 0$, figure (\ref{SLG20}).

\begin{figure}[htbp]
  \centering
  \subfloat[]{\label{ConcSLG20}\includegraphics[scale=0.5]{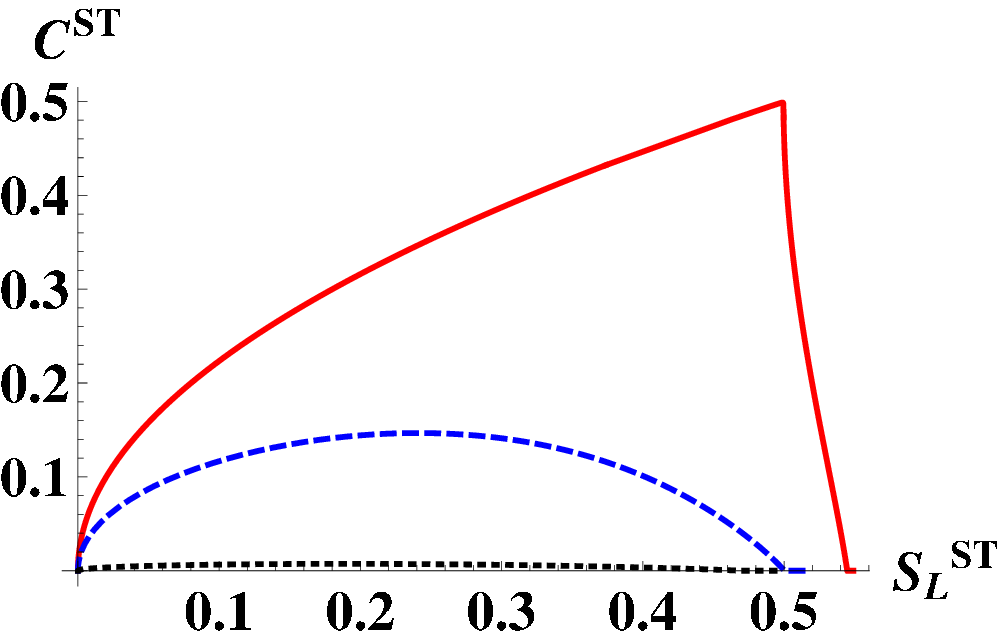}}\hspace{1cm}
  \subfloat[]{\label{ConcSLG1G2}\includegraphics[scale=0.5]{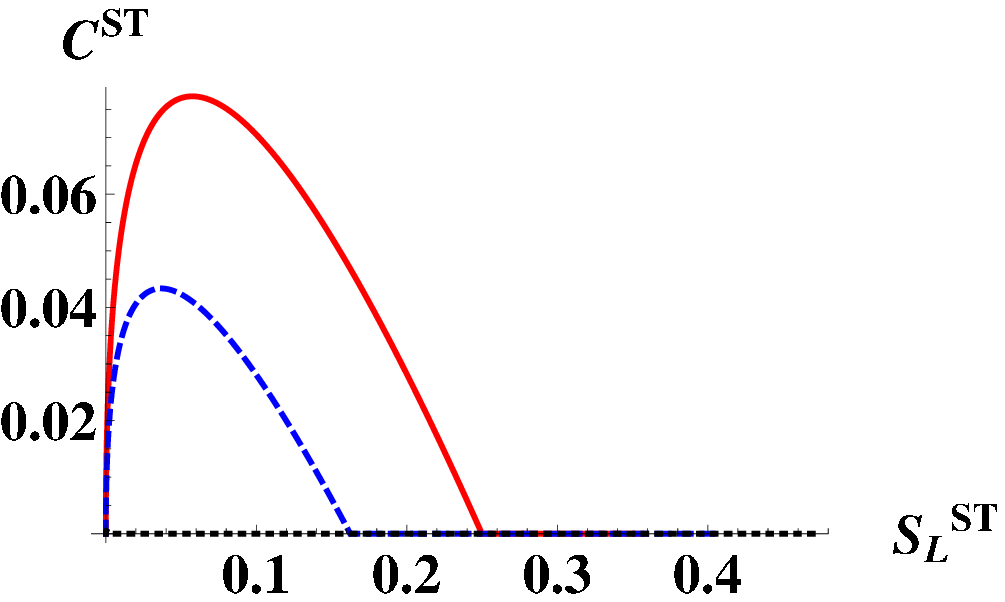}}\\
   \caption{(Color online) The figures (\ref{ConcSLG20}) and (\ref{ConcSLG1G2}) illustrate the steady-state concurrence $C^{ST}$ as a function of the steady-state linear entropy $S_{L}^{ST}$ for the cases $\bar{G}_{2} = 0$ and $\bar{G}_{1} = \bar{G}_{2}$, respectively. In these figures $d_{\bot}^2 = |\mathbf{d}_{01}|^{2}$.}
  \label{ConcSL}
\end{figure}

To end this Section we discuss the relation between the entanglement between the two atoms as measured by the concurrence $C^{ST}$ and the degree of mixed-ness as measured by the linear entropy $S_{L}^{ST}$. We observe that as state becomes more mixed, the degree of entanglement increases until $S_{L}^{ST}$ reaches a certain value, after which the degree of entanglement decays to zero, figure (\ref{ConcSLG20}). This behaviour corresponds to the fact that as the intensity of the laser field increases, the state of the two atoms becomes entangled and mixed due to the exchange of spontaneously emitted photons. Nevertheless, once the intensity of the electric field is strong enough to drive the populations of the triplet-singlet basis (\ref{acoplada}) near their asymptotic values, the entanglement begins to decrease while the state of the system still becomes more mixed. We also note that the behaviour shown in figure (\ref{ConcSLG20}) also illustrates the known facts that a mixed state of two qubits cannot contain an arbitrary amount of entanglement, and that the more mixed the state becomes, the less entanglement it has \cite{MaxEntanglement}.

\section{$G(\mathbf{r}_{1}) = G(\mathbf{r}_{2})$}

In this section we consider that the atoms are fixed at positions where the classical electric field (\ref{EL}) has the same value. Hence, $G_{1} = G_{2}$ which  can be varied by changing the intensity of the laser field (see the definition of $G(\mathbf{r})$ following (\ref{EL})). The value of the distance $|\mathbf{r}_{1}-\mathbf{r}_{2}|$ between the two atoms can also be varied independently. For example, such case would occur in the standing wave configuration with $g(\mathbf{r}_{j}) = \mbox{cos}(\mathbf{k}_{L}\cdot \mathbf{r}_{j})$ if the two atoms are located on the same line perpendicular to the wave vector $\mathbf{k}_{L}$ of the standing wave.

The solution $\rho_{12}^{ST}$ of (\ref{SEstacionaria}) has the following matrix representation in the \textit{triplet-singlet basis} $\mathsf{B}$ (\ref{acoplada}):
\begin{eqnarray}
\label{RhoAEstG1G2}
[\rho_{12}^{ST}]_{\mathsf{B}} \ = \ \cr
\frac{1}{\kappa_{2}}
\left(
\begin{array}{cccc}
4 \bar{G}_{1}^{4}   & ie^{-i \delta_{\mbox{\tiny{L}}}t}2 \sqrt{2} \bar{G}_{1}^{3}    \cr
-ie^{i \delta_{\mbox{\tiny{L}}}t}2 \sqrt{2} \bar{G}_{1}^{3}  &  4 \bar{G}_{1}^{4} + 2 \bar{G}_{1}^{2}   \cr
-e^{i 2 \delta_{\mbox{\tiny{L}}}t} (1 + \frac{3}{2}F_{12}) \bar{G}_{1}^{2}                 & -ie^{i\delta_{\mbox{\tiny{L}}}t}\frac{\sqrt{2}}{2}\bar{G}_{1}\left( 1 + \frac{3}{2}F_{12} + 4\bar{G}_{1}^{2} \right)   \cr
 0 & 0
\end{array}
\right. \cr
        \cr
        \cr
\left.
\begin{array}{cccc}
-e^{-i 2 \delta_{\mbox{\tiny{L}}}t} (1 + \frac{3}{2}F_{12}) \bar{G}_{1}^{2}     & 0  \cr
ie^{-i\delta_{\mbox{\tiny{L}}}t}\frac{\sqrt{2}}{2}\bar{G}_{1}\left( 1 + \frac{3}{2}F_{12} + 4\bar{G}_{1}^{2} \right)   & 0 \cr
4 \bar{G}_{1}^{4} + 2 \bar{G}_{1}^{2} + \frac{1}{4}(1+\frac{3}{2}F_{12})^{2}           & 0 \cr
0 & 4 \bar{G}_{1}^{4}
\end{array}
\right) \ . 
\end{eqnarray} 
Here
\begin{equation}
\label{kappa2}
\kappa_{2} \ = \ 16 \bar{G}_{1}^{4} + 4\bar{G}_{1}^{2} + \frac{1}{4}\left( 1 + \frac{3}{2}F_{12} \right)^{2} \ ,
\end{equation}
and $\bar{G}_{1} = G_{1}/\gamma_{1}$ again. We note that this solution is valid only if the distance $|\mathbf{r}_{1} - \mathbf{r}_{2}|$ between the two atoms is not zero. If the $|\mathbf{r}_{1} - \mathbf{r}_{2}| = 0$, then the solution is
\begin{eqnarray}
\label{RhoEstG1G2Est}
\rho_{12}^{ST} \ = \ (1-P_{00})\rho_{12,1}^{ST} + P_{00}|0,0\rangle\langle 0,0| \ ,
\end{eqnarray}
where $P_{00}$ is the population of the state $|0,0\rangle$ in the initial state, and $\rho_{12,1}^{ST}$ is obtained from (\ref{RhoAEstG1G2}) by replacing $\kappa_{2}$ and $\langle 0,0| \rho_{12}^{ST} |0,0\rangle$ by
 \begin{equation}
 \label{kappa3}
 \kappa_{3} \ = \ 12 \bar{G}_{1}^{4} + 4\bar{G}_{1}^{2} + 1 \ ,
 \end{equation}
and $0$, respectively, and by taking the limit $|\mathbf{r}_{1}-\mathbf{r}_{2}| \rightarrow 0^{+}$ in the resulting expression.

Note that some coherences depend explicitly on time; hence, in the IP we are working, a steady-state solution exists only for the resonant case ($\delta_{\mbox{\tiny{L}}} = 0$). The steady-state solution in (\ref{RhoAEstG1G2}) and (\ref{RhoEstG1G2Est}) would also be approximately valid for small enough  detuning, for instance $| \delta_{\mbox{\tiny{L}}}t | \ll 1$. In the following we will restrict to the case where the steady-state solution exists, that is, we will take $\delta_{\mbox{\tiny{L}}} = 0$. Moreover, we will consider only the case $|\mathbf{r}_{1}-\mathbf{r}_{2}| \not= 0$, although we will make some remarks about the case $|\mathbf{r}_{1}-\mathbf{r}_{2}| = 0$.

In complete analogy  to the case $\bar{G}_{2} = 0$, it can be shown that, given any initial state $\rho_{12}(0)$, the density operator $\rho_{12}(t)$ of the two atoms tends to the state (\ref{RhoAEstG1G2}) above if the distance between the two atoms is not zero and $\delta_{\mbox{\tiny{L}}} = 0$.

For the case $|\mathbf{r}_{1}-\mathbf{r}_{2}| = 0$ and $\delta_{\mbox{\tiny{L}}} = 0$ the argument is a bit different because the zero eigenvalue has both algebraic and geometric multiplicity equal to two, and the rest of the eigenvalues have negative real part. In a similar notation as before it follows that
\begin{eqnarray}
\mathbf{x}(t) \ \rightarrow \  \tilde{c}_{1}\tilde{\mathbf{v}}_{1} + \tilde{c}_{2}\tilde{\mathbf{v}}_{2} \ \ \ \mbox{if} \ t \rightarrow +\infty .
\end{eqnarray}
Now it can be shown that $\mathbf{e}_{16}$ (the vector which has a $1$ in the $16$-th component and $0$ in the rest) and the vector  associated with $\rho_{12,1}^{ST}$ in (\ref{RhoEstG1G2Est}) are elements of the null space of $\mathbb{A}$, and that $\mathbf{e}_{16}$ is orthogonal to the rest of the eigenvectors of $\mathbb{A}$. Hence, $\tilde{\mathbf{v}}_{1}$ and $\tilde{\mathbf{v}}_{2}$ can be chosen to be those vectors, and $\tilde{c}_{1}$ and $\tilde{c}_{2}$ must be respectively equal to $P_{00}$ and $1-P_{00}$, where $P_{00}$ is the population of $|0,0 \rangle$ in the initial state $\rho_{12}(0)$. Thus any initial state $\rho_{12}(0)$ tends to the density operator in (\ref{RhoEstG1G2Est}) when $|\mathbf{r}_{1}-\mathbf{r}_{2}| = 0$ and $\delta_{\mbox{\tiny{L}}}=0$.

We immediately observe from (\ref{RhoAEstG1G2}) that the effect of the collective interaction with the reservoir is much smaller than in the case $G_{2} = 0$ studied above. This is due to the fact that the term $F_{12}$ appears in combinations in which it rapidly becomes negligible when $\bar{G}_{1}$ increases (see (\ref{RhoAEstG1G2}) and (\ref{kappa2}) and the matrix elements there). Also notice that the collective interaction with the reservoir has an opposite effect in the populations of the triplet-singlet basis $\mathsf{B}$ (\ref{acoplada}) when compared with the case where $\bar{G}_{2} = 0$. In the latter the effect is to increase the populations of the states $|1,1\rangle$ and $|0,0\rangle$ and to decrease the populations of the states $|1,0\rangle$ and $|1,-1\rangle$ when the atoms are close. This is easily seen in the high laser field intensity limit (\ref{poblacionesG20inf}). On the other hand, in the case $\bar{G}_{1} = \bar{G}_{2}$ the effect is to decrease all the populations of the states of the triplet-singlet basis $\mathsf{B}$ (\ref{acoplada}), except for the state $|1,-1\rangle$. Nevertheless, the effect is small, since the terms in $\bar{G}_{1}$ quickly overwhelm those with $F_{12}$ as the laser field intensity increases ($\bar{G}_{1} = \bar{G}_{2}$ grows).

There are several limiting cases of interest. First, if $\bar{G}_{1} \rightarrow 0$, then it is seen from (\ref{RhoAEstG1G2}) that $\rho_{12}^{ST} \rightarrow |1,-1\rangle\langle 1,-1|$. This result is expected since the atoms are located at different positions and, without the driving field, the reservoir ultimately leaves the two atoms in their respective ground states. Notice that this result would have not been obtained if the atoms where exactly in the same position $|\mathbf{r}_{1}-\mathbf{r}_{2}| = 0$ because in that case the antisymmetric state $|0,0\rangle$ is invulnerable to dissipation.

When the laser field intensity is weak ($|\bar{G}_{1}| = |\bar{G}_{2}| \ll 1$) we find from (\ref{RhoAEstG1G2}) that the matrix representation of $\rho_{12}^{ST}$ in the triplet-singlet basis $\mathsf{B}$ (\ref{acoplada}) takes the form
\begin{eqnarray}
\label{RhoG1G2c}
[\rho_{12}^{ST}]_{\mathsf{B}} \ = \
\left(
\begin{array}{cccc}
 0 & 0 & -\frac{8 \bar{G}_{1}^{2}}{2 + 3F_{12}} & 0 \\
 0 & \frac{32\bar{G}_{1}^{2}}{(2 + 3F_{12})^{2}} & i\frac{4\sqrt{2}\bar{G}_{1}}{2 + 3F_{12}} & 0 \\
 -\frac{8 \bar{G}_{1}^{2}}{2 + 3F_{12}} & -i\frac{4\sqrt{2}\bar{G}_{1}}{2 + 3F_{12}} & 1-\frac{32\bar{G}_{1}^{2}}{(2 + 3F_{12})^{2}} & 0 \\
 0 & 0 & 0 & 0
\end{array}
\right) \ ,
\end{eqnarray}
to second order in $\bar{G}_{1}$. Notice that $\bar{G}_{1}$ must be sufficiently small in order for (\ref{RhoG1G2c}) to be valid. In fact, it must occur that
\begin{eqnarray}
\label{CotaG1G2}
\bar{G}_{1}^{2} \ \leq \ \frac{(2+3F_{12})^{2}}{32} \ ,
\end{eqnarray}
to have a population of the state $|1,0\rangle$ less than or equal to $1$. Hence, the symmetric state $|1,0\rangle$ is much more populated than the antisymmetric state $|0,0\rangle$ for a weak laser field intensity  ($\bar{G}_{1} = \bar{G}_{2}$ is small). Notice that this contrasts greatly with the case $G_{2} = 0$ (see (\ref{cotas00})). From (\ref{RhoG1G2c}) it is also observed that the collective interaction with the reservoir tends to decrease the population of the state $|1,0\rangle$ and to increase that of the state $|1,-1\rangle$ when the atoms are close. It is important to keep in mind that for very small distances between the atoms the effects of the coherent dipole-dipole interaction not incorporated in the model under consideration can be very important.

Another limiting case of more interest occurs when the electric field is very intense ($|\bar{G}_{1}| = |\bar{G}_{2}| \rightarrow + \infty$). From (\ref{RhoAEstG1G2}) we note that $\rho_{12}^{ST} = (1/4)\mathbb{I}$ with $\mathbb{I}$ the identity operator. This behaviour is expected, since for high field intensities the atom-laser field interaction overwhelms the atom-atom interaction through the reservoir and each atom approximately interacts independently with the laser field.  Also notice that in the limit of high field intensity, the populations of the states $|1,1\rangle$ and $|0,0\rangle$ take their absolute maximum value of $1/4$, while that of the state $|1,-1\rangle$ takes its absolute minimum value of $1/4$.

\begin{figure}[htbp]
  \centering
  \subfloat[]{\label{ConcG1G2a11}\includegraphics[scale=0.4]{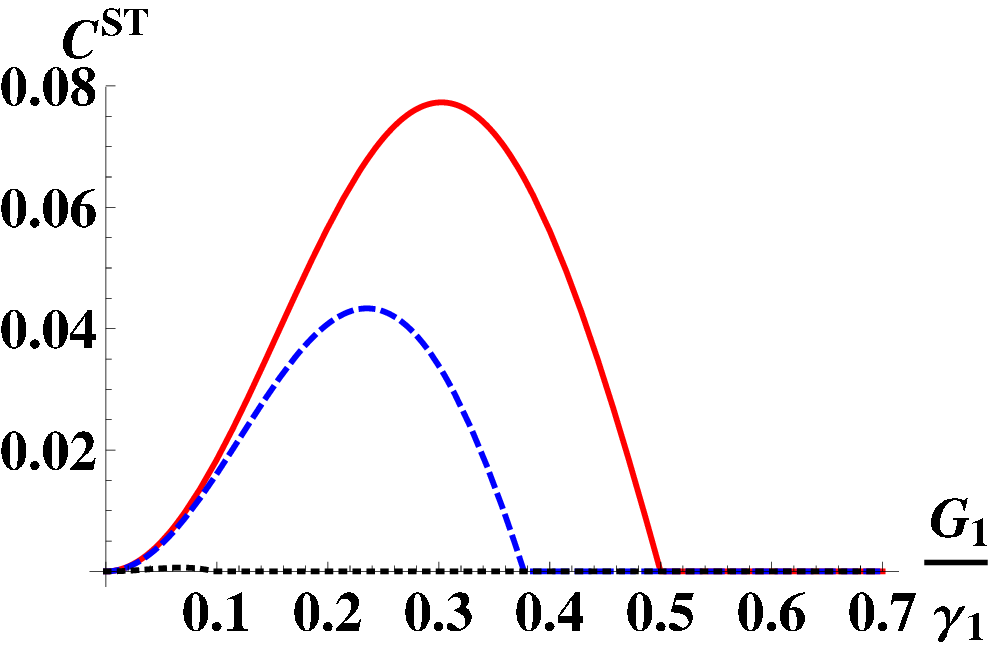}}\hspace{1cm}
  \subfloat[]{\label{ConcG1G2a10}\includegraphics[scale=0.4]{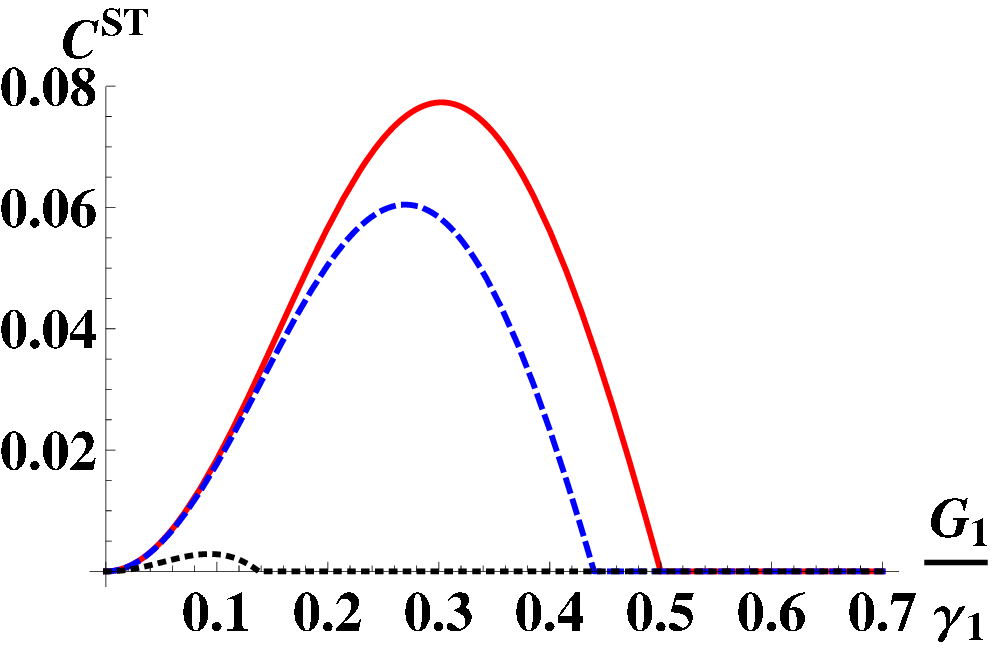}}\hspace{1cm}
  \subfloat[]{\label{CloseUpG1G2}\includegraphics[scale=0.4]{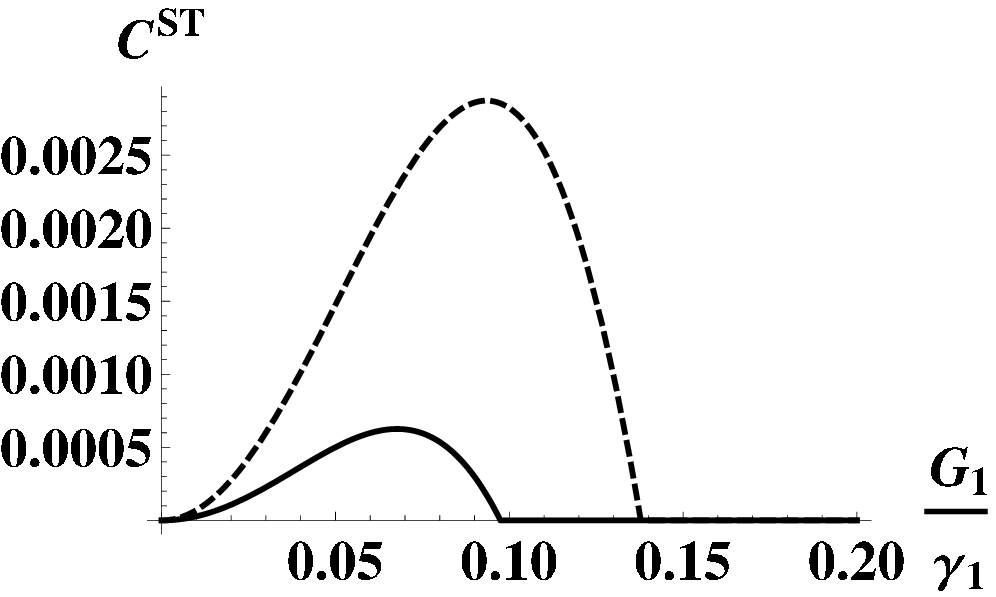}}\\
   \caption{(Color online) Case $G_{1} = G_{2}$: (\ref{ConcG1G2a11}) and (\ref{ConcG1G2a10}) show the steady-state concurrence $C^{ST}$ as a function of $\bar{G}_{1} = G_{1}/\gamma_{1}$ for the cases $d_{\bot}^2 = |\mathbf{d}_{01}|^{2}$ and $d_{\bot}^2 = 0$, respectively. Results are shown for distances between the two atoms equal to $\lambda_{A}/100$ (red-solid line), $\lambda_{A}/4$ (blue-dashed line), and $\lambda_{A}$ (black-dotted line) where $\lambda_{A}$ is the wavelength associated with the atomic transition. Figure (\ref{CloseUpG1G2}) shows a close-up of the cases $d_{\bot}^2 = |\mathbf{d}_{01}|^{2}$ (black-solid line) and $d_{\bot}^2 = 0$ (black-dashed line) when the distance between atoms is $\lambda_{A}$.}
  \label{concurrenciaCasosG1G2}
\end{figure}

\begin{figure}[htbp]
  \centering
  \subfloat[]{\label{ConcRho10a11G1G2}\includegraphics[scale=0.4]{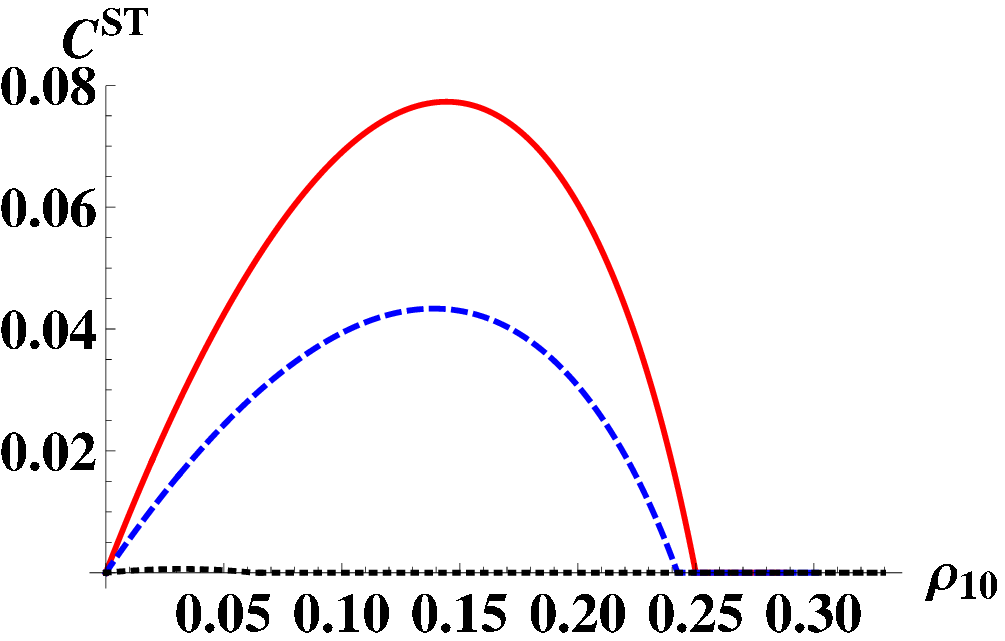}}\hspace{1cm}
  \subfloat[]{\label{ConcRho10a10G1G2}\includegraphics[scale=0.4]{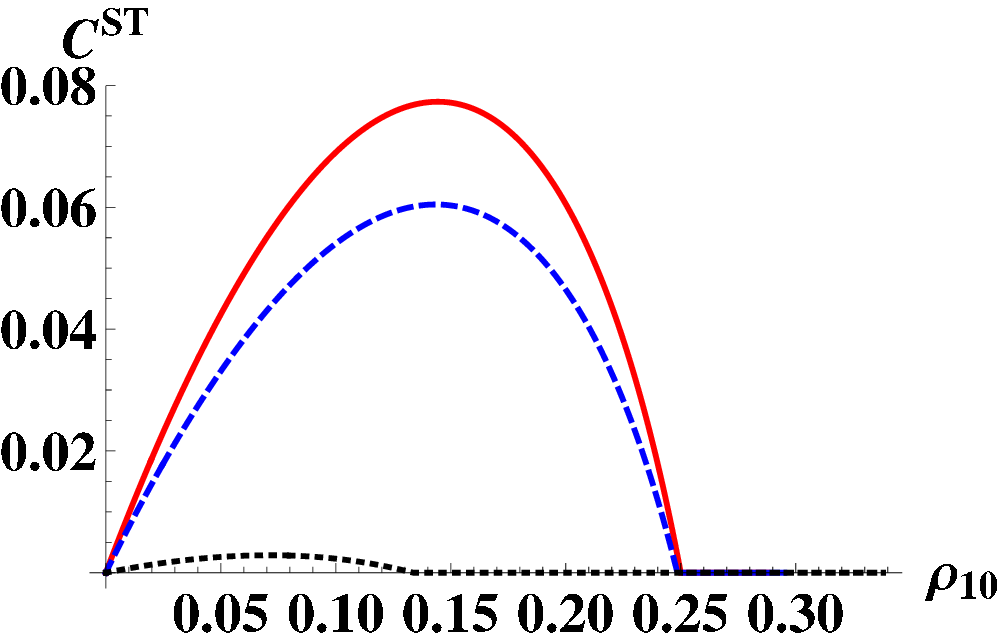}}\hspace{1cm}
  \subfloat[]{\label{ConcRho10G1G2closeup}\includegraphics[scale=0.4]{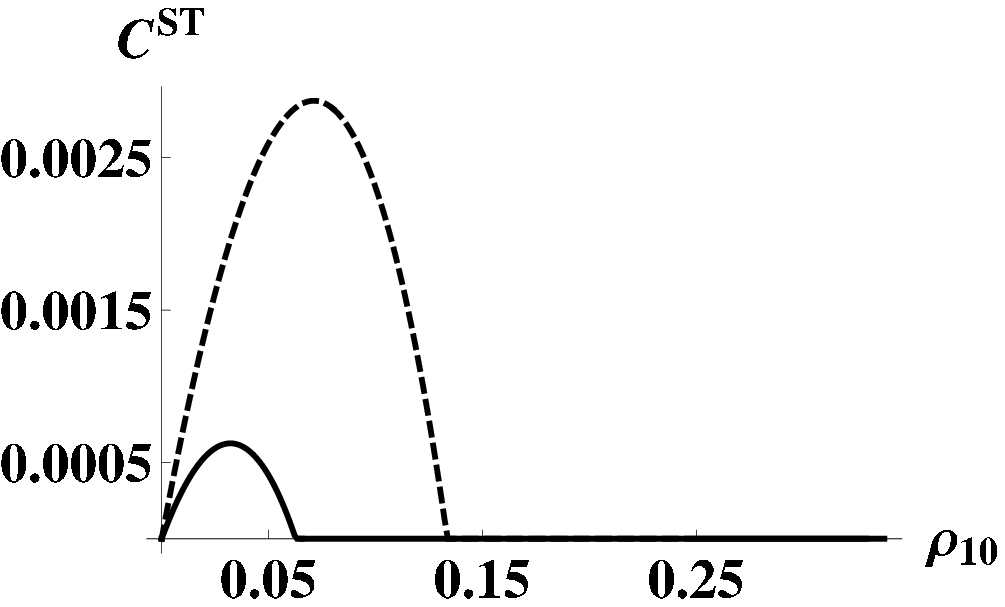}}\\
   \caption{(Color online) Case $G_{1} = G_{2}$: (\ref{ConcRho10a11G1G2}) and (\ref{ConcRho10a10G1G2}) show the steady-state concurrence $C^{ST}$ as a function of the population $\rho_{10}$ of the state $|1,0\rangle$ for the cases $d_{\bot}^2 = |\mathbf{d}_{01}|^{2}$ and $d_{\bot}^2 = 0$, respectively.  Results are shown for distances between the two atoms equal to $\lambda_{A}/100$ (red-solid line), $\lambda_{A}/4$ (blue-dashed line), and $\lambda_{A}$ (black-dotted line) where $\lambda_{A}$ is the wavelength associated with the atomic transition. Figure (\ref{ConcRho10G1G2closeup}) shows a close-up of the cases $d_{\bot}^2 = |\mathbf{d}_{01}|^{2}$ (black-solid line) and $d_{\bot}^2 = 0$ (black-dashed line) when the distance between atoms is $\lambda_{A}$.}
  \label{ConcRho10}
\end{figure}

\begin{figure}[htbp]
  \centering
  \vspace{2cm}
  \subfloat[]{\label{Contoura11G1G2}\includegraphics[scale=0.5]{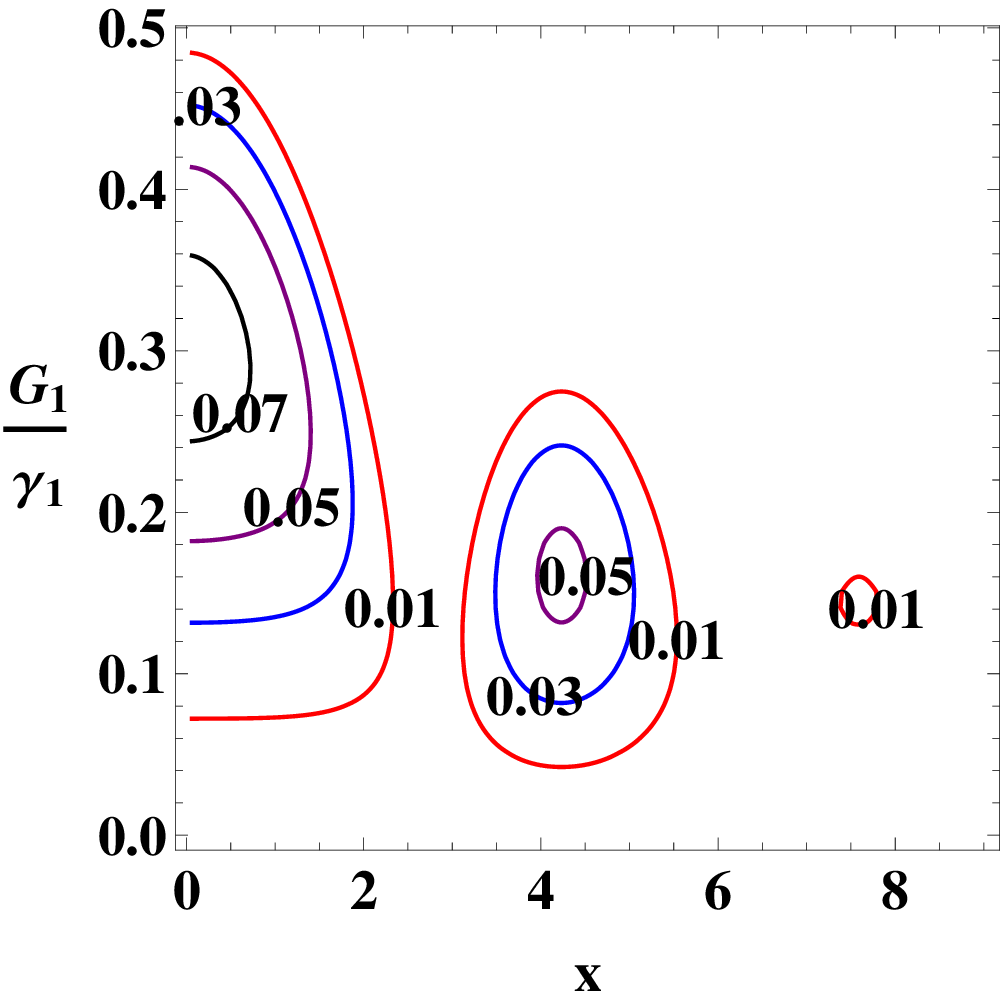}}\hspace{1cm}
  \subfloat[]{\label{Contoura10G1G2}\includegraphics[scale=0.5]{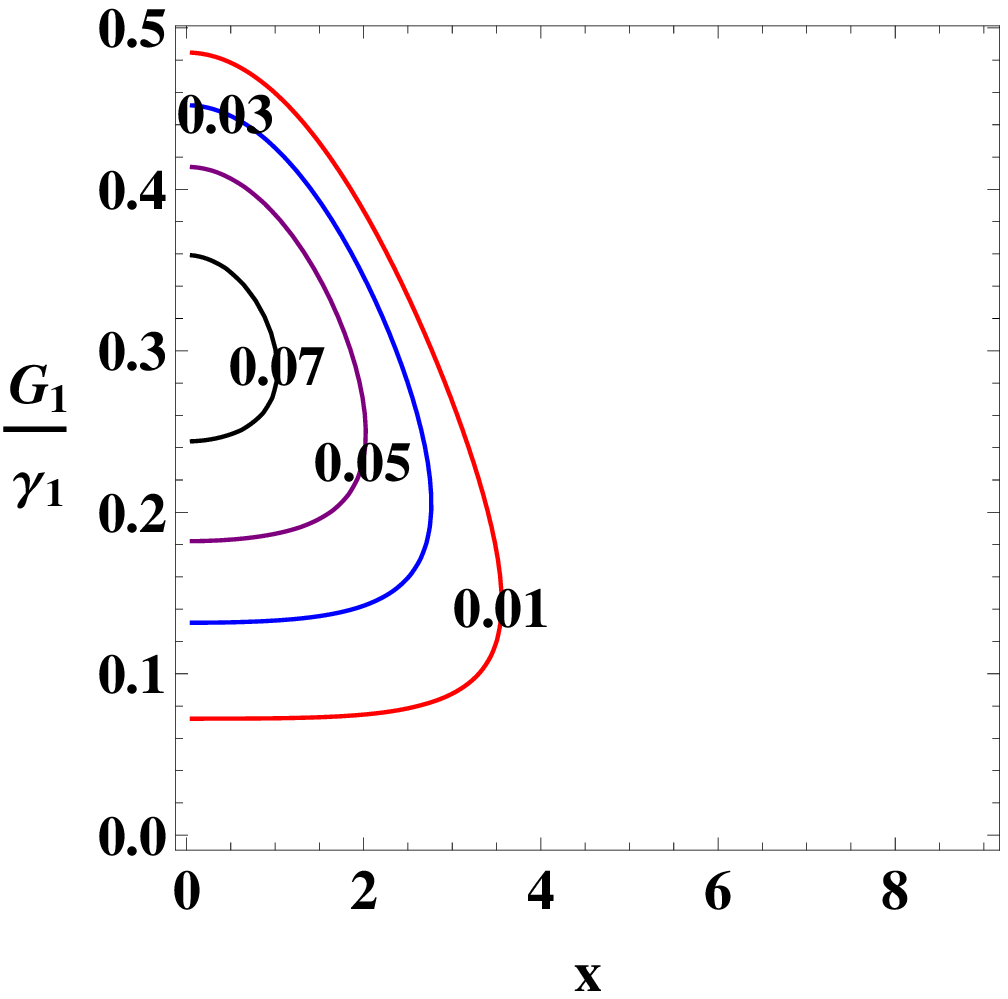}}\\
  \caption{(Color online) Case $G_{1} = G_{2}$: (\ref{Contoura11G1G2}) and (\ref{Contoura10G1G2}) show contour plots of the steady-state concurrence $C^{ST}$ as a function of $x=\omega_{A}|\mathbf{r}_{1}-\mathbf{r}_{2}|/c$ (horizontal axis)  and $\bar{G}_{1}$ (vertical axis) for the cases $d_{\bot}^2 = |\mathbf{d}_{01}|^{2}$ and $d_{\bot}^2 = 0$, respectively. Contours for $C^{ST} = 0.07, 0.05, 0.03, 0.01$ are drawn.}
  \label{contourG1G2}
\end{figure}

 We now discuss the degree of entanglement of $\rho_{12}^{ST}$ as measured by the concurrence $C^{ST}$, figure (\ref{concurrenciaCasosG1G2}). We observe that, as the intensity of the laser field increases ($\bar{G}_{1} = \bar{G}_{2}$ increases), $C^{ST}$ increases, takes on a maximum value, then decreases and finally dies abruptly, figures (\ref{concurrenciaCasosG1G2}) and (\ref{contourG1G2}). This \textit{sudden death} of entanglement as a function of the laser field intensity is found for most values of the distance between the atoms. Also notice that $C^{ST}$ takes its largest values when the distance between the atoms is less than a wavelength associated with the atomic transition and when the intensity of the laser field is small, although in this case the largest values of $C^{ST}$ are much smaller than those of the case $G_{2} = 0$, figures (\ref{concurrenciaCasosG20}) and (\ref{concurrenciaCasosG1G2}). It thus appears that significant entanglement between the two atoms cannot be generated when they interact only through the reservoir and if $G_{1} = G_{2}$ (recall that we are only considering the dissipative interaction). This is physically understandable when the the laser field intensity is large, since each atom interacts independently (approximately) with the laser field due to the fact that this interaction overwhelms the atom-atom interaction through the reservoir. This is confirmed by the weak dependence of $\rho_{12}^{ST}$ on $F_{12}$ in (\ref{RhoAEstG1G2}) and (\ref{RhoEstG1G2Est}). On the other extreme, when the laser field intensity is small, from (\ref{RhoG1G2c}) we find to second order in $\bar{G}_{1} = \bar{G}_{2}$ that the concurrence of $\rho_{12}^{ST}$ takes the form
 \begin{eqnarray}
 \label{ConcChico}
 C^{ST} \ = \ \mbox{max}\left\{ 0, \ -\frac{3}{2}F_{12}\langle 1,0|\rho_{12}^{ST}|1,0 \rangle \right\} \ .
 \end{eqnarray}
 Now $F_{12}$ is an oscillatory function bounded above by $2/3$ and bounded below by $-0.2237$ which tends to $0$ as the distance between atoms increases. Hence, the concurrence to second order in $\bar{G}_{1}$ in (\ref{ConcChico}) will be non-zero only for certain values of the distance between the two atoms. The equation in (\ref{ConcChico}) might lead one to think that the behaviour of the concurrence is determined to a great extent by the population of the state $|1,0\rangle$. From figure (\ref{ConcRho10}) we see that this is not the case, since the population still increases considerably after the concurrence has already vanished.

 Also observe that $C^{ST}$ decreases its value when the distance between the atoms increases, and that this decay with distance is slower when $d_{\bot}^2 = 0$ (that is, the dipoles of the two atoms are parallel to the line joining them), figures (\ref{concurrenciaCasosG1G2}) and (\ref{contourG1G2}). Finally notice that $C^{ST}$ exhibits an oscillatory behaviour when $d_{\bot}^2 = |\mathbf{d}_{01}|^{2}$ (that is, the dipoles of the two atoms are orthogonal to the line joining them), and that this phenomenon is smoothed out when $d_{\bot}^2 = 0$. Again, the reason for this is the function $F_{12}$ that has oscillations of greater amplitude when $d_{\bot}^2 = |\mathbf{d}_{01}|^{2}$.

As before, we will now study the behaviour of the quantum discord of $\rho_{12}^{ST}$ as a function of both the distance between the atoms and of $\bar{G}_{1}$. We first observe that in this case $D_{1}^{Q}(\rho_{12}^{ST}) = D_{2}^{Q}(\rho_{12}^{ST})$, since the two atoms are in equivalent positions, that is, both atoms are in a position where $\bar{G}_{1} = \bar{G}_{2}$. Note that this contrasts with the case considered in the previous section where the atoms were in non-equivalent positions and, as a result, $D_{1}^{Q}(\rho_{12}^{ST}) \not= D_{2}^{Q}(\rho_{12}^{ST})$.

We observed above that $\rho_{12}^{ST} = (1/4)\mathbb{I}$ in the limit $\bar{G}_{1} \rightarrow + \infty$. Hence, we conclude from (\ref{QMI}), (\ref{Discordia1}), and (\ref{Discordia2}) that $\rho_{12}^{ST}$ has no correlations in this limit, that is, $C^{ST}=I(\rho_{12}^{ST}) = C_{j}^{cl}(\rho_{12}^{ST}) = D_{j}^{Q}(\rho_{12}^{ST}) = 0$ ($j=1,2$). This is explained intuitively by the fact that in the high laser field intensity limit (that is, $|\bar{G}_{1}| \rightarrow + \infty$) each atom interacts independently (approximately) with the laser field and the atoms are in different positions (if the atoms were exactly at the same position, the quantum discord would have to be calculated from (\ref{RhoEstG1G2Est})).

\begin{figure}[htbp]
  \centering
  \subfloat[]{\label{QDVa11G1G2}\includegraphics[scale=0.5]{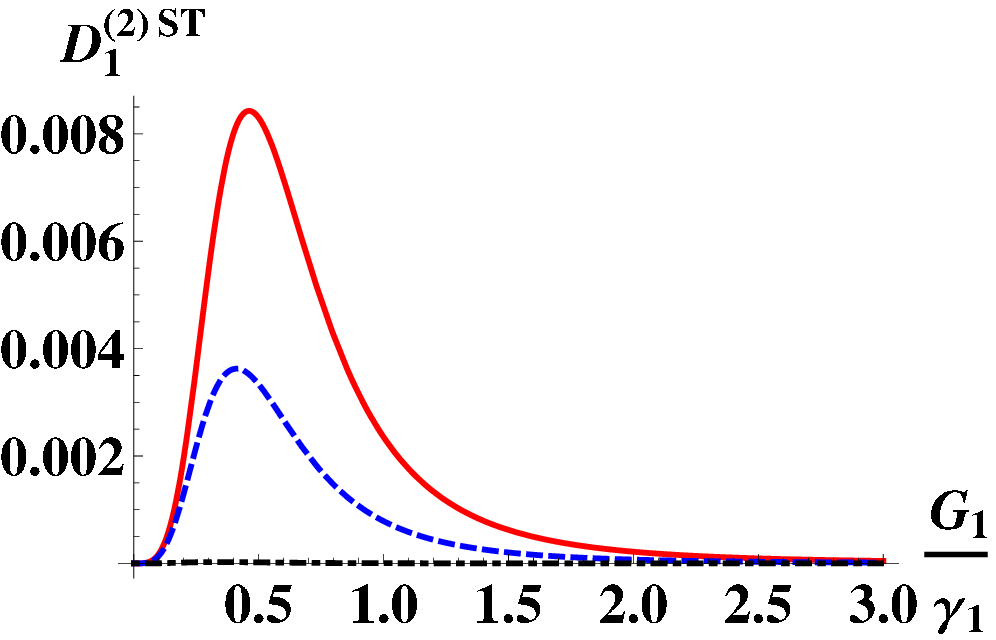}}\hspace{1cm}
  \subfloat[]{\label{QDVa10G1G2}\includegraphics[scale=0.5]{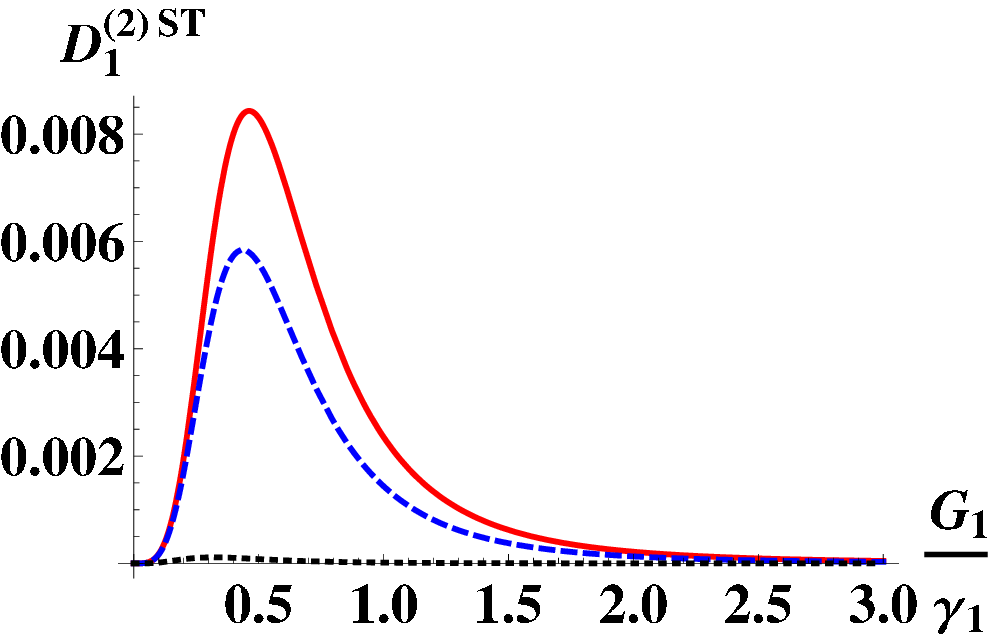}}\\
   \caption{(Color online) Case $G_{1} = G_{2}$: (\ref{QDVa11G1G2}) and (\ref{QDVa10G1G2}) show the steady-state geometric discord $D_{1}^{(2)ST}$ as a function of $\bar{G}_{1} = G_{1}/\gamma_{1}$ for the cases $d_{\bot}^2 = |\mathbf{d}_{01}|^{2}$ and $d_{\bot}^2 = 0$, respectively.  Results are shown for distances between the two atoms equal to $\lambda_{A}/100$ (red-solid line), $\lambda_{A}/4$ (blue-dashed line), and $\lambda_{A}$ (black-dotted line), where $\lambda_{A}$ is the wavelength associated with the atomic transition.}
  \label{QDVG1G22}
\end{figure}

\begin{figure}[htbp]
  \centering
  \vspace{1.5cm}
  \subfloat[]{\label{QDVa11G1G2CP}\includegraphics[scale=0.5]{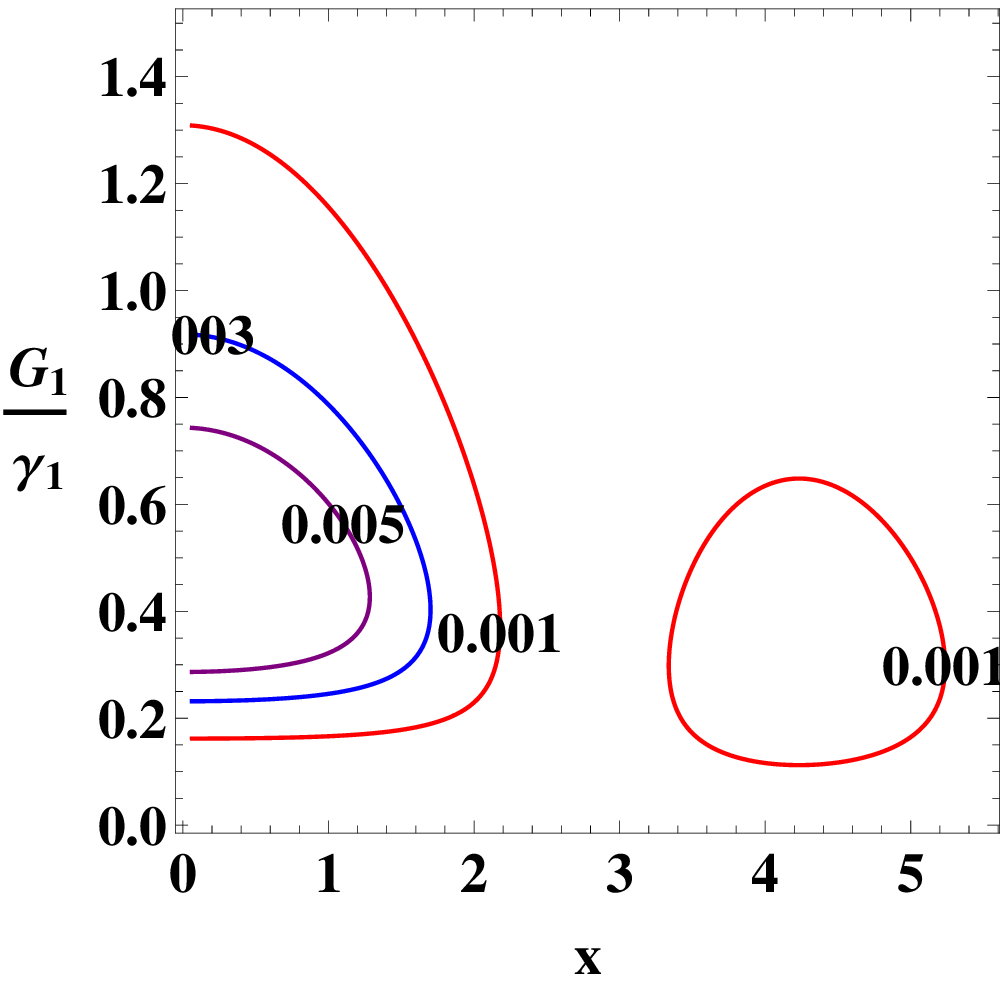}}\hspace{1cm}\\
   \caption{(Color online) Case $G_{1} = G_{2}$: (\ref{QDVa11G1G2CP}) shows a contour plot of the geometric discord $D_{1}^{(2)ST}$ of $\rho_{12}^{ST}$ as a function of $x=\omega_{A}|\mathbf{r}_{1}-\mathbf{r}_{2}|/c$ (horizontal axis)  and $\bar{G}_{1}$ (vertical axis) for the case $d_{\bot}^2 = |\mathbf{d}_{01}|^{2}$. Contours for $D_{1}^{(2)ST} = 0.005, 0.003, 0.001$ are drawn.}
  \label{QDVG1G2contour}
\end{figure}

Using (\ref{DiscordiaQVedral}) we can study the geometric discord $D_{1}^{(2)}(\rho_{12}^{ST})$ in the general case (\ref{RhoAEstG1G2}). From figure (\ref{QDVG1G22}) we observe that the $D_{1}^{(2)}(\rho_{12}^{ST})$ tends to zero as $\bar{G}_{1}$ increases, a result expected due to the fact that $D_{1}^{Q}(\rho_{12}^{ST}) = 0$  in the limit $|\bar{G}_{1}| \rightarrow + \infty $. Moreover, notice that $D_{1}^{(2)}(\rho_{12}^{ST})$ decreases more slowly as the distance between atoms increases when $d_{\bot}^2 = 0$. This is clearly seen in figure (\ref{QDVG1G2contour}) where it is also observed that $D_{1}^{(2)}(\rho_{12}^{ST})$ has oscillations of a greater magnitude when $d_{\bot}^2 = |\mathbf{d}_{01}|^{2}$. As before, the oscillatory behaviour of $D_{1}^{(2)}(\rho_{12}^{ST})$ is due to $F_{12}$, a result which is known due to the analytic but unwieldy expression obtained from the use of (\ref{DiscordiaQVedral}).

From the results presented above it thus appears that driving the two atoms by the laser field is ineffective in creating quantum correlations, since both the entanglement and the quantum discord are considerably smaller when compared to the case where only one atom is being driven (the case $\bar{G}_{2} = 0$ considered above). This decrease of correlations is even more dramatic in the limit of high laser field intensity ($|\bar{G}_{1}| \rightarrow + \infty$), because it was shown that all correlations disappear in the case $\bar{G}_{1} = \bar{G}_{2}$.

\begin{figure}[htbp]
  \centering
  \subfloat[]{\label{SLa11G1G2}\includegraphics[scale=0.5]{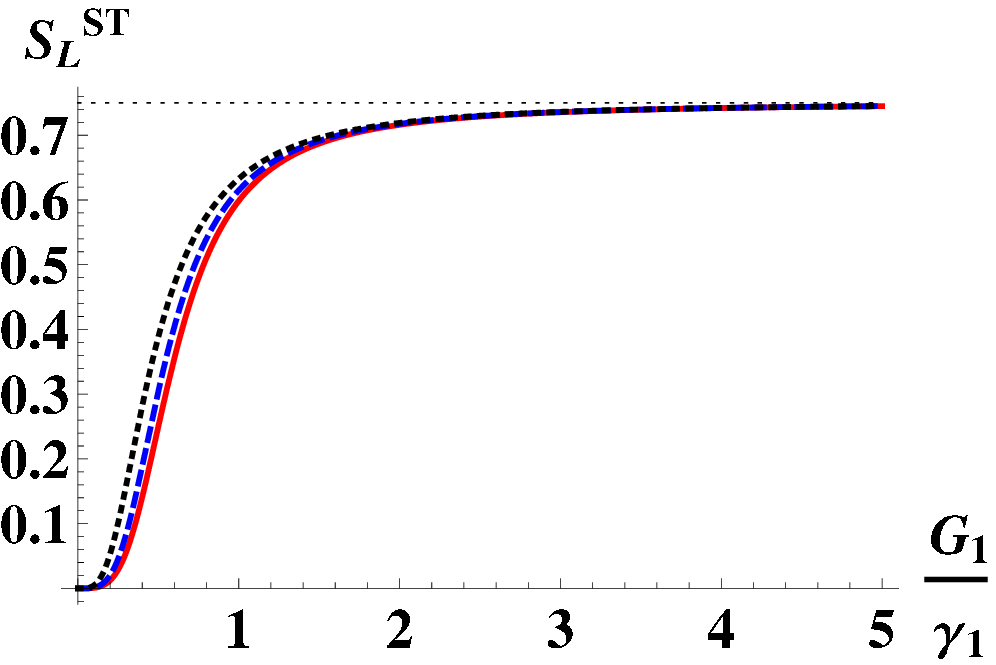}}\hspace{1cm}
  \subfloat[]{\label{SLa10G1G2}\includegraphics[scale=0.5]{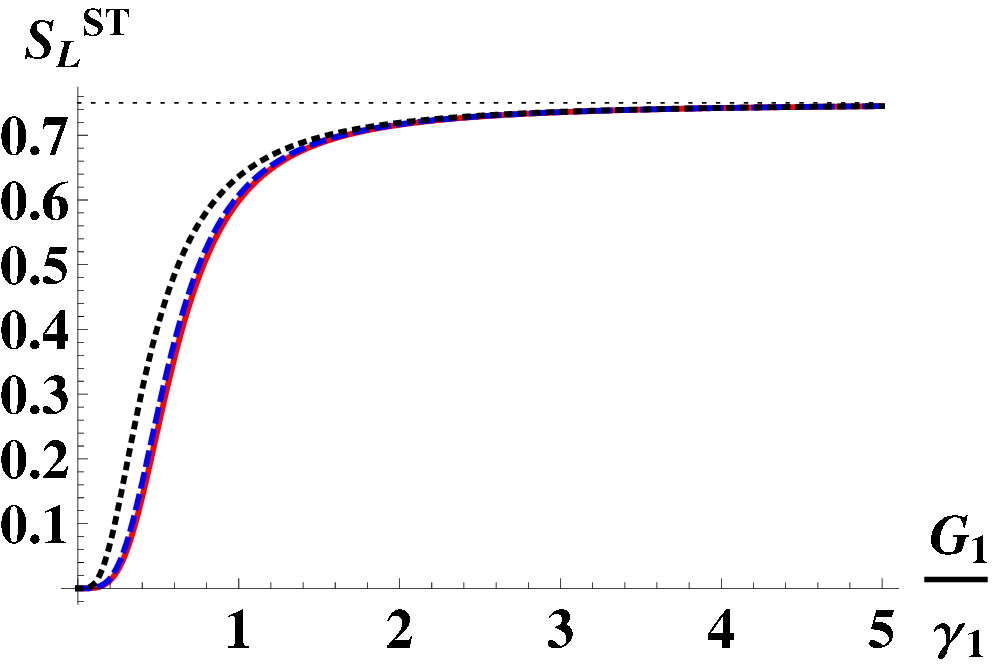}}\\
   \caption{(Color online) Case $G_{1} = G_{2}$: (\ref{SLa11G1G2}) and (\ref{SLa10G1G2}) show the steady-state linear entropy $S_{L}^{ST}$ as a function of $\bar{G}_{1} = G_{1}/\gamma_{1}$ for the cases $d_{\bot}^2 = |\mathbf{d}_{01}|^{2}$ and $d_{\bot}^2 = 0$, respectively.  Results are shown for distances between the two atoms equal to $\lambda_{A}/100$ (red-solid line), $\lambda_{A}/4$ (blue-dashed line), and $\lambda_{A}$ (black-dotted line) where $\lambda_{A}$ is the wavelength associated with the atomic transition.}
  \label{SLG1G2}
\end{figure}

We now turn to discuss the degree of mixed-ness of $\rho_{12}^{ST}$, figure (\ref{SLG1G2}). As the laser field intensity increases ($\bar{G}_{1}$ grows), $S_{L}^{ST}$ increases until it acquires an asymptotic value of $3/4$. In fact, we observed above that $\rho_{12}^{ST} = (1/4)\mathbb{I}$ when $|\bar{G}_{1}| \rightarrow + \infty$. Hence, $\rho_{12}^{ST}$ tends to a maximum mixed state when the laser field intensity increases. Again, this behaviour is expected, since for high field intensities each atom approximately interacts independently with the laser field. Finally, notice that $S_{L}^{ST}$ depends weakly on the distance between the two atoms and on the value of $d_{\bot}^2$, figure (\ref{SLG1G2}). This is expected, since it was observed above that $\rho_{12}^{ST}$ depends weakly on $F_{12}$.

To end this Section we discuss the relation of the steady-state entanglement between the two atoms as measured by the concurrence $C^{ST}$ with the steady-state degree of mixed-ness as measured by the linear entropy $S_{L}^{ST}$. We observe that, as $S_{L}^{ST}$ becomes larger, the degree of entanglement increases until a maximum value. After this the degree of entanglement decays to zero, figure (\ref{ConcSLG1G2}). This behaviour corresponds to the fact that, as the intensity of the electric field increases, the state of the two atoms becomes entangled and mixed due to the exchange of spontaneously emitted photons. Nevertheless, once the intensity of the electric field is strong enough to dominate over the interaction between the atoms by means of the reservoir, the entanglement begins to decrease while the state of the system still becomes more mixed. Figure (\ref{ConcSLG1G2}) also illustrates the known facts that a mixed state of two qubits cannot contain an arbitrary amount of entanglement, and that the more mixed the state becomes, the less entanglement it has \cite{MaxEntanglement}.

\section{Conclusions}

In this article we considered two two-level atoms (qubits) fixed at different positions, driven by a monochromatic laser field, and interacting collectively with the vacuum electromagnetic field. A Born-Markov-secular master equation was used to describe the dynamics of the two atoms and their steady-state was studied for two configurations. In the first, one atom was located in a region where the driving electric field is zero while the other was in a position where it is not zero. In the second configuration both atoms were located at different positions where the driving electric field was the same. We  refer to the first case as $G_{2} = 0$ and to the second case as $G_{1} = G_{2}$.

We neglected the coherent dipole-dipole interaction between the two atoms that results from the collective interaction with the vacuum reservoir, and we only kept the dissipative atom-atom interaction through the reservoir. This led us to a model of interest in itself: two driven qubits at different fixed positions interacting with a vacuum reservoir whose Born-Markov-secular master equation is (\ref{EcMaestraCFIP}). Furthermore, this model allowed us to identify and understand which effects are due solely to the dissipative interaction between the two atoms. For this model the steady-state density operator exists in an interaction picture only when the laser field frequency is resonant with the atomic transition frequency (the resonance condition). The analytic steady-state density operator $\rho_{12}^{ST}$ was obtained for both configurations of the atoms. This allowed us to study several limiting cases that resulted in a much deeper understanding of the steady-state of the system and the mechanisms responsible for the formation of correlations between the two atoms.

It was shown for both configurations that any initial state of the two atoms tends to the respective $\rho_{12}^{ST}$ under the resonance condition. These steady-state density operators are entangled and have non-zero left and right quantum discord if the laser field intensity is not very high. High laser field intensities turn $\rho_{12}^{ST}$ into a maximum mixed non-correlated state $(1/4)\mathbb{I}$ in the case $G_{1} = G_{2}$, and into a separable $X$-state that has non-zero right quantum discord (measurements are made on the un-driven atom) in the case $G_{2} = 0$.

It was found that driving both atoms with the laser field (case $G_{1} = G_{2}$) is not efficient for the generation of steady-state quantum correlations (entanglement and quantum discord), since all steady-state quantum correlations decreased their values in the case $G_{1} = G_{2}$ and only the steady-state degree of mixed-ness was increased. In both configurations, steady-state quantum correlations (entanglement and quantum discord) were maximized for weak laser field intensities. Increasing the laser field intensity led in all cases to steady-state entanglement sudden death, but the quantum discord still survived and, in some cases, did not decrease to zero. We also found that the distance between atoms is fundamental for the build up of steady-state quantum correlations. The farther the atoms are apart, the less quantum correlations they can have. This is physically expected since the atoms can only interact by interchanging spontaneously emitted photons and this interaction is less effective the farther the atoms are apart.

In general, all correlations exhibited an oscillatory behaviour that tended to zero as the distance between the two atoms was increased. The oscillations were more marked when the dipoles of the atoms were orthogonal to the line joining them and the decay with distance was slower when the dipoles were parallel to the aforementioned line. It was found that this behaviour was due to the function $F_{12}$ that comes about from the collective interaction of the atoms with the reservoir.

The configurations considered also allowed us to study the behaviour of the quantum discord when the atoms are in non-equivalent and equivalent positions. In particular, a method to calculate numerically the left and right quantum discords was presented, and it was used in the case of a high laser field intensity. This allowed a direct comparison with the geometric measure of quantum discord. It was found that the geometric discord reproduced the behaviour of the quantum discord and there was good qualitative agreement between both. The geometric measure was also used to study the cases where the laser field intensity was not high and showed that the quantum discord survives long after the steady-state entanglement has disappeared.

Due to the non-equivalent position of the atoms in the $G_{2}=0$ configuration, it was found that the right geometric discord $D_{2}^{(2)}(\rho_{12}^{ST})$ (obtained when measurements are made on the un-driven atom) is very different from the left geometric discord $D_{1}^{(2)}(\rho_{12}^{ST})$ (obtained when measurements are made on the driven atom). $D_{2}^{(2)}(\rho_{12}^{ST})$ rapidly acquires a stationary non-negligible value as the intensity of the laser field is increased and has an oscillatory behaviour as a function of the distance between atoms with discrete zeros determined by the function $F_{12}$. On the other hand, $D_{1}^{(2)}(\rho_{12}^{ST})$ tends to zero as the laser field intensity is increased. Hence, the information that cannot be extracted from measurements only on the driven/un-driven atom is very different in the $G_{2} = 0$ configuration. This is not the case of the $G_{1} = G_{2}$ configuration where the atoms are in equivalent positions.

Analytic expressions for the steady-state populations of the eigenstates of the free Hamiltonian of the two atoms where derived and we established their simple behaviour in the limits of high and low laser field intensities. We determined that the populations of the antisymmetric $|0,0\rangle$ and symmetric $|1,0\rangle$ states explained the behaviour of the concurrence to a great extent in both configurations. This was much more dramatic in the case $G_{2} = 0$ were we showed that the concurrence increased/decreased with the population of the antisymmetric state almost linearly when the distance between the two atoms was less than a wavelength of the atomic transition.

In this work we have characterized the quantum correlations between the two atoms when they interact only through the dissipative collective interaction. It is very interesting to study the effects of the dipole-dipole interaction between the two atoms, not taken into consideration in this work, to see what new dynamics will be introduced or if the system will behave similarly. This will be investigated in future work.


\begin{thebibliography}{30}
\bibitem{notacion} Notice that we have chosen a slightly different notation than the one in \cite{tanas}. In particular, that article uses the component of $\mathbf{d}_{01}$ along $\mathbf{r}_{1} - \mathbf{r}_{2}$ and $(3/2)F(|\mathbf{r}_{1}-\mathbf{r}_{2}|)$ instead of $d_{\bot}$ and $F(|\mathbf{r}_{1}-\mathbf{r}_{2}|)$, respectively.
\end{thebibliography}
\end{document}